\pdfoutput=1
\documentclass[11pt,twoside]{report}
\usepackage{longtable,tabularx,array,url,ifthen,multicol}
\usepackage{graphpap}
\usepackage[fit]{truncate}

\newcommand{\BLKP}{
  \ifthenelse{\isodd{\value{page}}}{\relax}{\mbox{}\thispagestyle{empty}\newpage}}
\newcommand{\CLDP}{\newpage 
  \ifthenelse{\isodd{\value{page}}}{\relax}{\mbox{}\thispagestyle{empty}\newpage}}
\usepackage{geometry}
\usepackage{fancyhdr}
\newcommand{\ARTauthor}{~}
\newcommand{\ARTtitle}{~}
\usepackage{pdfpages}
\newenvironment{papers}{\clearpage}{\clearpage}
\newcommand*\coltoctitle[1]{\def\CTIT{#1}}
\newcommand*\coltocauthor[1]{\def\CAUT{#1}}
\makeatletter
\renewcommand{\l@section}{\@dottedtocline{1}{2em}{0em}}
\renewcommand{\@dotsep}{1000}
\makeatother
\newcommand{\Includeart}[4][]{%
\def\AAA{#2}\def\TTT{#3}%
\renewcommand{\ARTauthor}{~}
\renewcommand{\ARTtitle}{~}
   \includepdf[
               pages=1,
               noautoscale,
               pagecommand={\pagestyle{fancy}},
               offset=0mm 0mm,
               addtotoc={1, subsubsection, 3, ~,  S#4},
               trim=19mm 21mm 19mm 27mm, clip]
               {#4.pdf}%
\addtocontents{toc}{\protect\contentsline{chapter}{\textbf{\TTT}}{\textbf{\pageref{S#4}}}}
\addtocontents{toc}{\protect\medskip}
\addtocontents{toc}{\protect\contentsline{section}{\AAA}{~}}
\ifthenelse{\equal{#1}{OnePage}}{
                                }{
  \renewcommand{\ARTauthor}{\truncate{.9\linewidth}{\AAA}}
  \renewcommand{\ARTtitle}{\truncate{.9\linewidth}{\TTT}}
    \includepdf[
                pages=2-,
                noautoscale,
                pagecommand={\pagestyle{fancy}},
                offset=0mm 0mm,
                trim=19mm 21mm 19mm 27mm, clip]
                {#4.pdf}}
}
\begin{document}
\pagestyle{empty}
\setlength{\fboxsep}{0pt}
\setlength{\fboxrule}{0.04pt}
\thispagestyle{empty}
\setlength{\unitlength}{1mm}
\begin{picture}(0.001,0.001)
\put(110,3){DESY--PROC--2007--02}
\put(-20,-250){\includegraphics[bb=0 0 550 652,width=18cm]{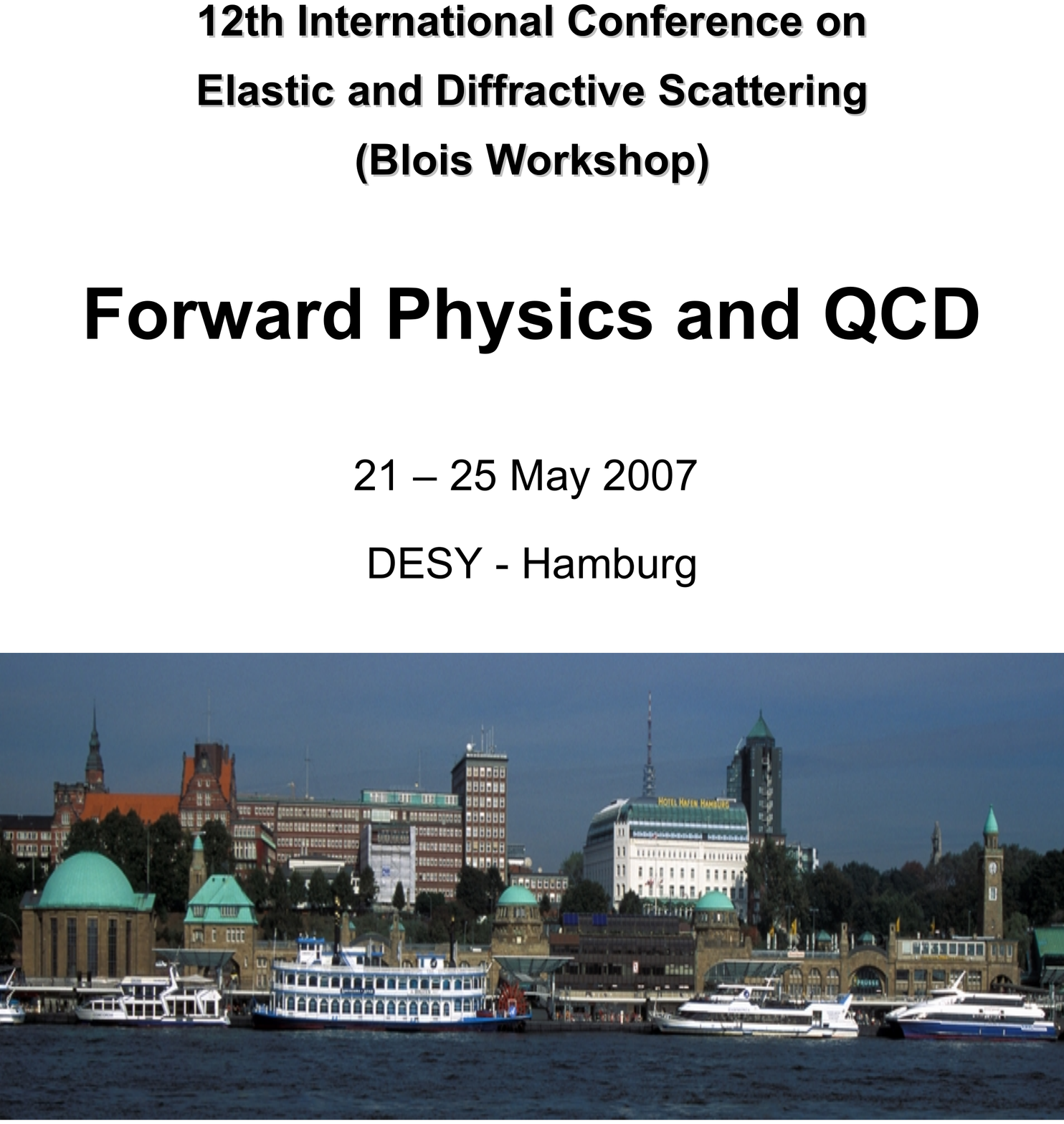}}
\end{picture}
\CLDP
{\parindent 0mm
\vspace*{2cm}
{\sffamily\LARGE
Proceedings of the \\
\Huge
12th International Conference on \\ Elastic and Diffractive Scattering\\
Forward Physics and QCD
\vspace*{1cm}

EDS\,2007
}

\vspace*{7mm}

{\Large\sffamily
May 21-25, 2007\\[0.5ex]
DESY, Hamburg, Germany}

\vfill

{\Large
Editors: J. Bartels, K. Borras, M. Diehl and H. Jung\\

\vspace{1cm}

Verlag Deutsches Elektronen - Synchrotron
}
\newpage

\vspace*{5cm}
{\bfseries\large Impressum}

\vspace*{6ex}
{\bfseries
Proceedings of the 
12th International Conference on Elastic and Diffractive Scattering\\
Forward Physics and QCD
\,2007
}

\vspace*{3ex}
 
Conference homepage\\
\verb$http://www.desy.de/~eds07/$
\\[2ex]
Slides at\\
\verb$https://indico.desy.de/conferenceDisplay.py?confId=372$
\vfill

\parbox{12cm}{
The copyright is governed by the Creative Commons agreement, which allows
for free use and distribution of the articles for non-commertial activity,
as long as the title, the authors' names and the place of the original
are referenced.}

\vspace*{3ex}

Editors: J. Bartels, K. Borras, M. Diehl and H. Jung\\
Cover Photo: \copyright Rainer Mankel, DESY\\
Photo of Participants: Marta Mayer (DESY, Hamburg) \\
December 2007\\
DESY-PROC-2007-02 \\
ISBN 978-3-935702-24-9\\
ISSN 1435-8077\\

\vspace*{3ex}

Published by\\
Deutsches Elektronen - Synchrotron, DESY\\
Notkestra\ss e 85\\
22607 Hamburg\\
Germany

}

\newpage
\begin{flushleft} 
\mbox{}\\[1cm]
{\bfseries\large Organizing Committee:}\\[3mm]
J. Bartels, K. Borras, M. Diehl and H. Jung\\[6mm]
{\bfseries\large Conveners:}\\[3mm]
{\bfseries lepton-proton collisions:} 
X. Janssen (Universite Libre de Bruxelles) and L. Motyka (DESY)\\ 	 	 
{\bfseries $pp$ and $p\bar{p}$ collisions:} 
S. Gieseke (Karlsruhe), M. Grothe (Wisconsin and Torino) and C. Royon (Saclay) \\		
{\bfseries heavy-ion collisions:}
K. Tuchin (Iowa State and RIKEN/BNL) and S. White (BNL) \\
{\bfseries opportunities at future colliders:}
U. Klein (Liverpool) 		\\
{\bfseries cosmic rays and astroparticle physics:}
R. Engel (Karlsruhe) 		\\
{\bfseries theoretical developments in high-energy QCD:}
V. Khoze (Durham) and A. Sabio Vera (CERN)\\[6mm]
{\bfseries\large Advisory Committee:}\\[3mm]
M.~Albrow~(Fermilab),
S.~J.~Brodsky~(SLAC), 
J.~Dainton~(Liverpool), 
A.~De~Roeck~(CERN), 
Yu.~Dokshitzer~(Paris), 
A.~Donnachie~(Manchester), 
K.~Goulianos~(New York), 
M.~Haguenauer~(Palaiseau), 
E.~Levin~(Tel Aviv), 
A.~Levy~(Tel Aviv), 
L.~Lipatov~(St. Petersburg/Hamburg), 
A.~D.~Martin~(Durham), 
L.~McLerran~(Brookhaven), 
J.~Mnich~(DESY), 
O.~Nachtmann~(Heidelberg), 
B.~Nicolescu~(Paris), 
R.~Peschanski~(Saclay), 
A.~Santoro~(Rio de Janeiro), 
M.~Strikman~(PennState), 
J.~Tr\^{a}n~Thanh~V\^{a}n~(Orsay), 
M.~Rijssenbeek~(Stony Brook), 
A.~Vorobiev~(St. Petersburg), 
A.~White~(Argonne),
G.~Wolf~(DESY) \\[6mm]
{\bfseries\large Permanent Committee of the EDS Conferences:}\\[3mm]
B.~Aubert~(Annecy),
G.~Belletini~(Pisa), 
D.~Denegri~(Saclay), 
G.~Giacomelli~(Bologna), 
A.~Krisch~(Michigan), 
V.~Kundrat~(Prague), 
N.~N.~Khuri~(New York), 
A.~Martin~(CERN), 
G.~Matthiae~(Rome), 
B.~Nicolescu~(Paris),
R.~Orava~(Helsinki),
J.~Orear~(Cornell),
J.~Peoples~(Fermilab),
E.~Predazzi~(Torino),
C.-I.~Tan~(Providence),
J.~Tr\^{a}n~Thanh~V\^{a}n~(Orsay), 
C.~N.~Yang~(Stony Brook) \\[6mm]
{\bfseries\large Supported by:}\\[3mm]
Deutsches Elektronen-Synchroton DESY\\
Deutsche Forschungsgemeinschaft \\
SFB 676 (Particles, Strings and the Early Universe)
\end{flushleft}

\pagestyle{plain}
\pagenumbering{roman}
\setcounter{page}{3}

\begin{center}
\mbox{}\\[5mm]
{\bfseries\Large Preface }\\[1cm]
\end{center}

The {\it 12th International Conference on Elastic and Diffractive Scattering}, 
held  on 21 -- 25 May 2007 at DESY in Hamburg, had the subtitle {\it
Forward Physics and QCD}.  Besides discussing in detail elastic and diffractive
scattering in $ep$ and $pp$ collisions, emphasis was put on the QCD description
of diffractive processes, which becomes more and more important in the
preparation for the start of the highest-energy proton-proton collider 
ever build, the LHC.
Especially the issue of the underlying event and multi-parton interaction, which
is relevant not only for QCD processes but also for all searches for new
phenomena, was discussed in detail.

With more than 100 participants this workshop showed the increasing interest and
importance of this area in high energy physics. Many young
researchers participated in this conference, showing that this is an
active, challenging and exciting field.

Unfortunately, not all presentations during the workshop appear as a writeup in
these proceedings: K. Below and  G. Marchesini were not able 
to deliver a written version of their contribution. All the talks with
transparcenies are available at: 
\verb$https://indico.desy.de/conferenceDisplay.py?confId=372$

We thank all the participants for making this conference so interesting and
lively. We also thank 
A. Grabowksy, S. Platz and L. Seskute for their continuous help and support during
all the meeting weeks. 
We are grateful to the DESY directorate for financial support
of this workshop and for the hospitality which they extended to all
the participants of the workshop. 

\begin{flushleft} 
{The Organizing Committee:}\\
J. Bartels, K. Borras, M. Diehl and H. Jung
\end{flushleft}

\begin{picture}(0.001,0.001)
\put(-5,-90){\includegraphics[width=17.5cm]{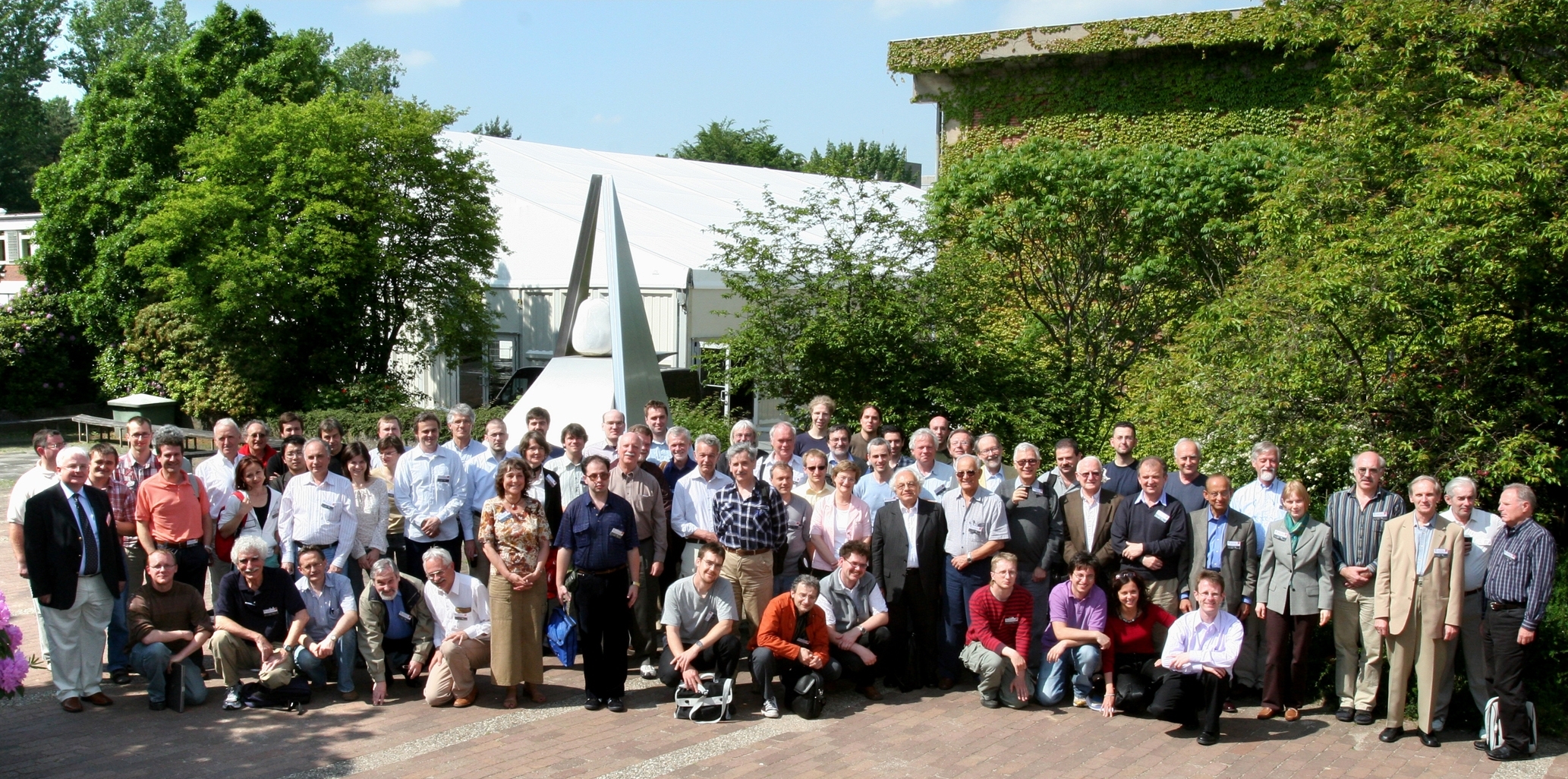}}
\end{picture}



\tableofcontents 
\CLDP
\pagestyle{fancy}
\setcounter{page}{1}
\pagenumbering{arabic}

\part{Lepton-Proton Collisions} 
\label{ep} 
\newpage

\begin{papers} 

\coltoctitle{Exclusive vector meson electroproduction at HERA} 
\coltocauthor{A Levy}
\Includeart{\CAUT}{\CTIT}{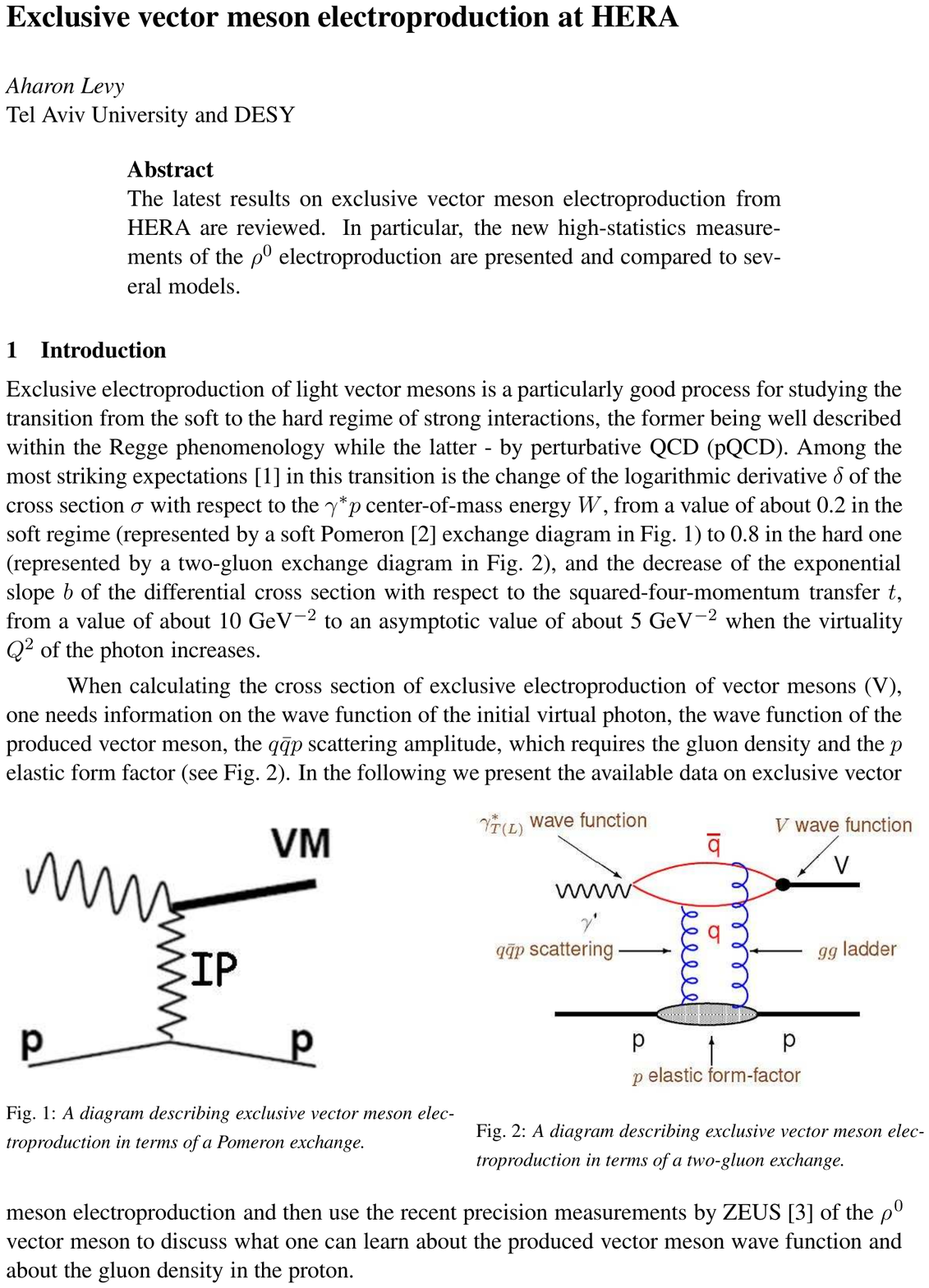}

\coltoctitle{Exclusive Processes at HERMES} 
\coltocauthor{R Fabbri}
\Includeart{\CAUT}{\CTIT}{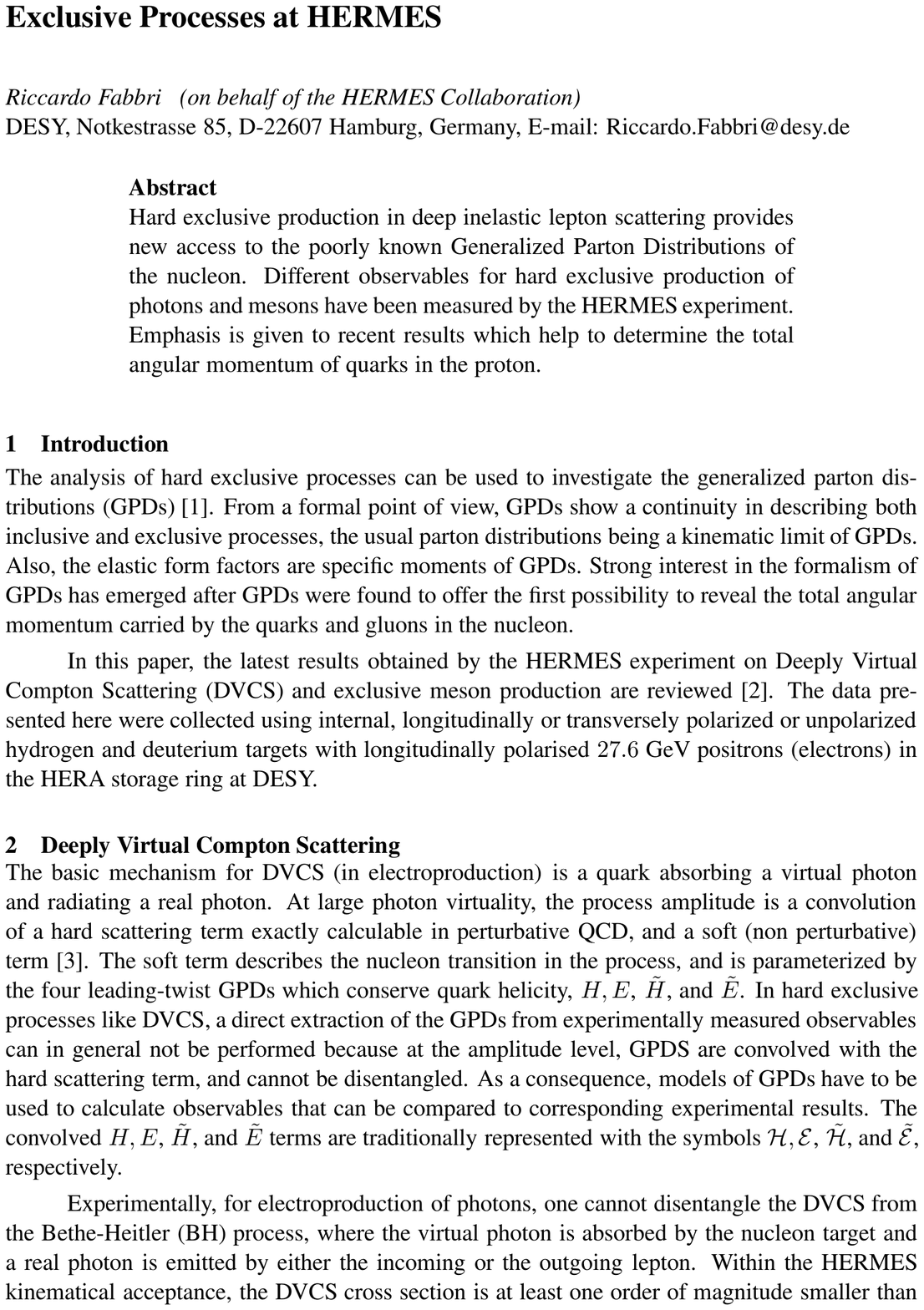} 

\coltoctitle{Fitting DVCS at NLO and beyond} 
\coltocauthor{K Kumeri\v{c}ki,
 D M\"{u}ller,
 K Passek-Kumeri\v{c}ki}
\Includeart{\CAUT}{\CTIT}{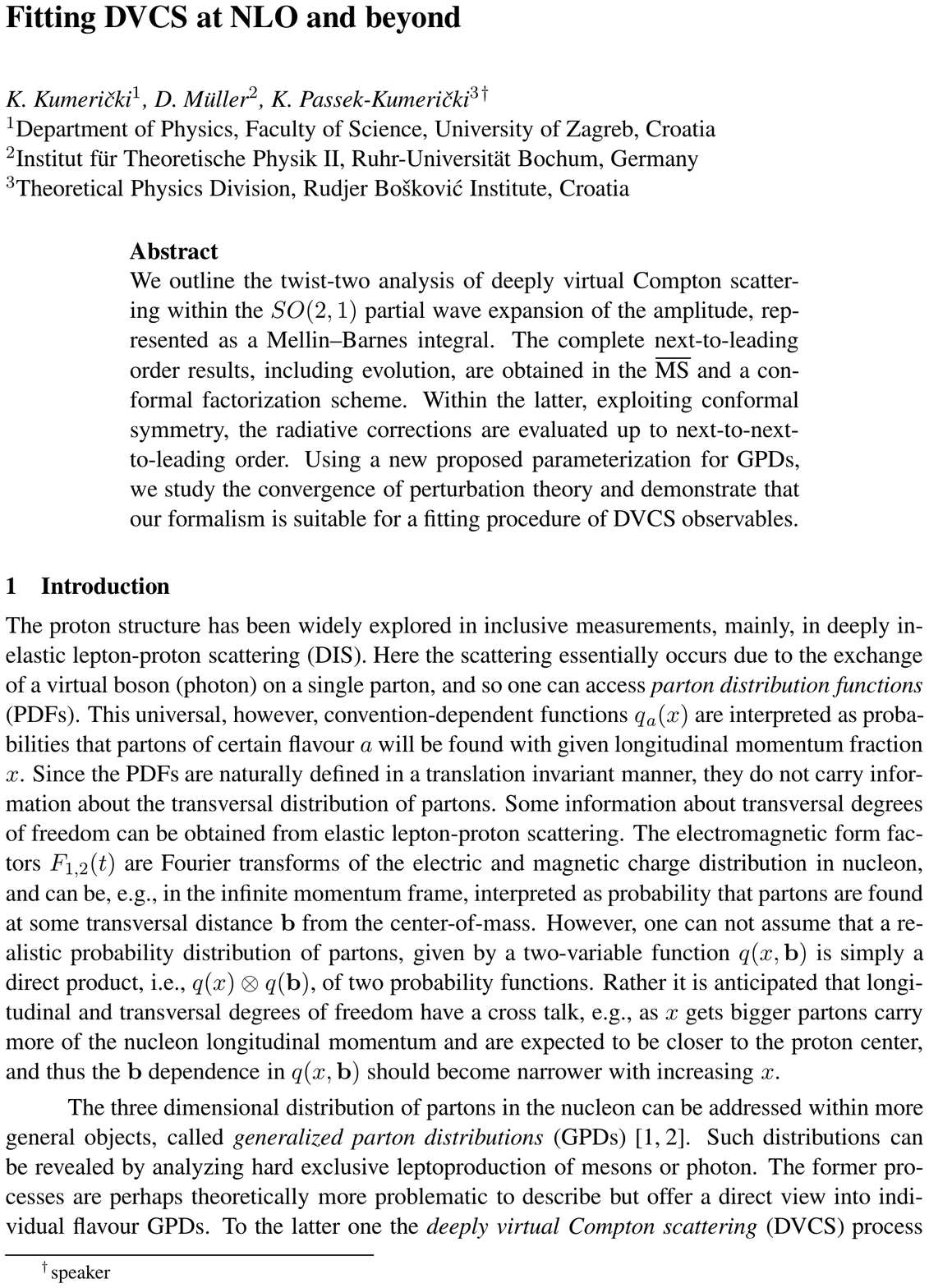} 

\coltoctitle{Exclusive vector meson electroproduction} 
\coltocauthor{D Ivanov}
\Includeart{\CAUT}{\CTIT}{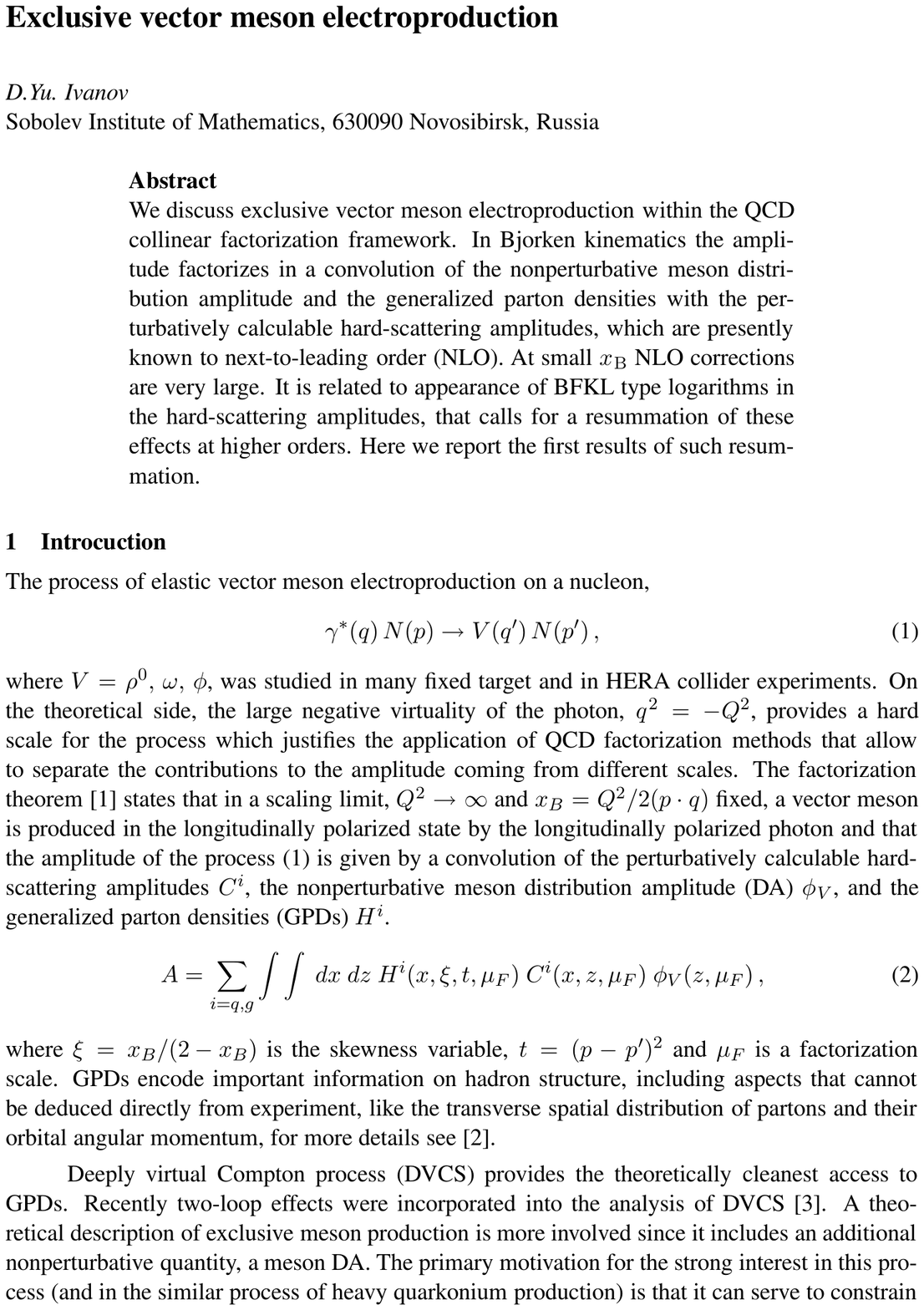} 

\coltoctitle{Photo and electroproduction of vector mesons: a unified nonperturbative  treatment} 
\coltocauthor{E Ferreira, H. G Dosch}
\Includeart{\CAUT}{\CTIT}{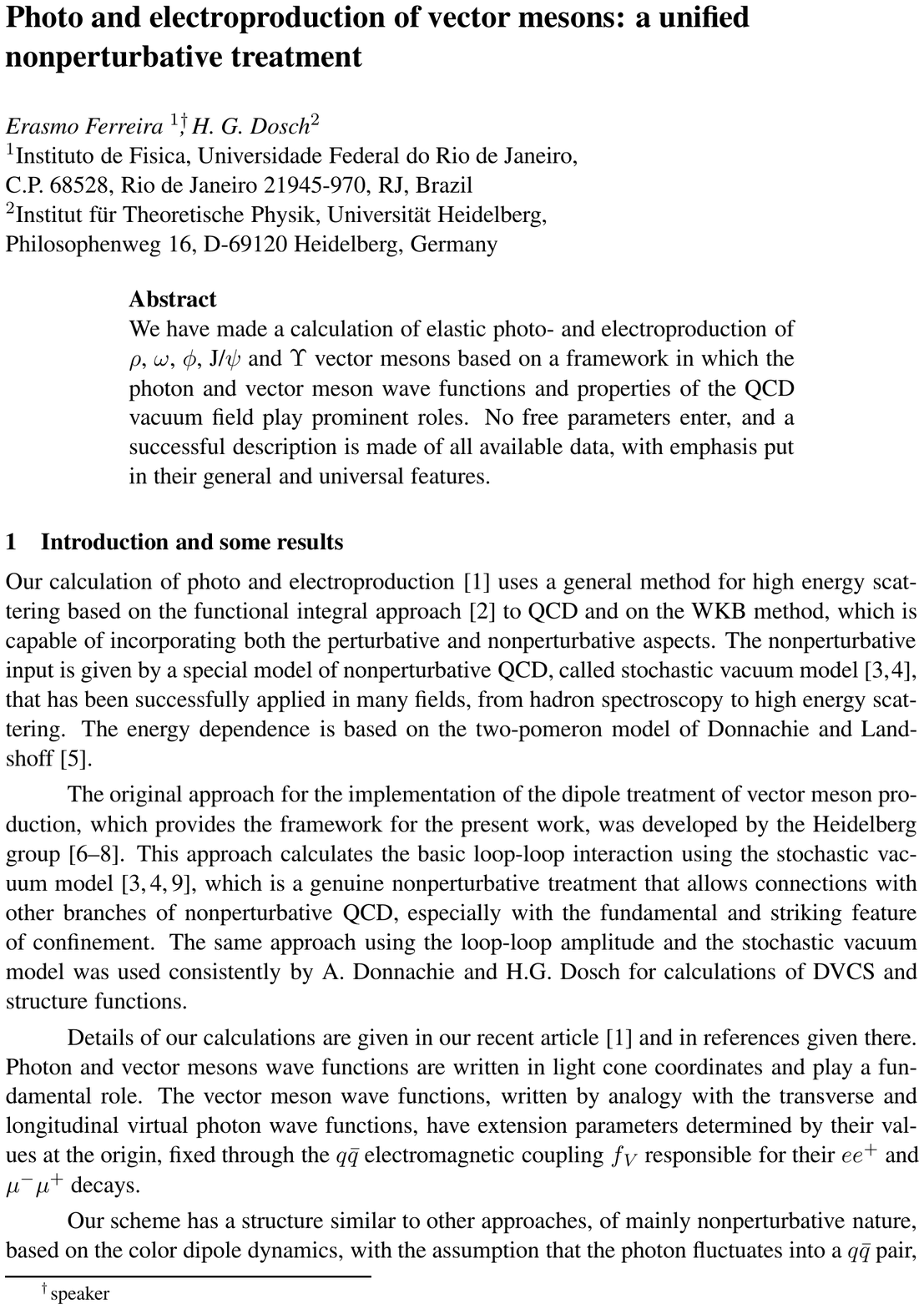}

\coltoctitle{Vector meson electroproduction within GPD approach} 
\coltocauthor{S Goloskokov }
\Includeart{\CAUT}{\CTIT}{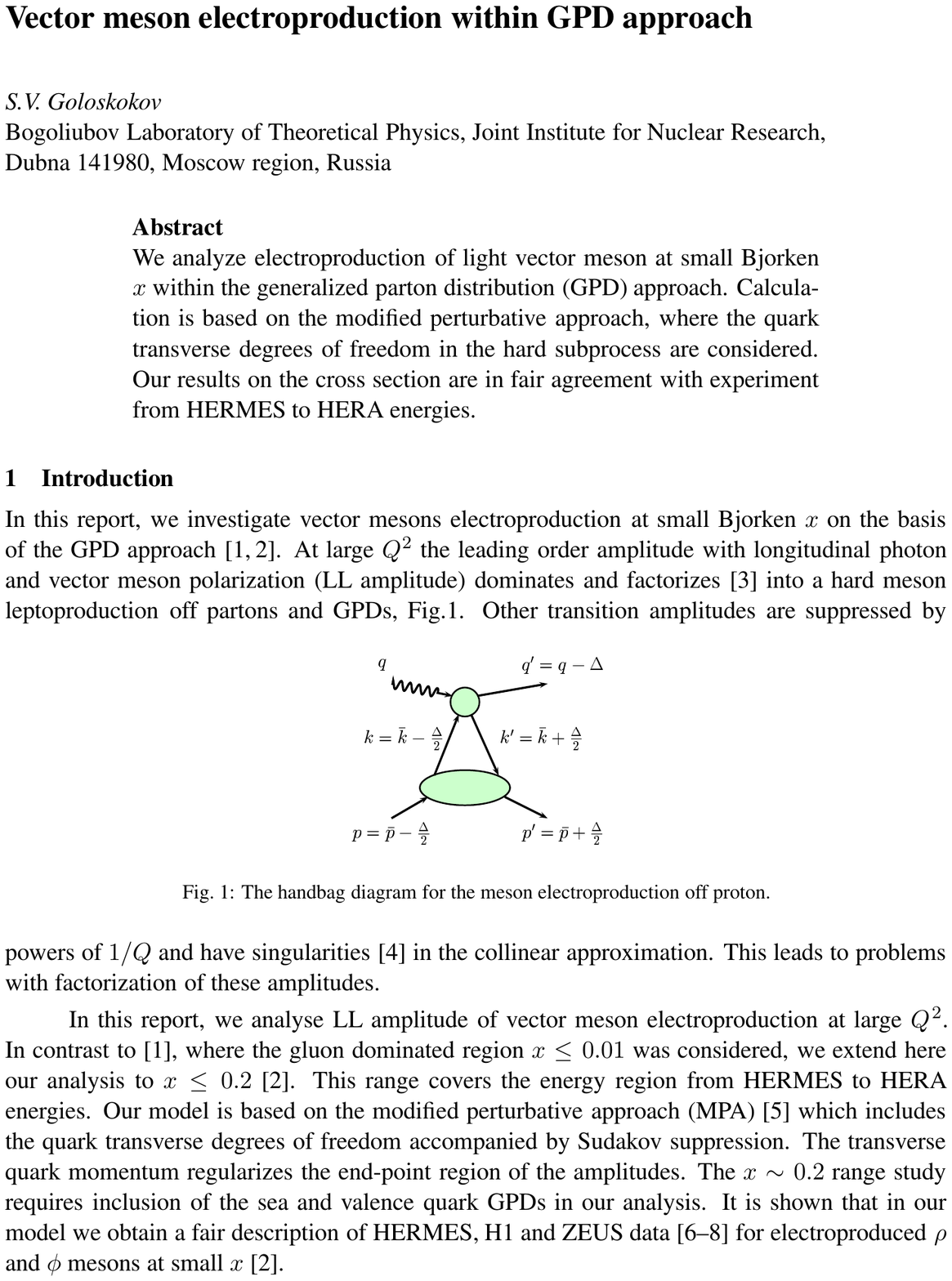} 

\coltoctitle{On the limitations of the color dipole picture} 
\coltocauthor{C Ewerz, A von Manteuffel, O Nachtmann}
\Includeart{\CAUT}{\CTIT}{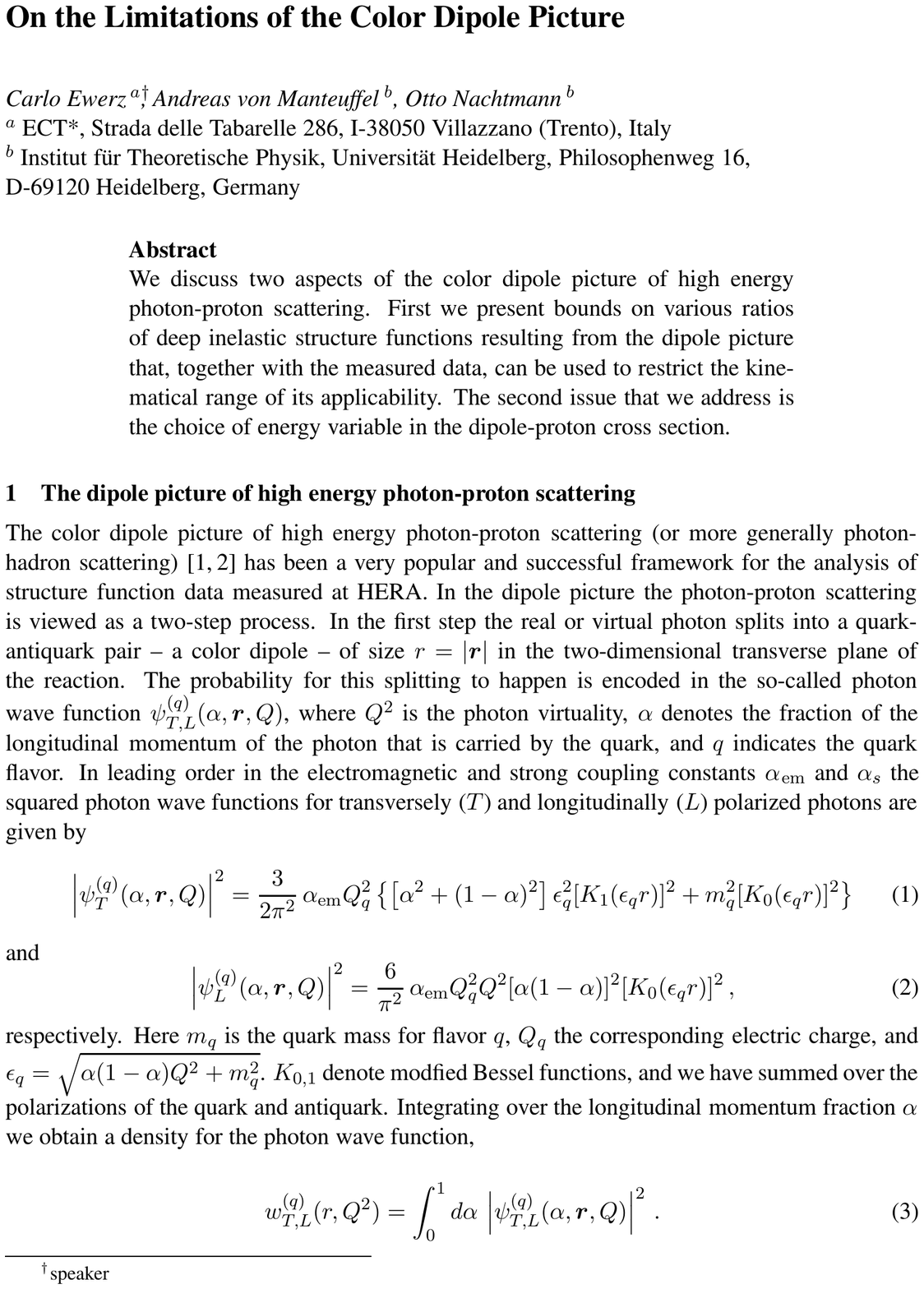} 

\coltoctitle{What is measured in hard exclusive processes?} 
\coltocauthor{J Londergan, A.P Szczepaniak}
\Includeart{\CAUT}{\CTIT}{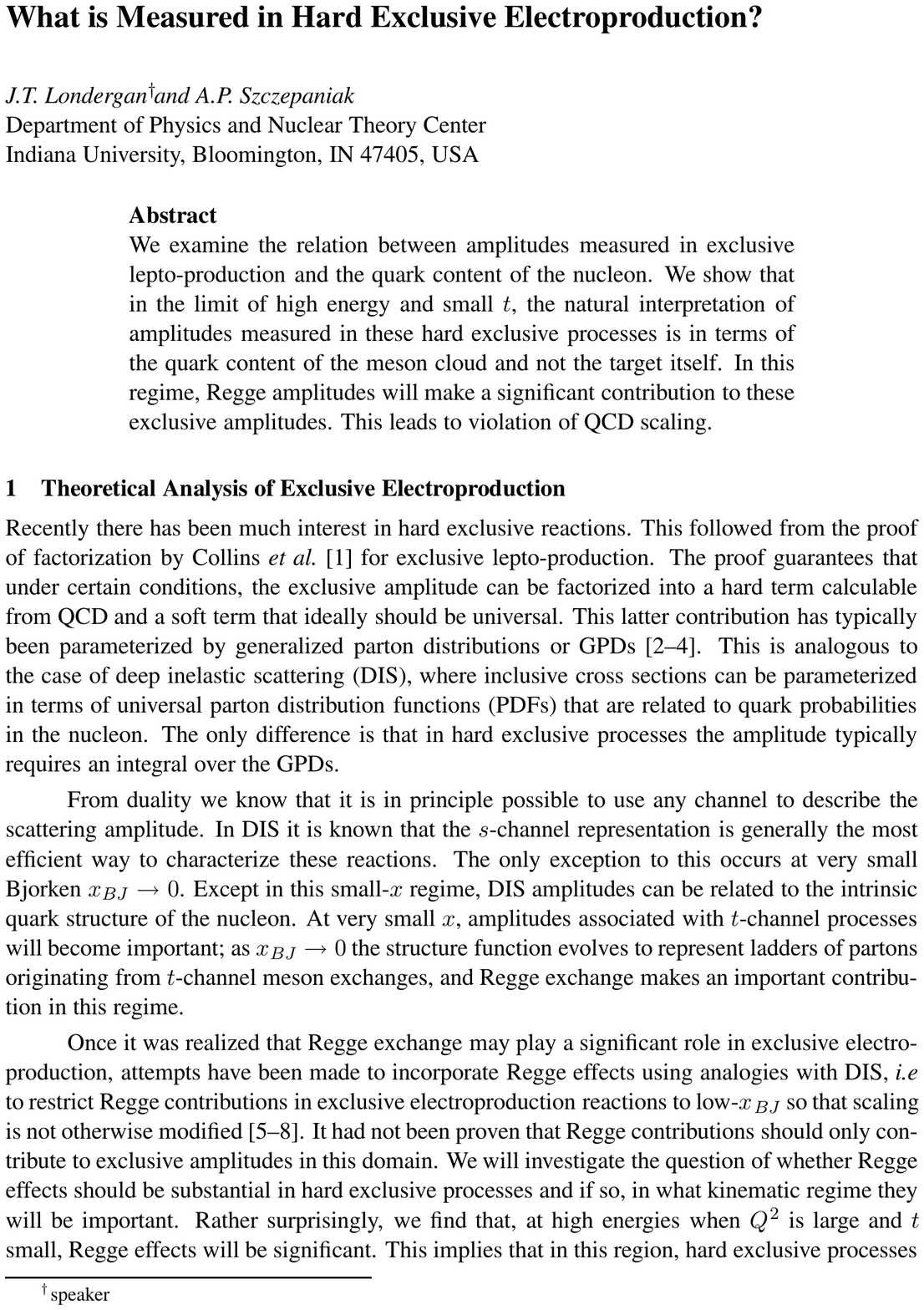} 

\coltoctitle{A Regge-pole model for DVCS} 
\coltocauthor{L Jenkovsky}
\Includeart{\CAUT}{\CTIT}{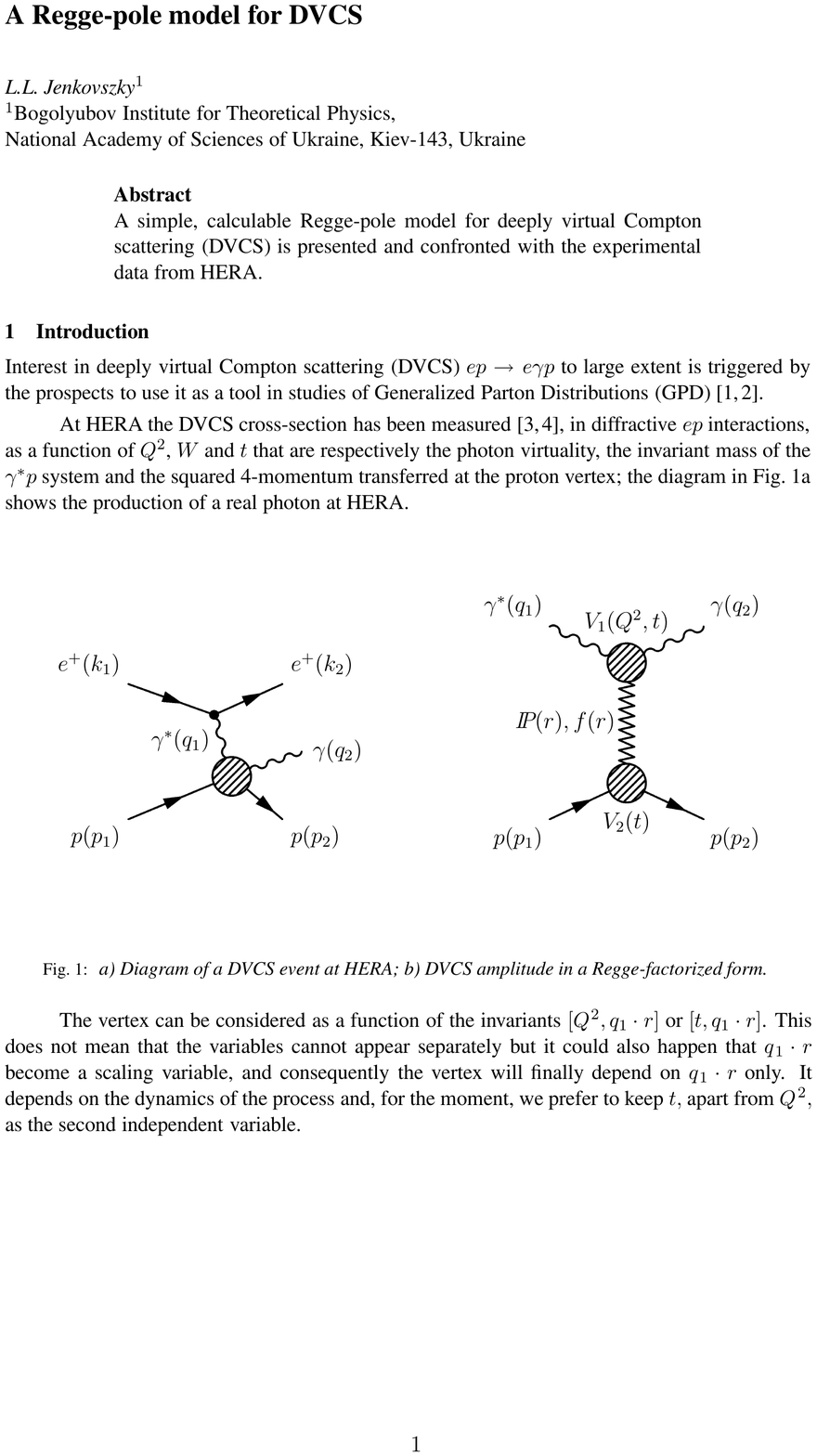} 

\coltoctitle{Glue drops inside hadrons} 
\coltocauthor{B.Z Kopeliovich, 
B Povh, 
I Schmidt}
\Includeart{\CAUT}{\CTIT}{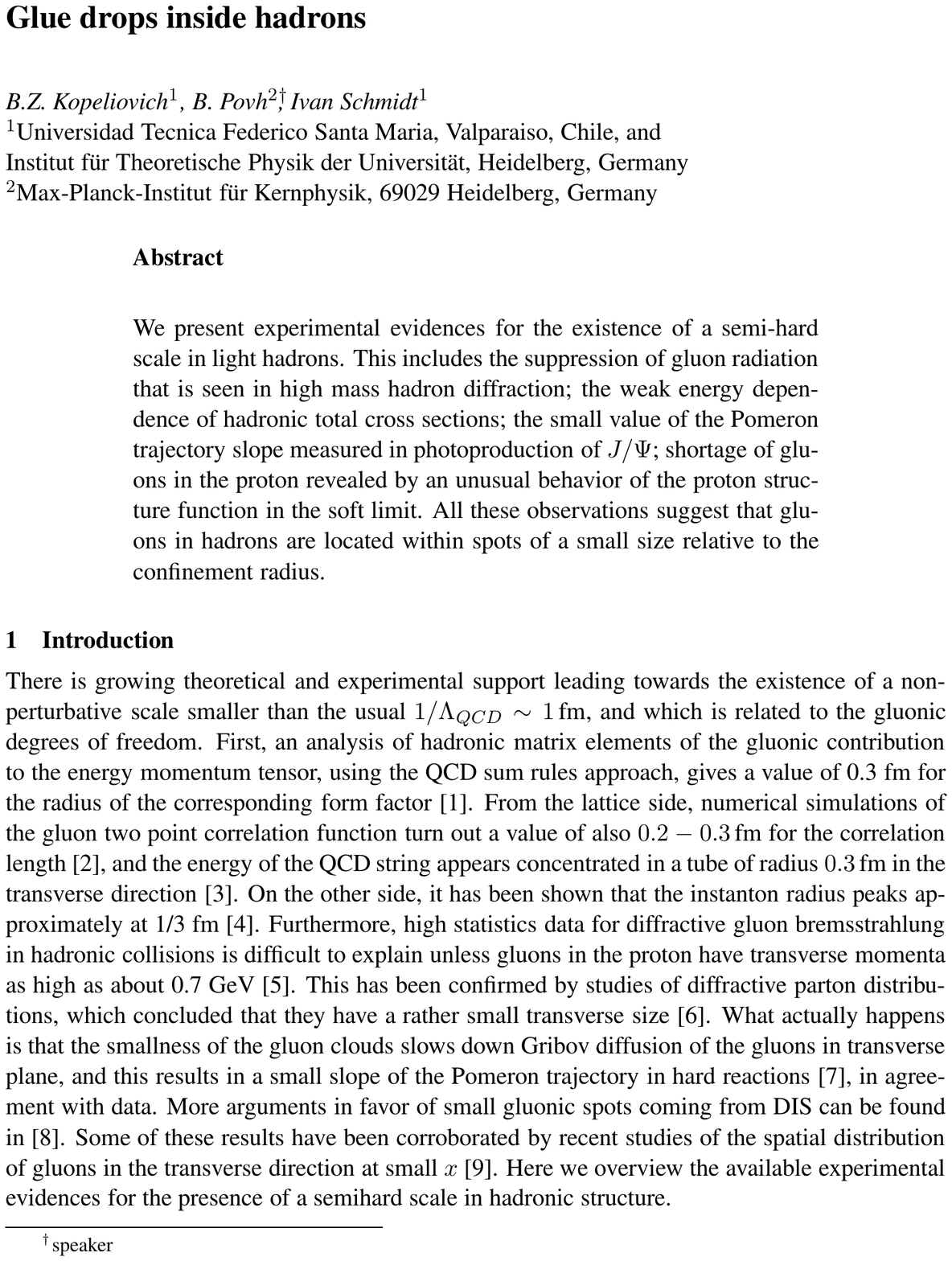} 

\coltoctitle{Studying QCD factorizations in exclusive $\gamma^* \gamma^* \to \rho^0_L 
\rho^0_L$ } 
\coltocauthor{S Wallon, B Pire, M Segond, L Szymanowski}
\Includeart{\CAUT}{\CTIT}{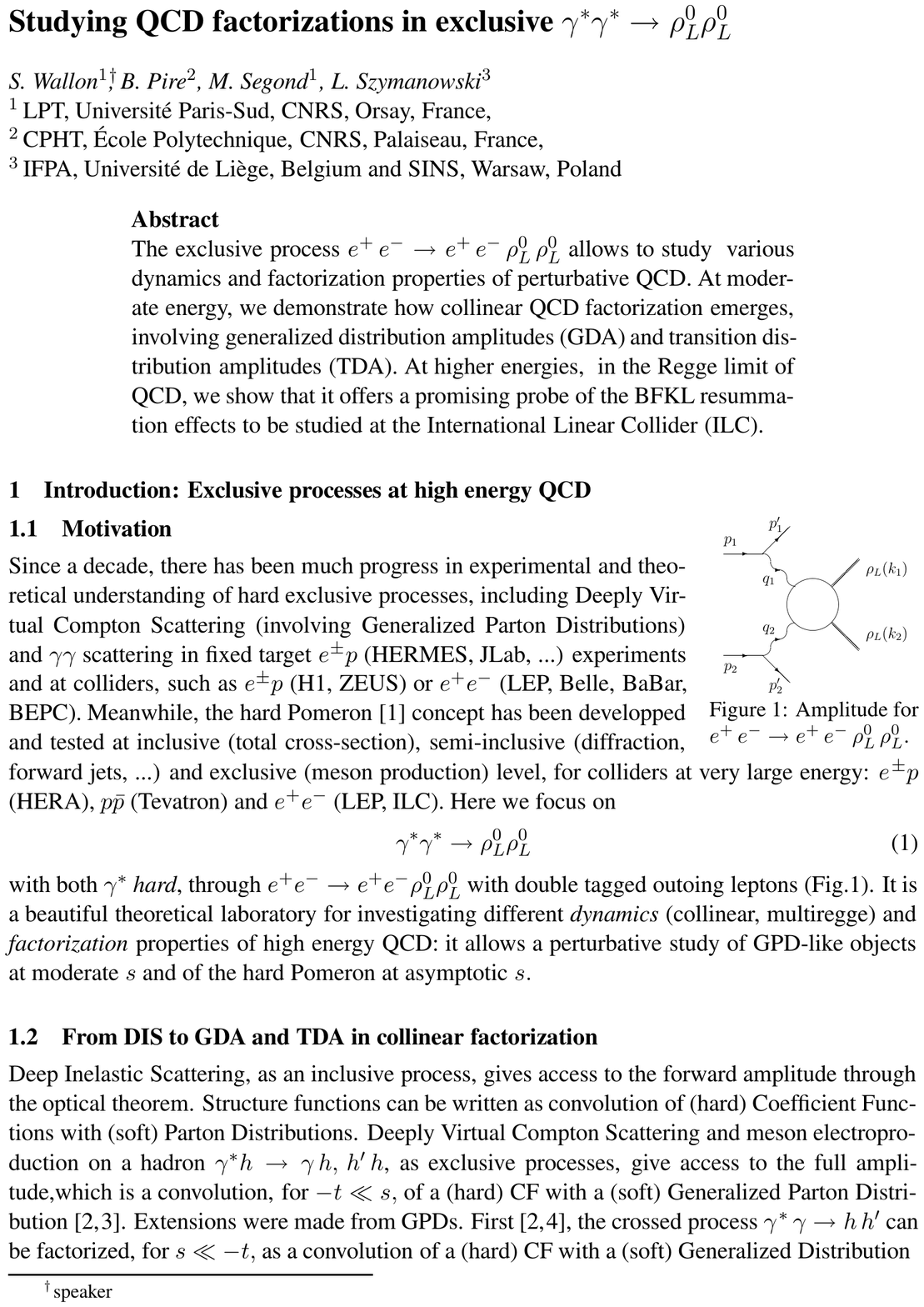} 

\coltoctitle{Low $Q^2$ and High $y$ Inclusive Cross Section Measurements
  from the HERA Experiments ZEUS and H1} 
\coltocauthor{J Kretzschmar}
\Includeart{\CAUT}{\CTIT}{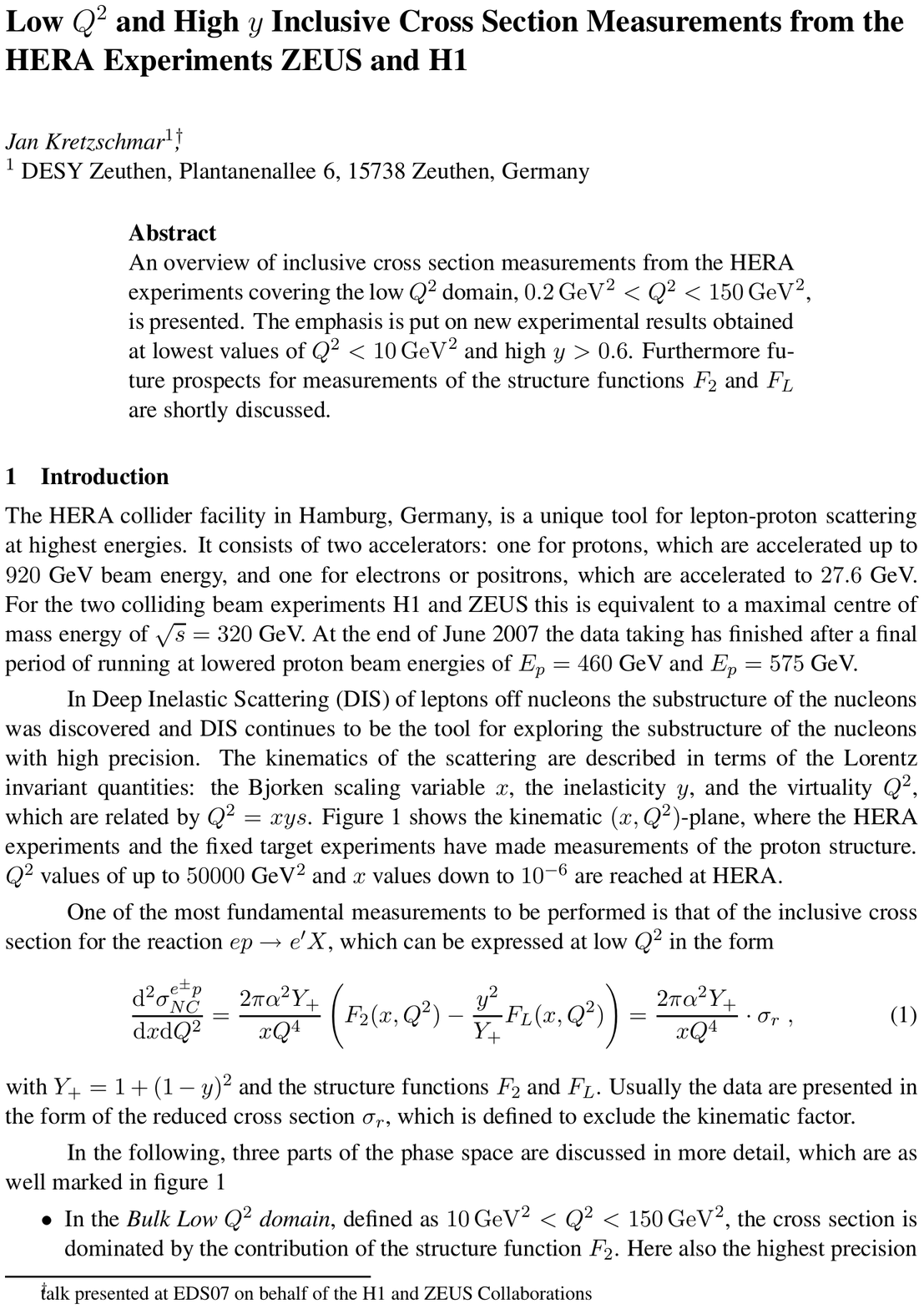} 

\coltoctitle{Status of Deeply Inealstic Parton Distributions} 
\coltocauthor{J Bl\"umlein}
\Includeart{\CAUT}{\CTIT}{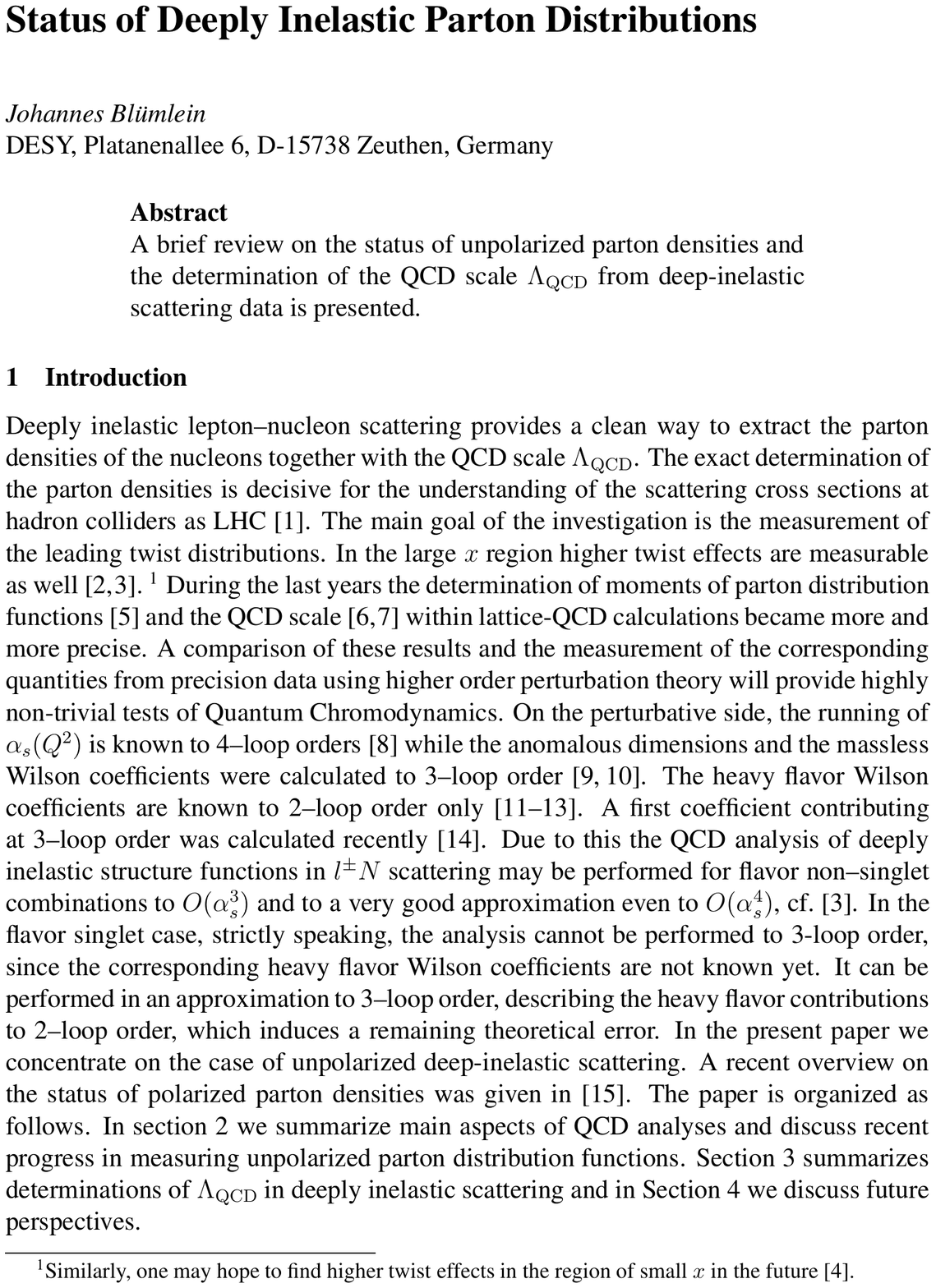} 

\coltoctitle{Diffractive PDFs} 
\coltocauthor{P Laycock}
\Includeart{\CAUT}{\CTIT}{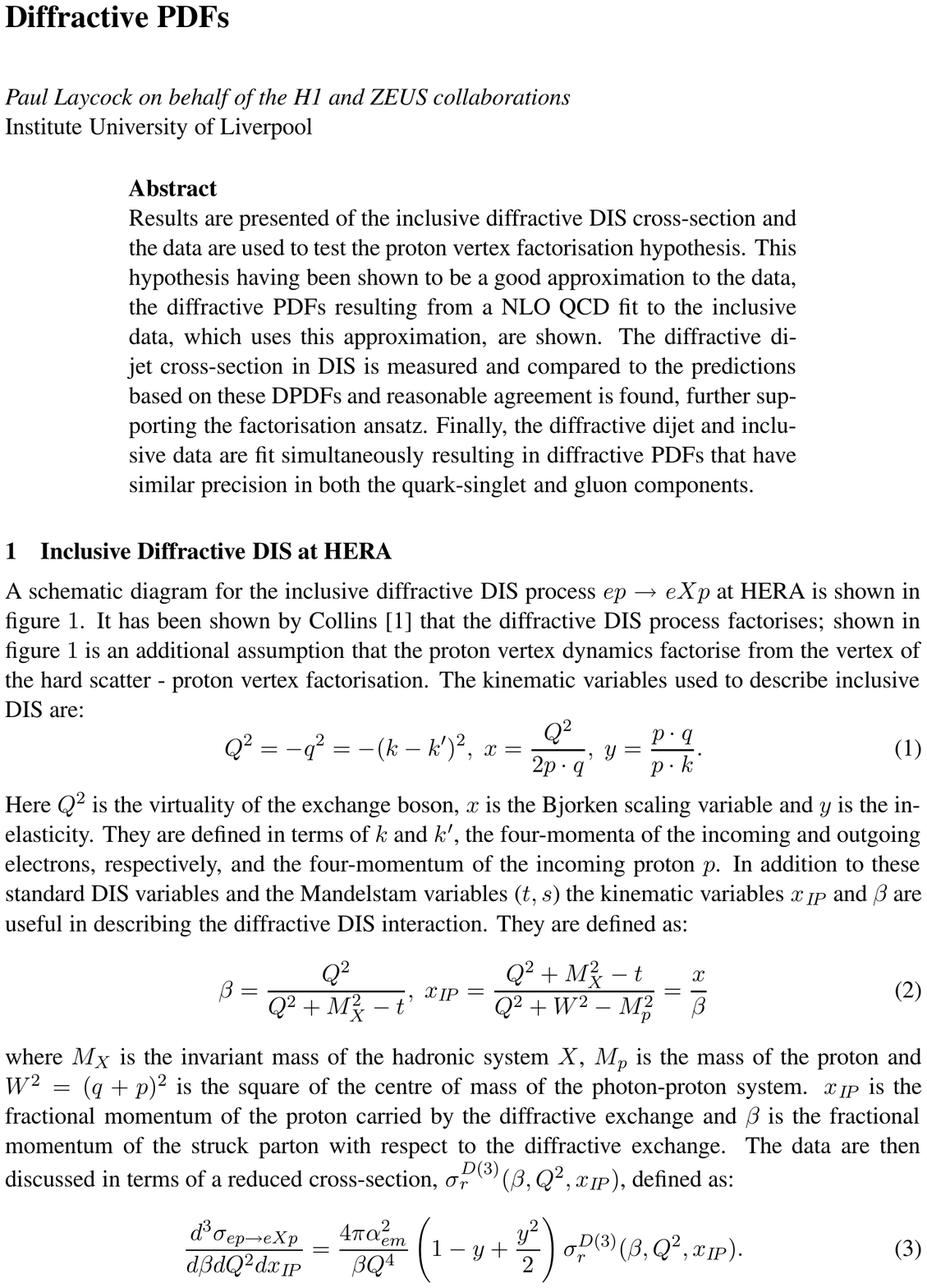} 

\coltoctitle{Factorisation breaking in diffraction} 
\coltocauthor{A Bonato}
\Includeart{\CAUT}{\CTIT}{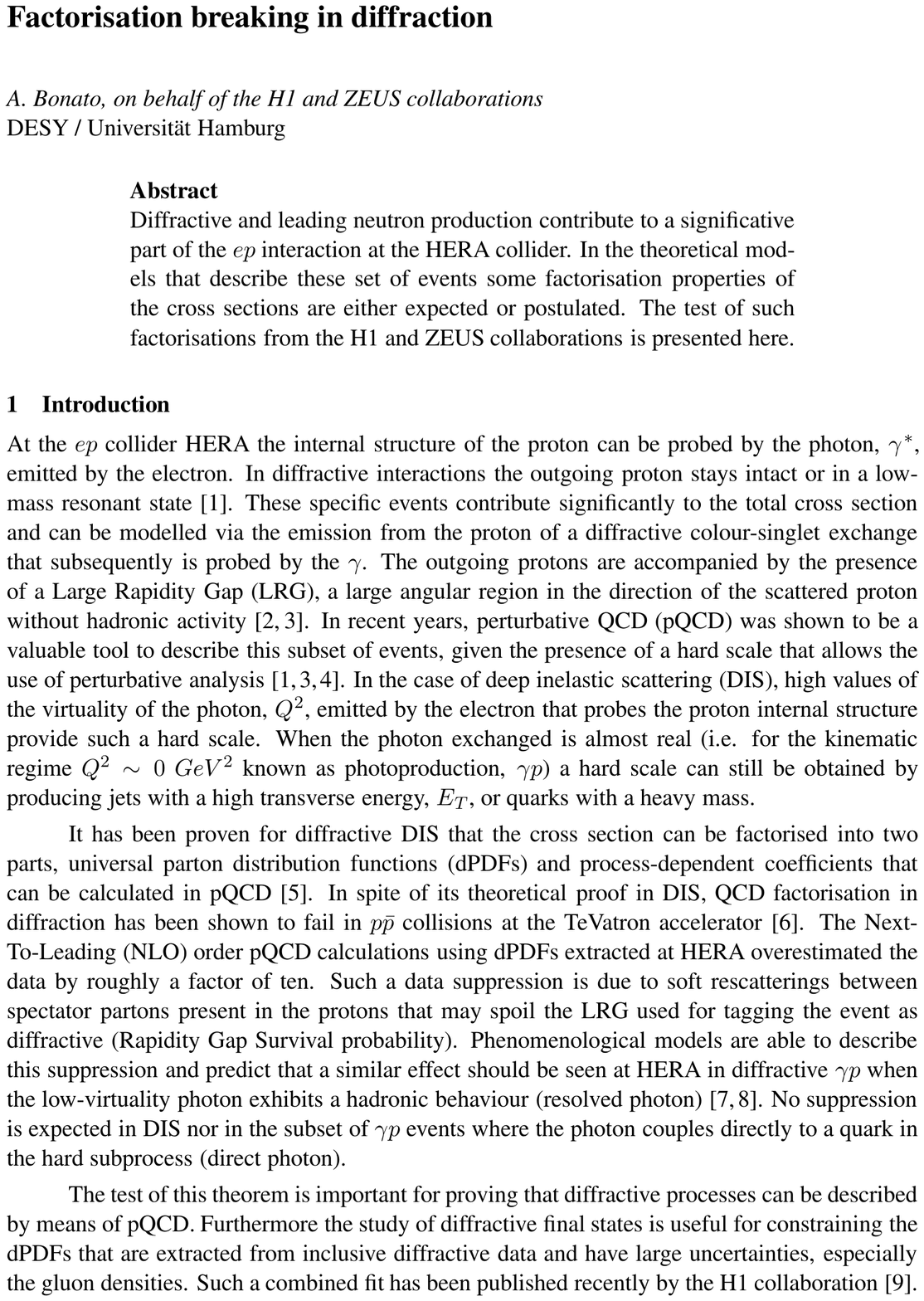} 

\coltoctitle{Diffractive structure function  $F_L^D$ from fits with higher twist} 
\coltocauthor{K Golec-Biernat, A \L{}uszczak}
\Includeart{\CAUT}{\CTIT}{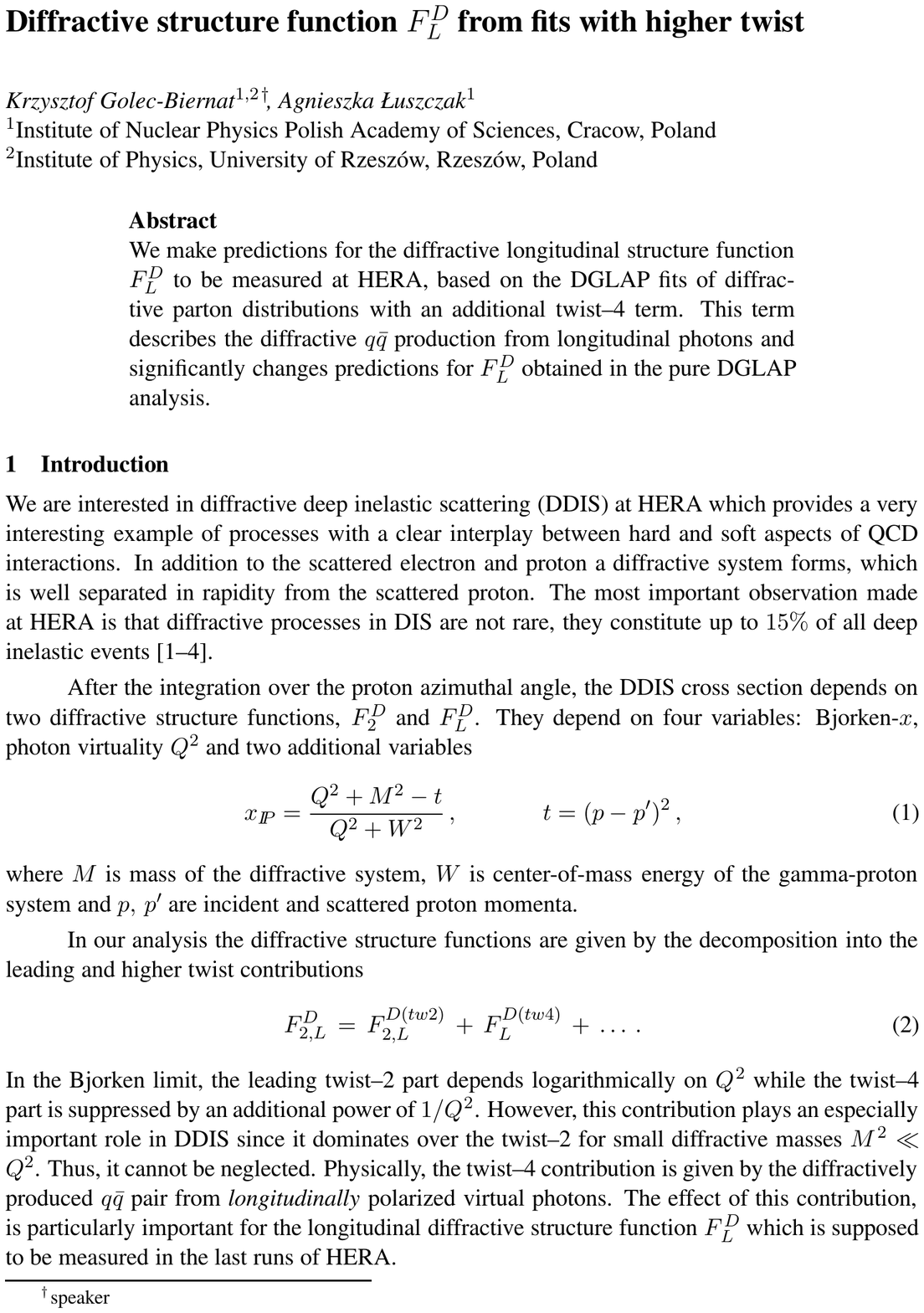} 

\coltoctitle{Diffractive parton distributions: the role of the perturbative Pomeron} 
\coltocauthor{G Watt, A.D Martin, M Ryskin}
\Includeart{\CAUT}{\CTIT}{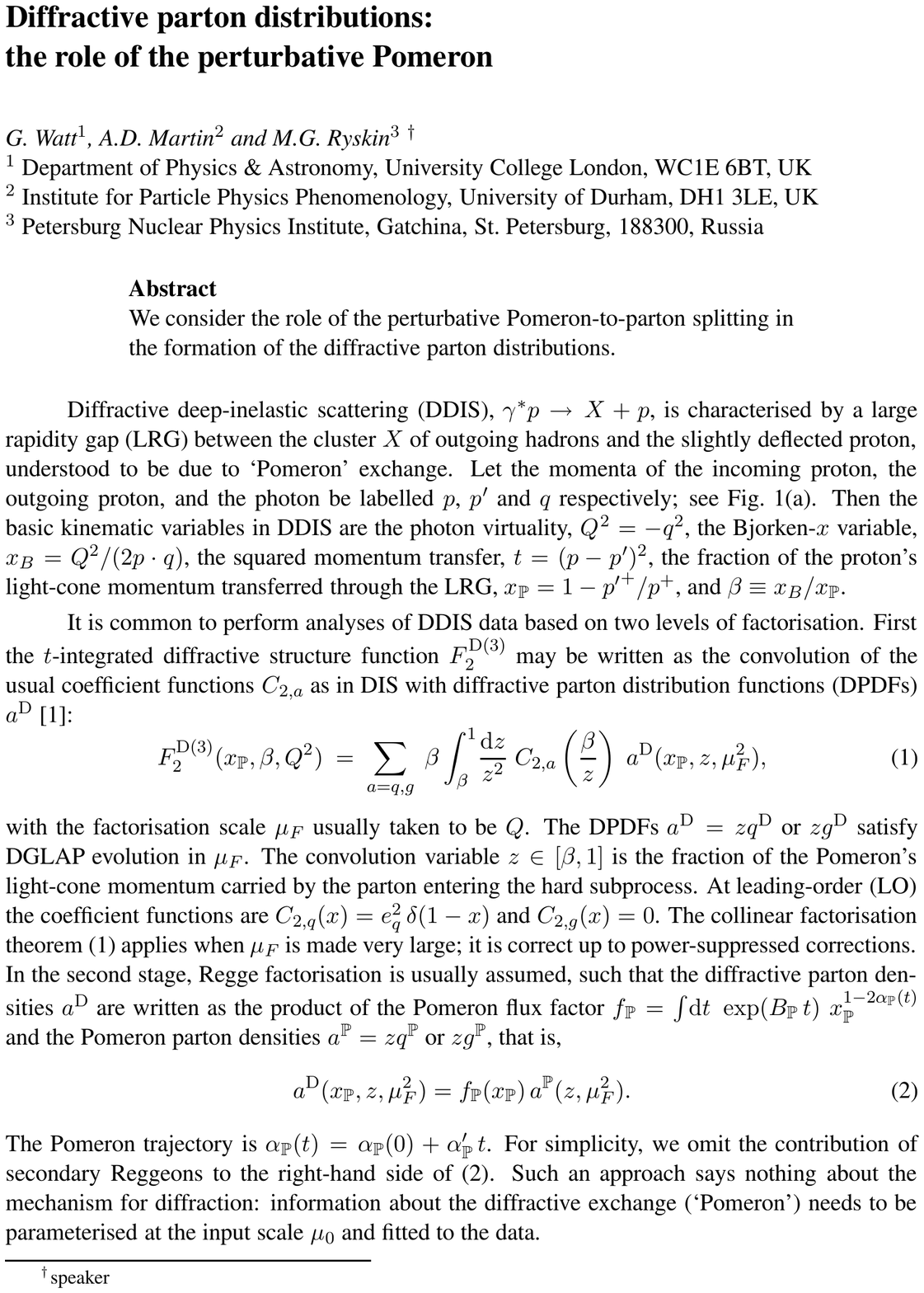} 

\coltoctitle{Multiple interactions in H1 and ZEUS} 
\coltocauthor{T Namsoo}
\Includeart{\CAUT}{\CTIT}{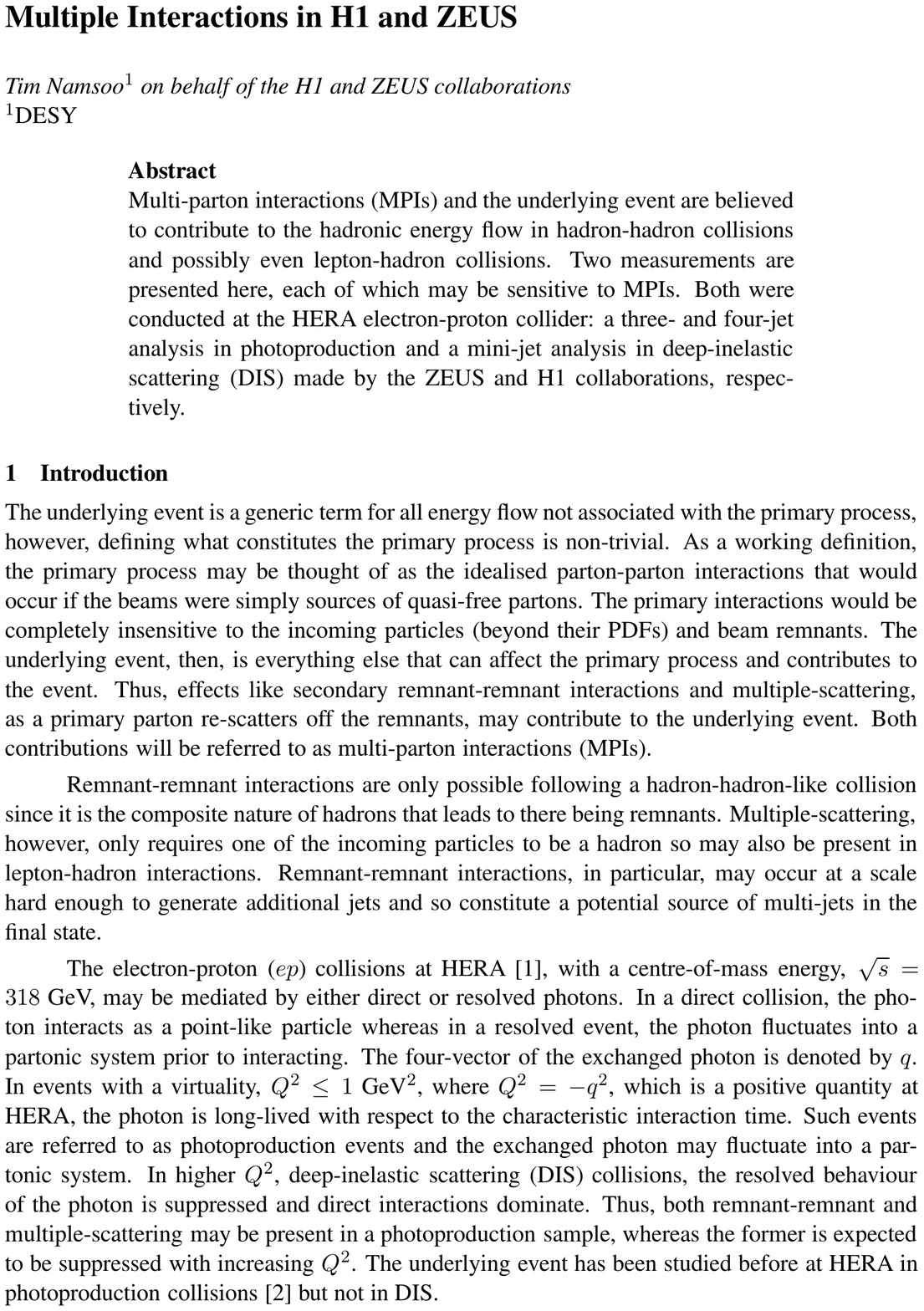} 

\coltoctitle{Diffractive neutral pion production, chiral symmetry and the odderon} 
\coltocauthor{C Ewerz, O Nachtmann}
\Includeart{\CAUT}{\CTIT}{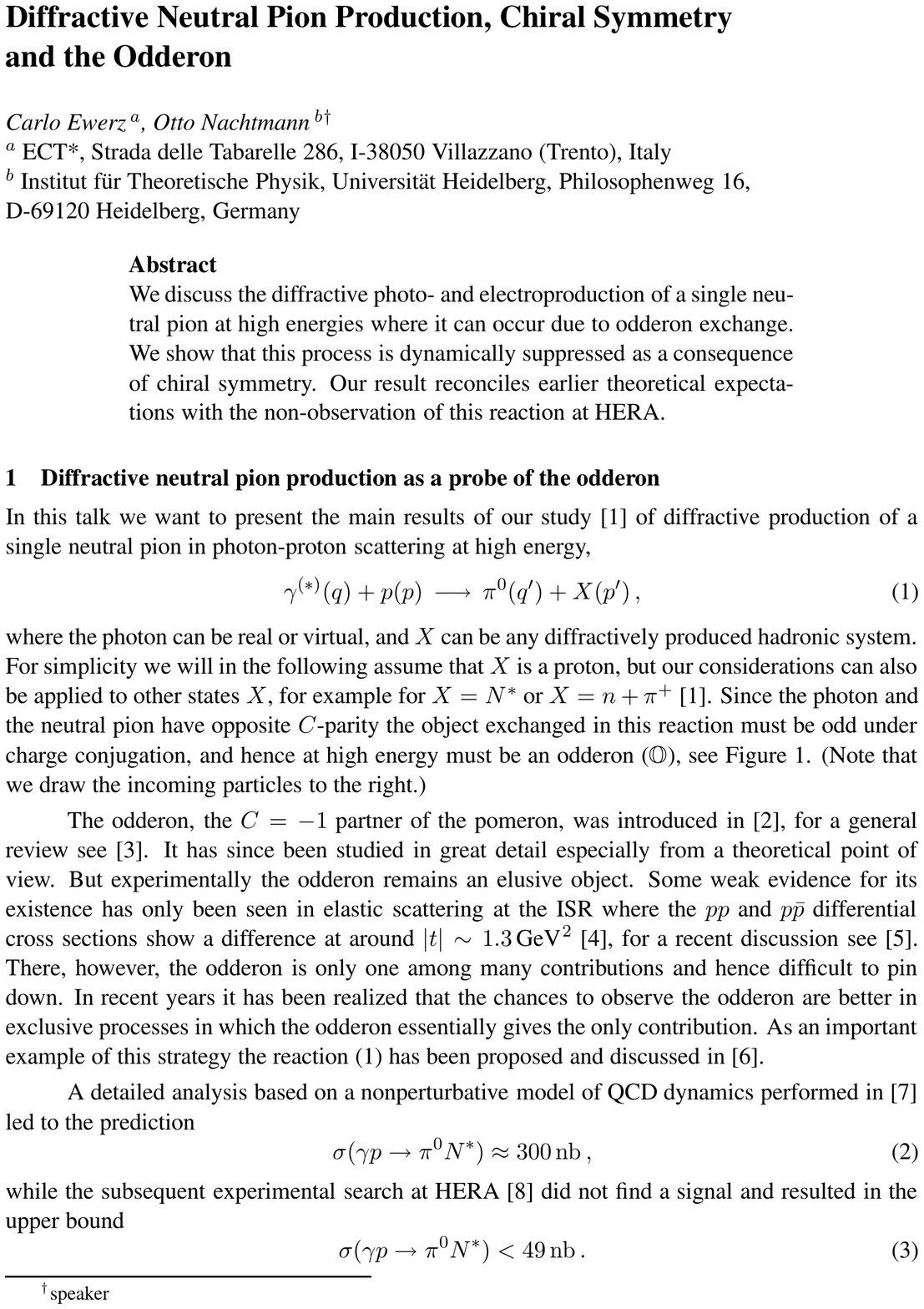} 

\end{papers} 

\CLDP 

\part{$pp$ and $p\bar{p}$ Collisions} 
\label{pp} 
\newpage

\begin{papers} 
\coltoctitle{Central production of new physics} 
\coltocauthor{J Forshaw, A Pilkington}
\Includeart{\CAUT}{\CTIT}{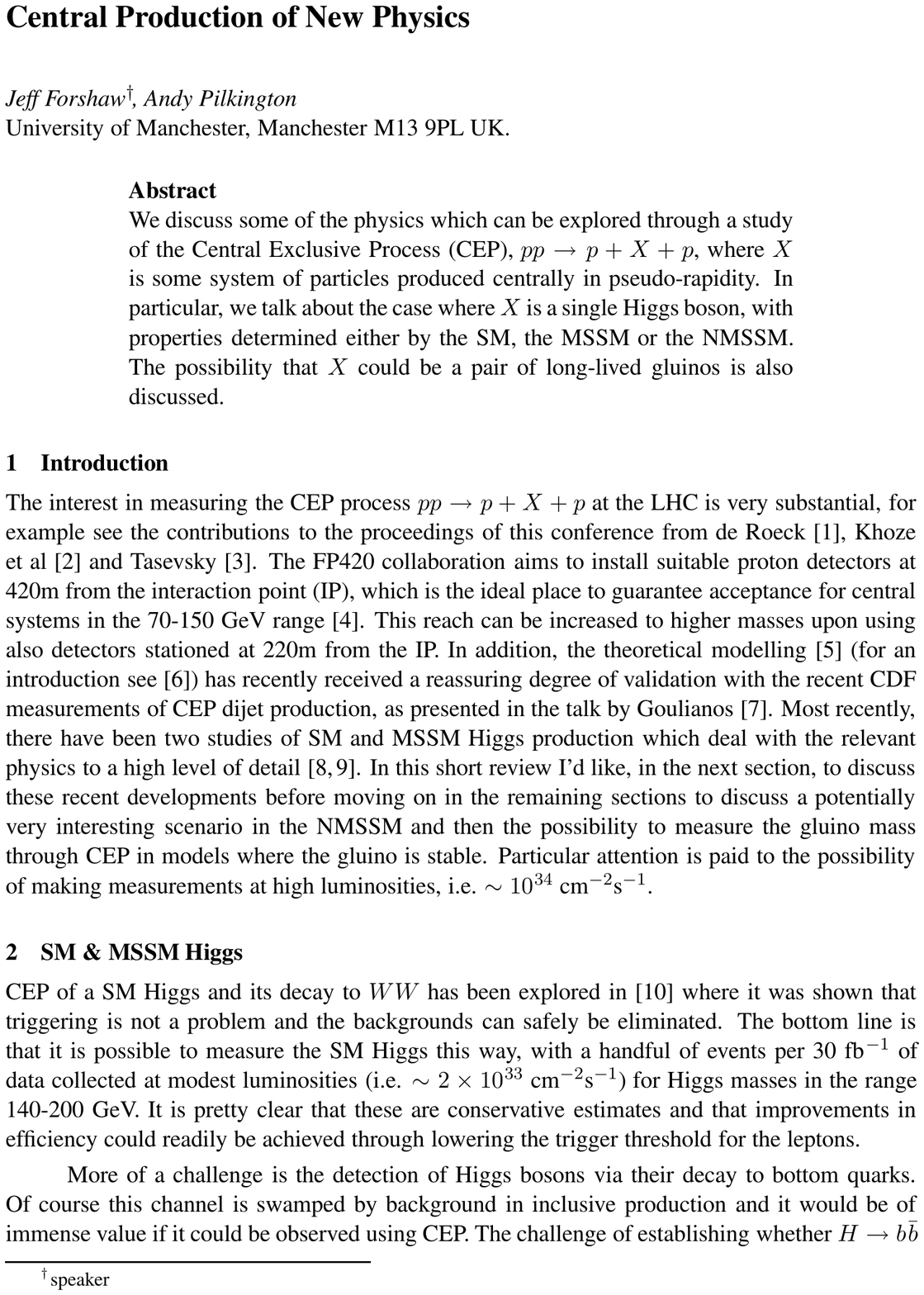} 

\coltoctitle{Diffraction at CDF} 
\coltocauthor{K Goulianos}
\Includeart{\CAUT}{\CTIT}{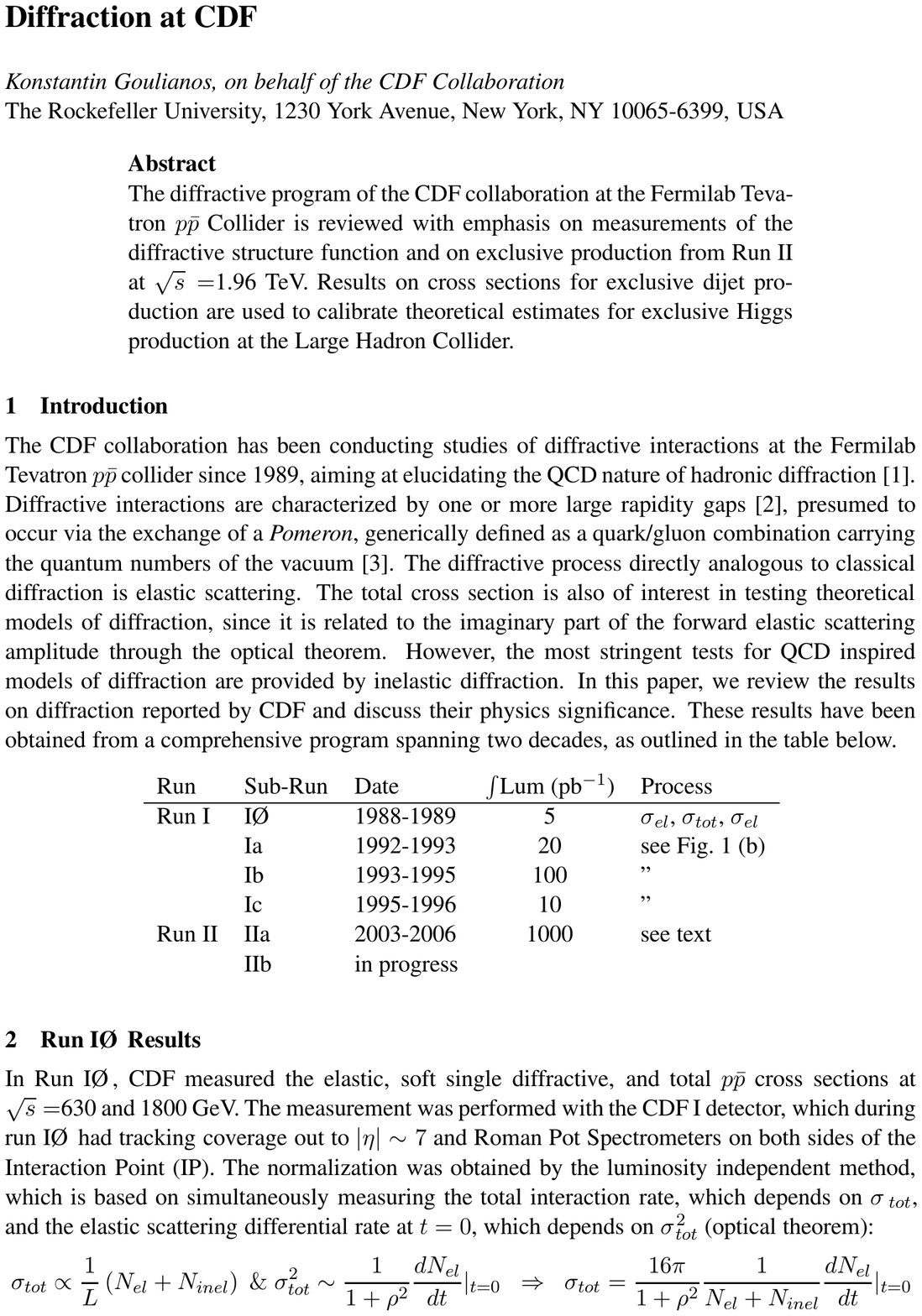} 

\coltoctitle{Prospects for proton tagging at high luminosities at the LHC} 
\coltocauthor{M Tasevsky}
\Includeart{\CAUT}{\CTIT}{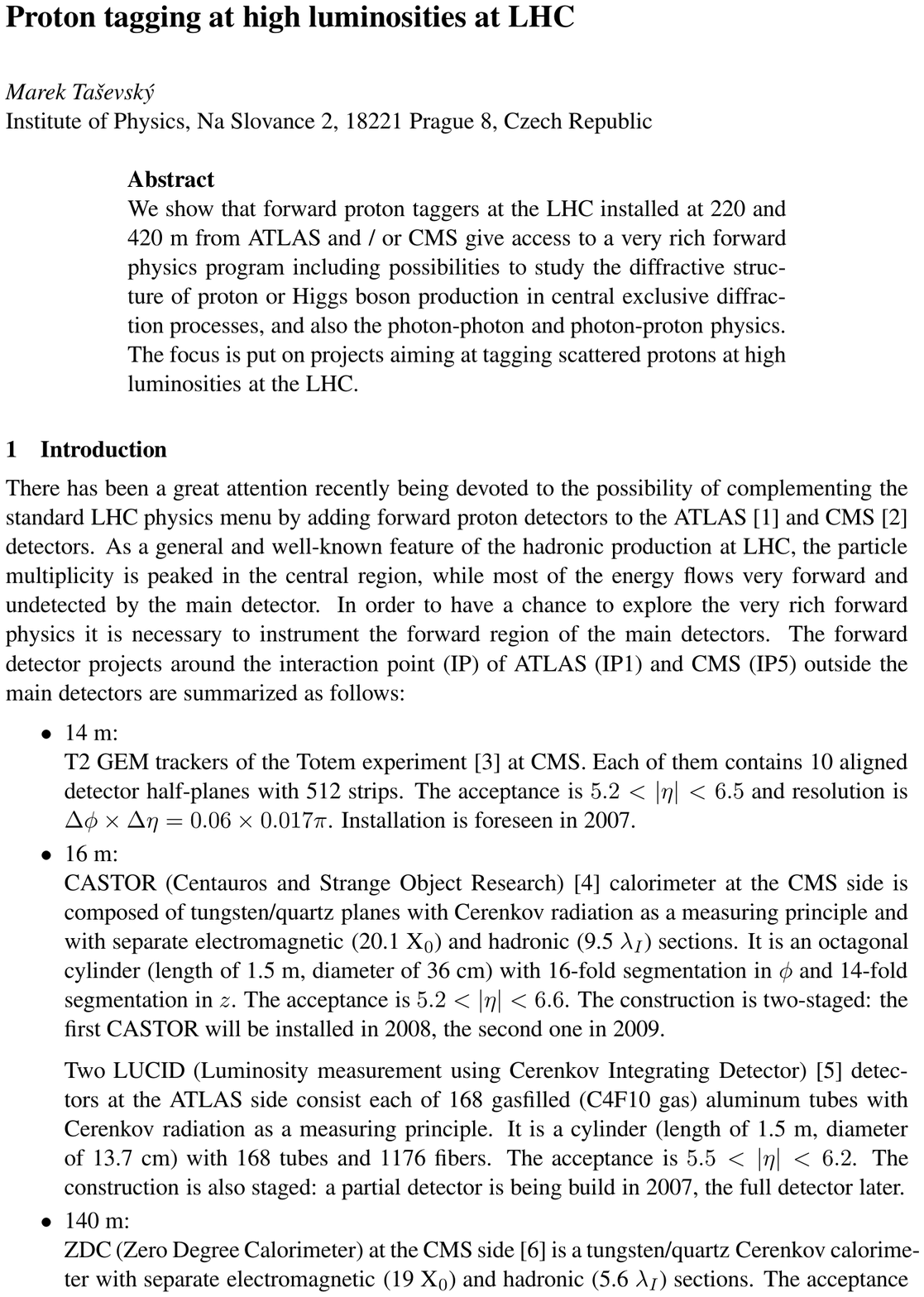} 

\coltoctitle{Total cross section measurement and diffractive physics with Totem} 
\coltocauthor{M Deile}
\Includeart{\CAUT}{\CTIT}{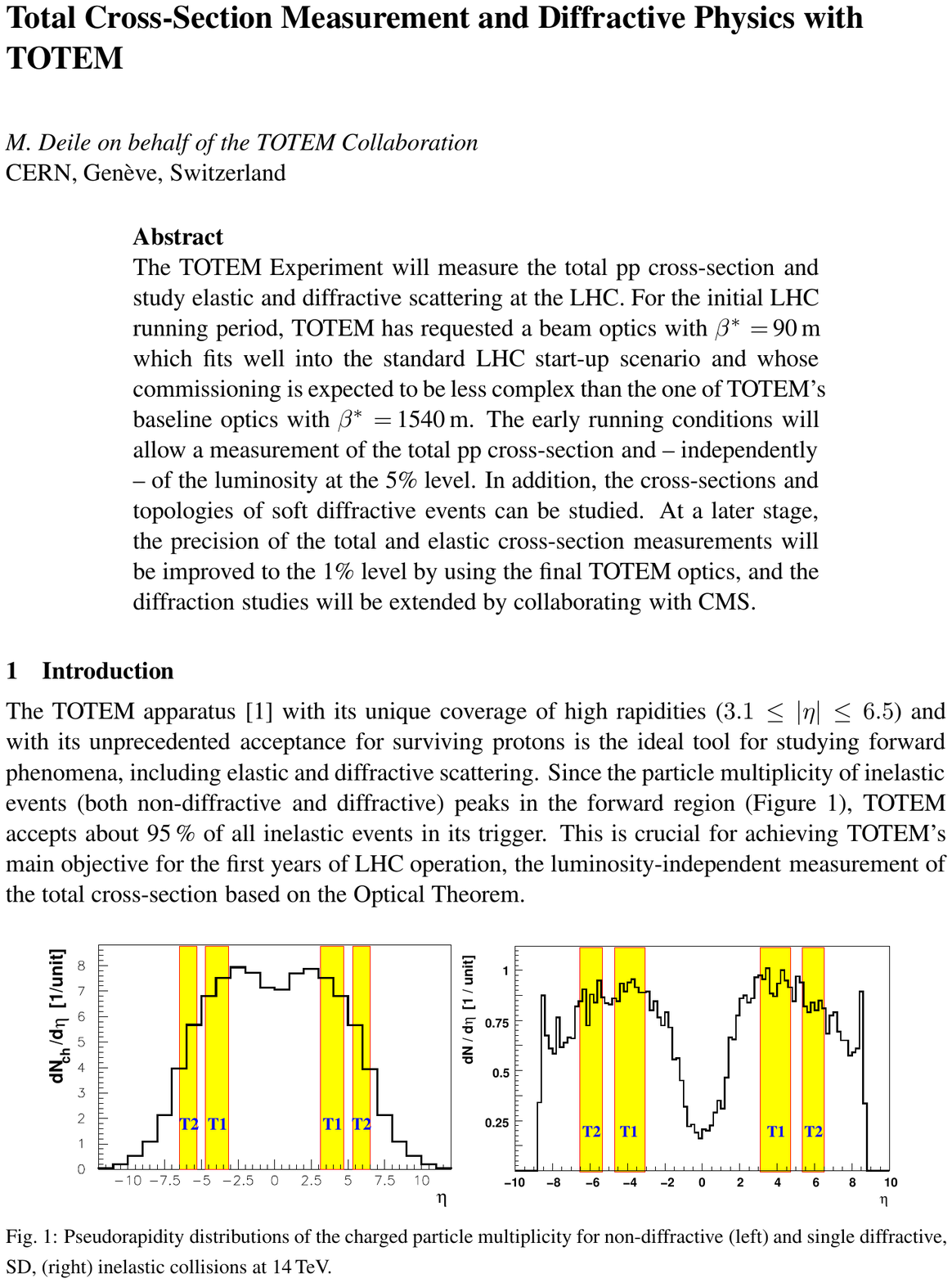} 

\coltoctitle{Prospects of forward energy flow and low-x physics at the LHC} 
\coltocauthor{A Hamilton}
\Includeart{\CAUT}{\CTIT}{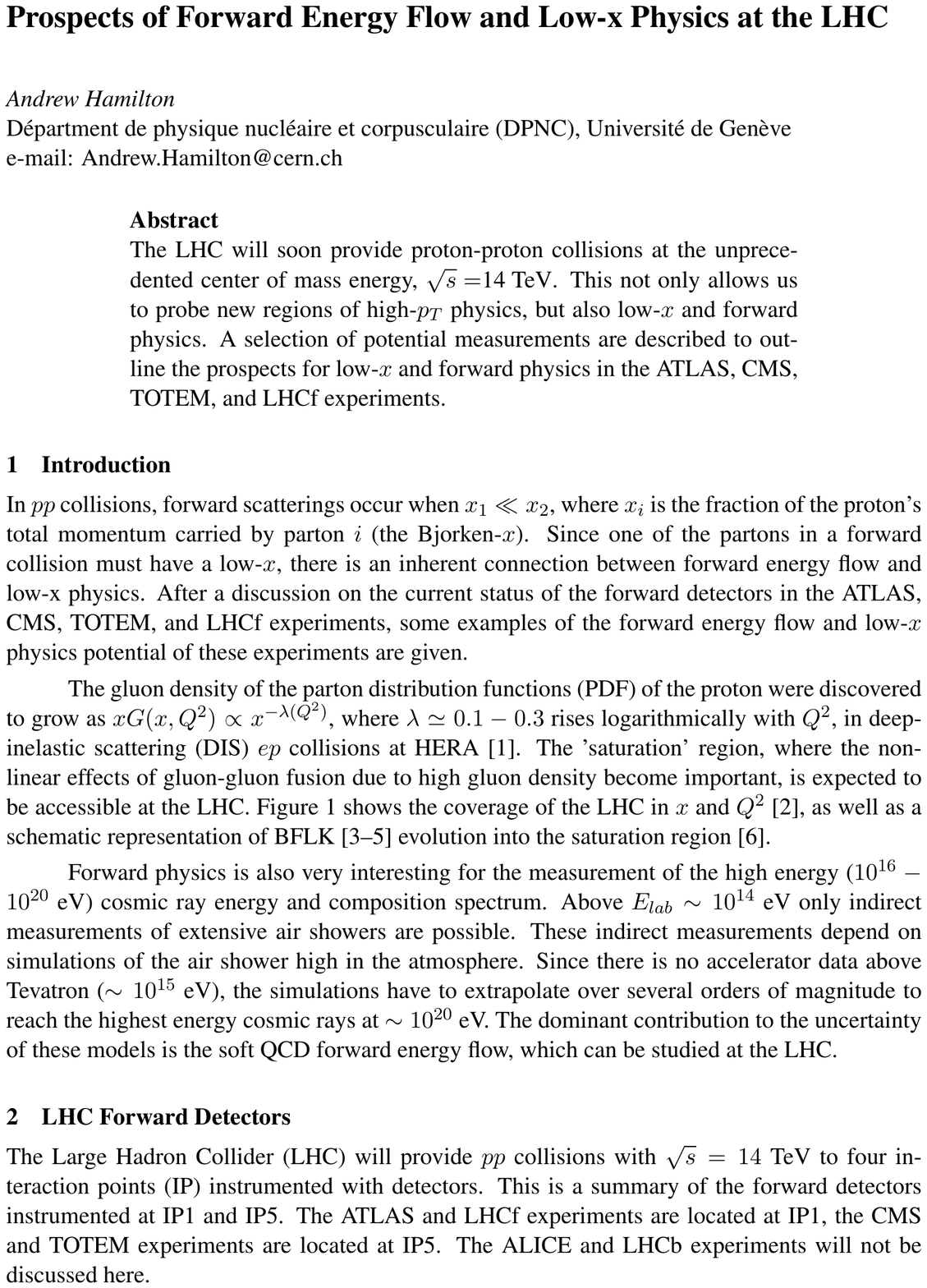} 

\coltoctitle{Diffractive physics in ALICE} 
\coltocauthor{R Schicker}
\Includeart{\CAUT}{\CTIT}{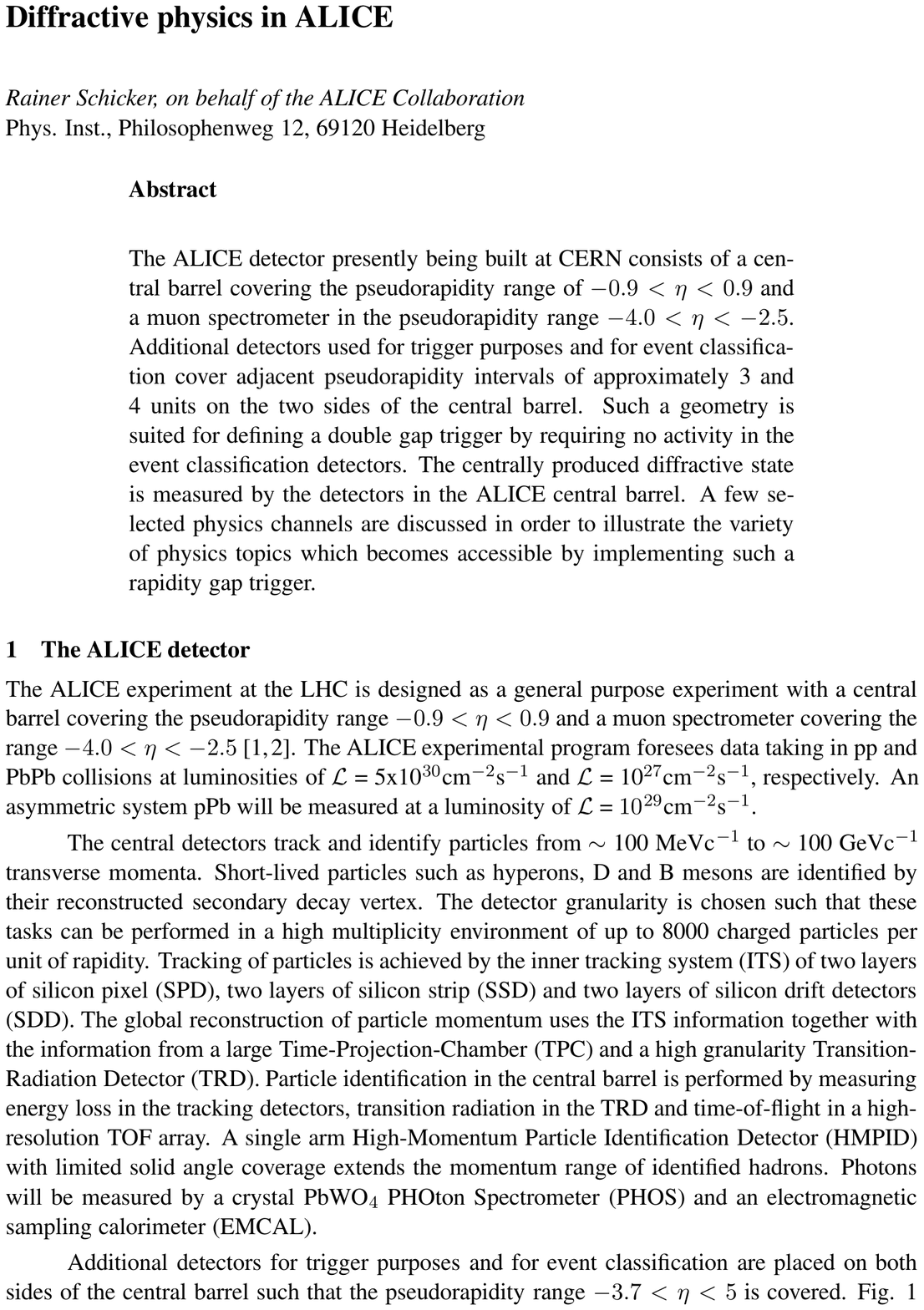} 

\coltoctitle{The odderon at RHIC and LHC} 
\coltocauthor{B Nicolescu}
\Includeart{\CAUT}{\CTIT}{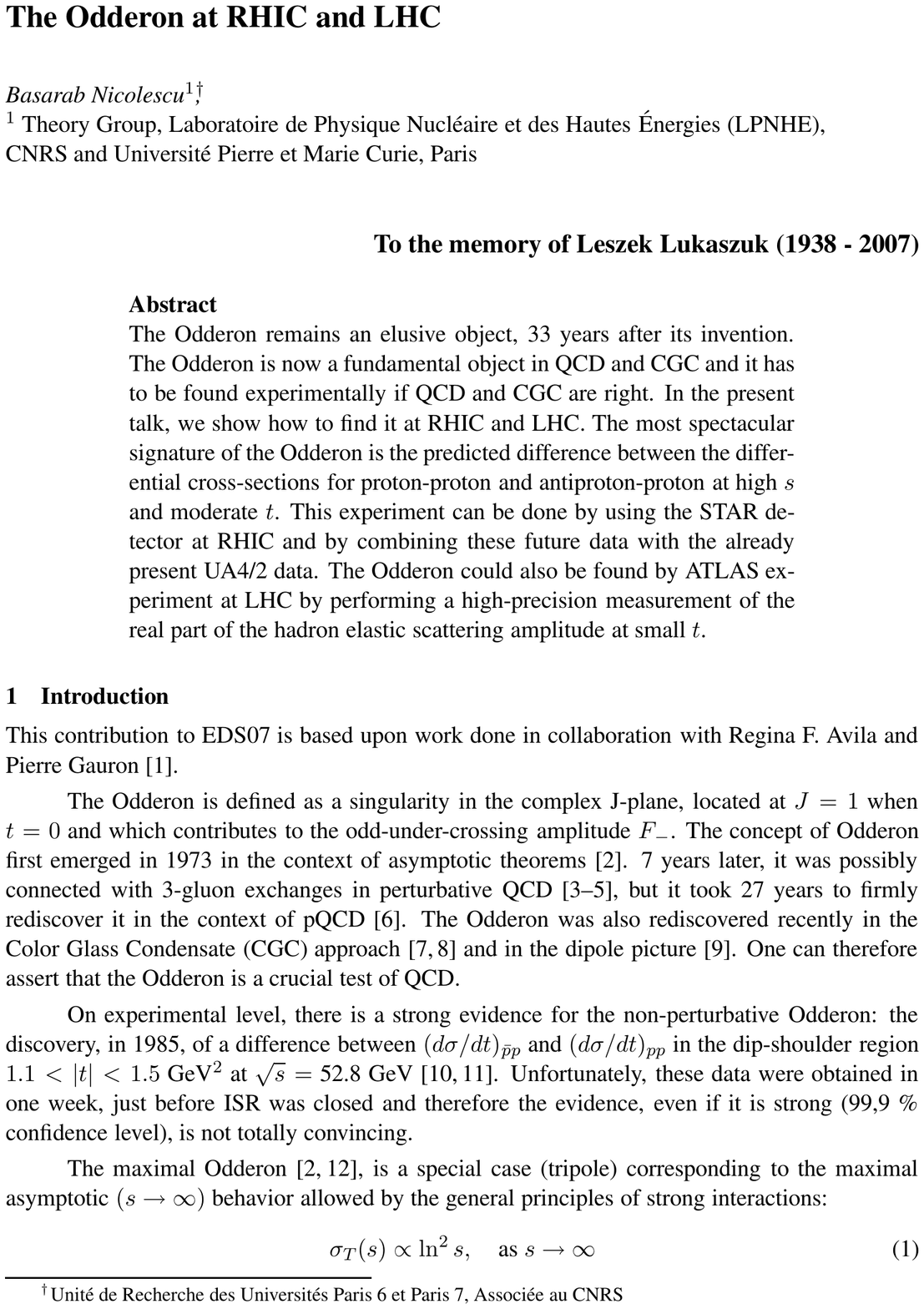} 

\coltoctitle{Prospects for diffraction at the LHC} 
\coltocauthor{A De Roeck}
\Includeart{\CAUT}{\CTIT}{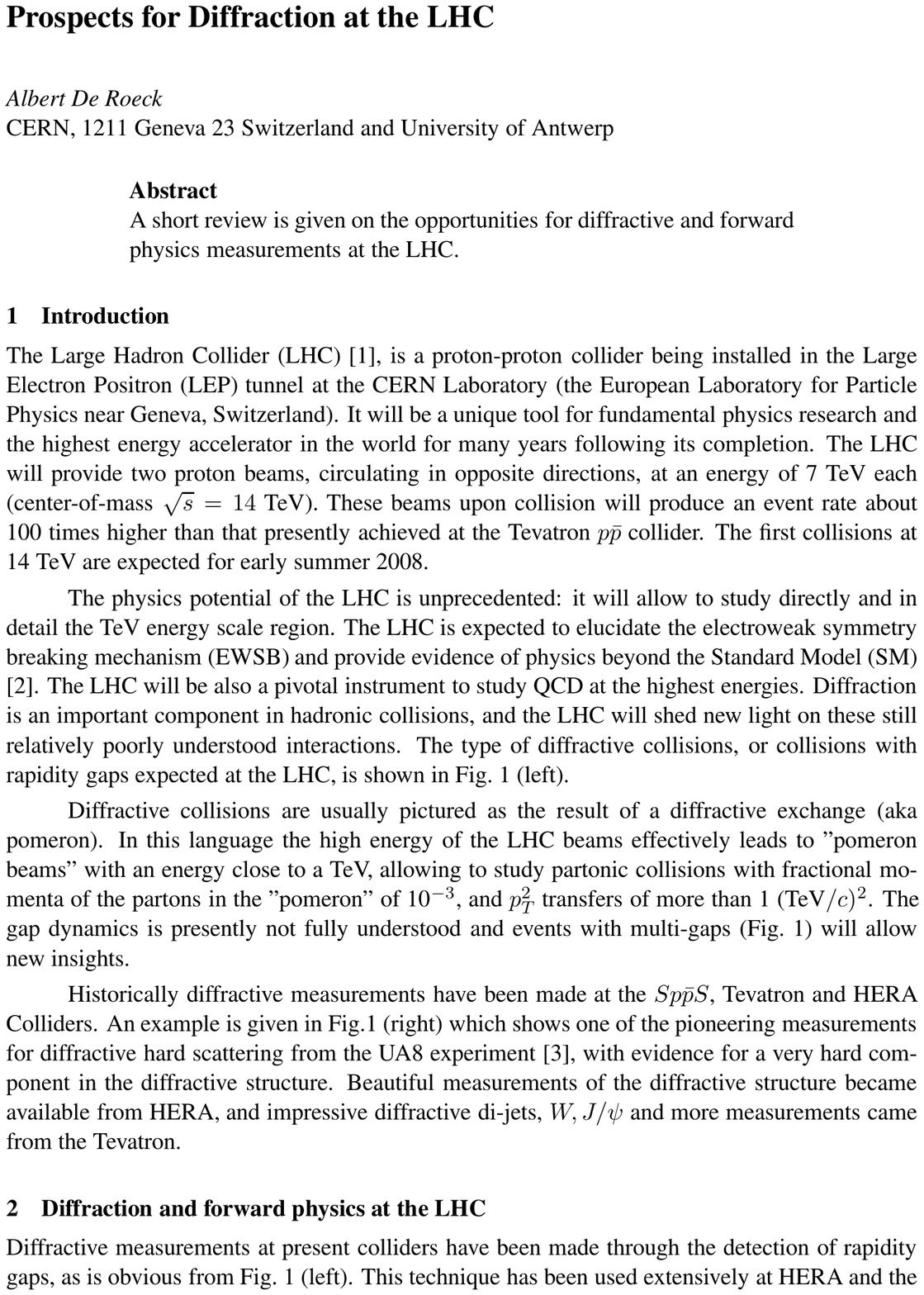} 

\coltoctitle{Multiple scattering, underlying event, and minimum bias} 
\coltocauthor{G Gustafson}
\Includeart{\CAUT}{\CTIT}{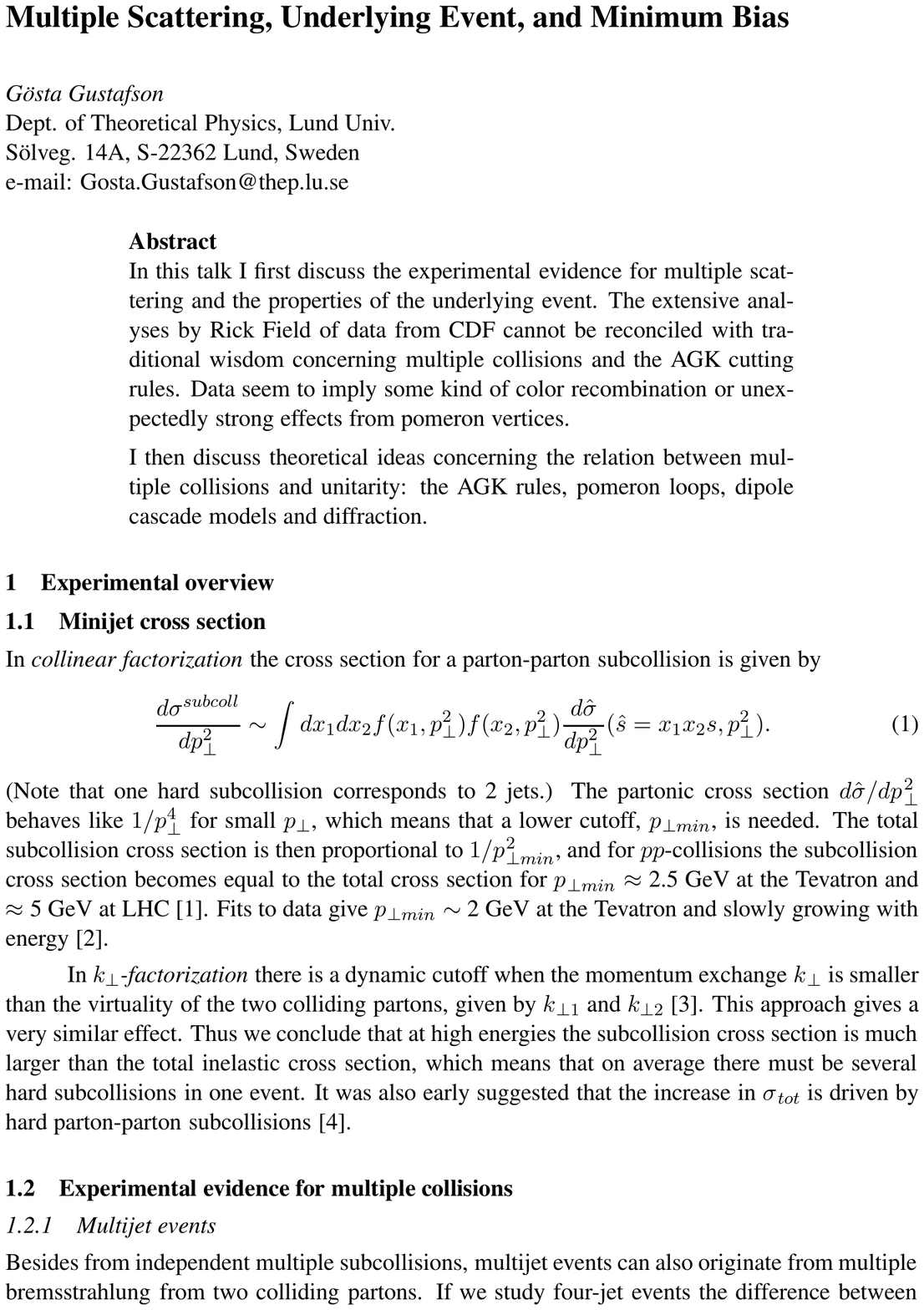} 

\coltoctitle{Multiple interactions and AGK rules in pQCD} 
\coltocauthor{M Salvadore,
J  Bartels, G. P Vacca}
\Includeart{\CAUT}{\CTIT}{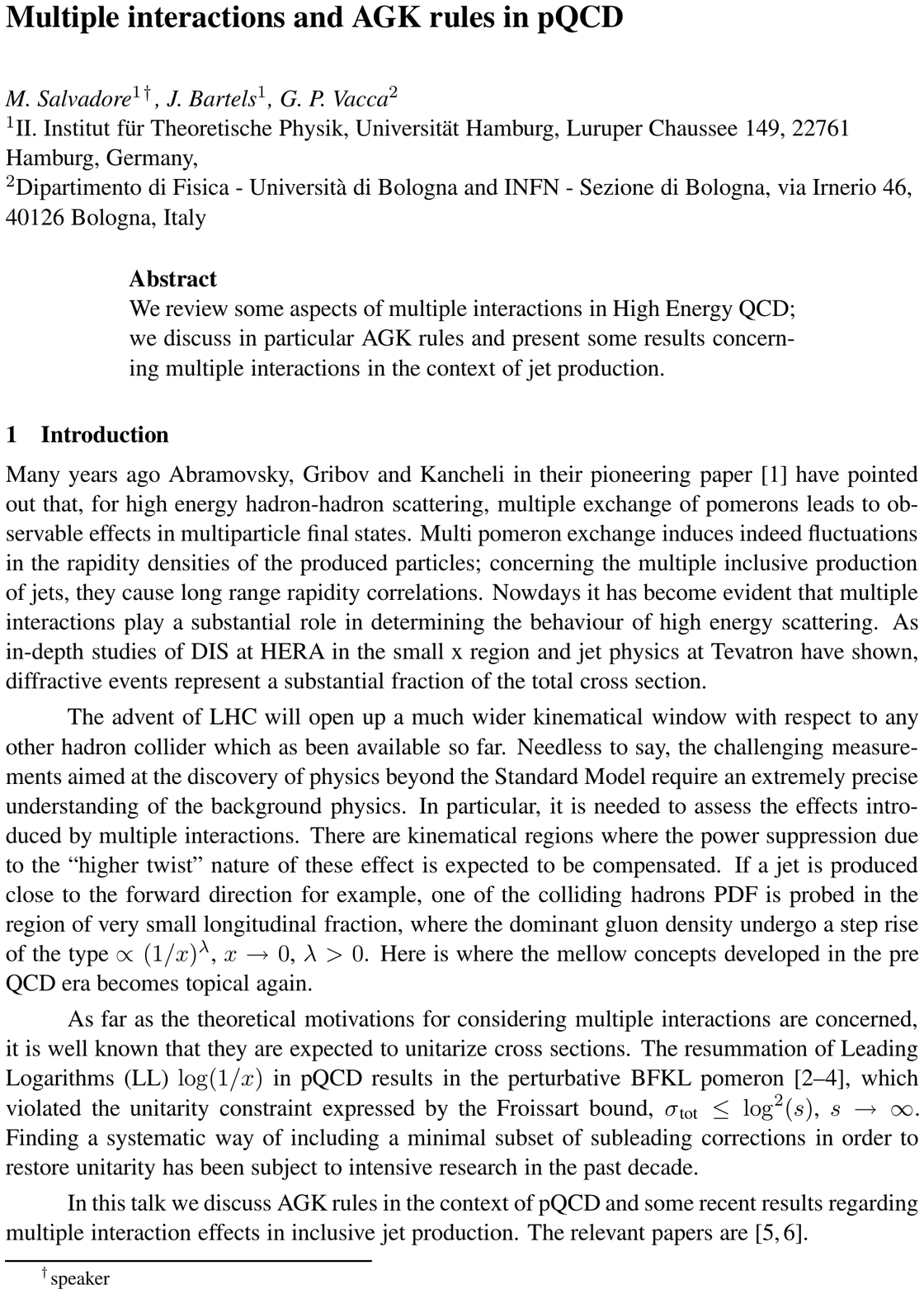} 

\coltoctitle{Multiple parton interactions, underlying event and forward physics at LHC} 
\coltocauthor{L Fan\`o}
\Includeart{\CAUT}{\CTIT}{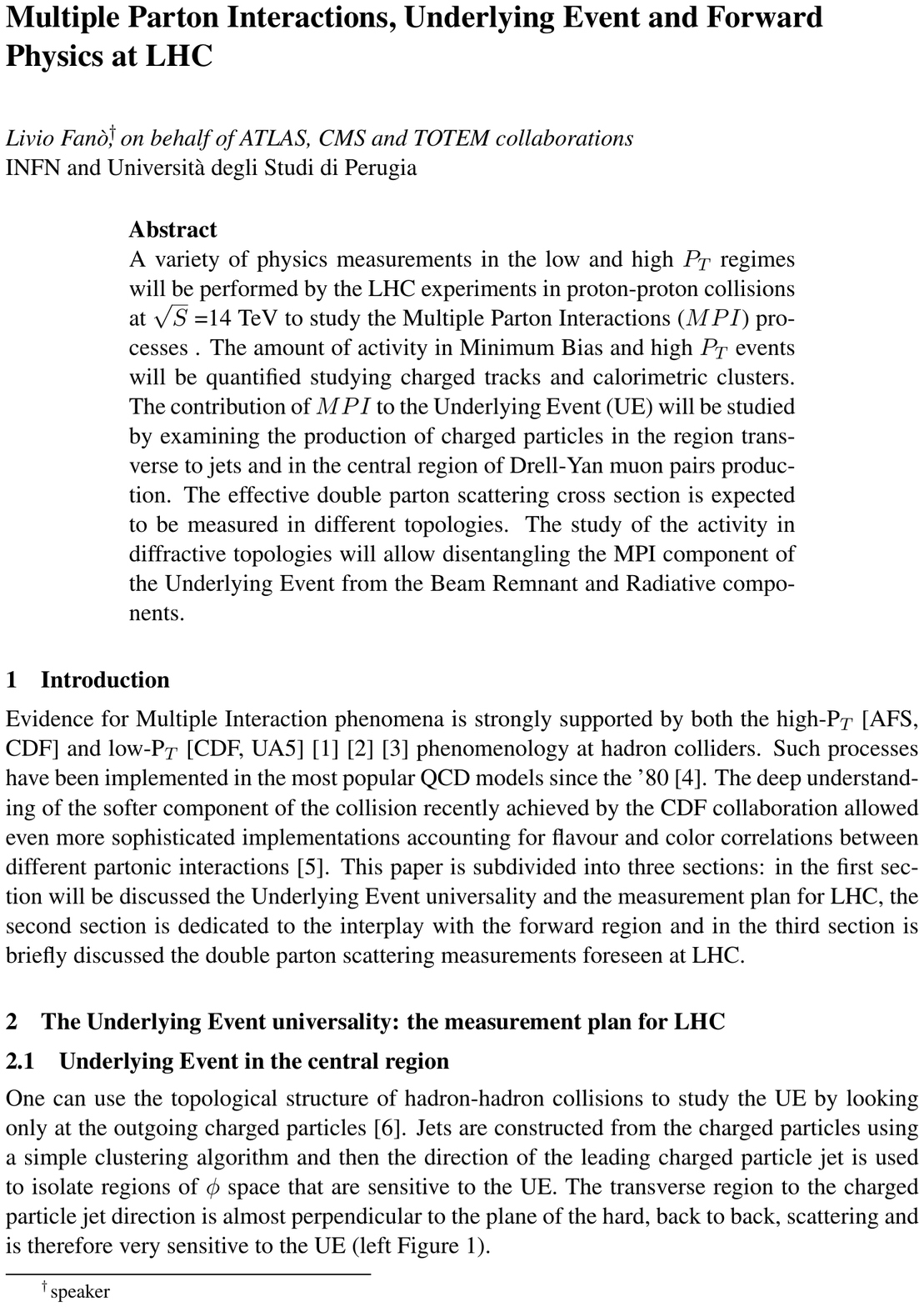} 

\coltoctitle{Elastic scattering, total cross section and luminosity measurements at ATLAS} 
\coltocauthor{C Sbarra}
\Includeart{\CAUT}{\CTIT}{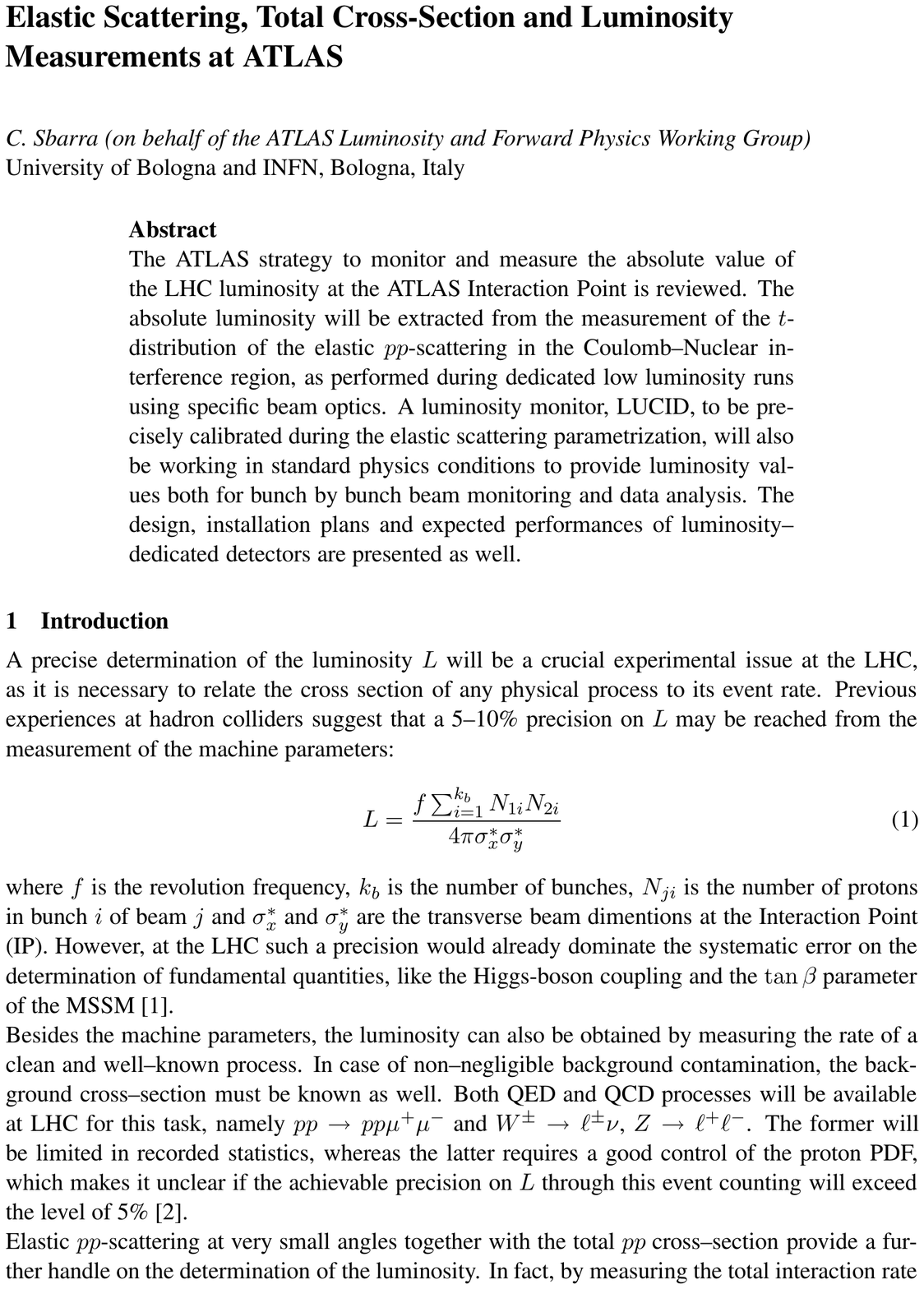} 

\coltoctitle{Diffraction at HERA and implications for Tevatron and LHC} 
\coltocauthor{L Schoeffel}
\Includeart{\CAUT}{\CTIT}{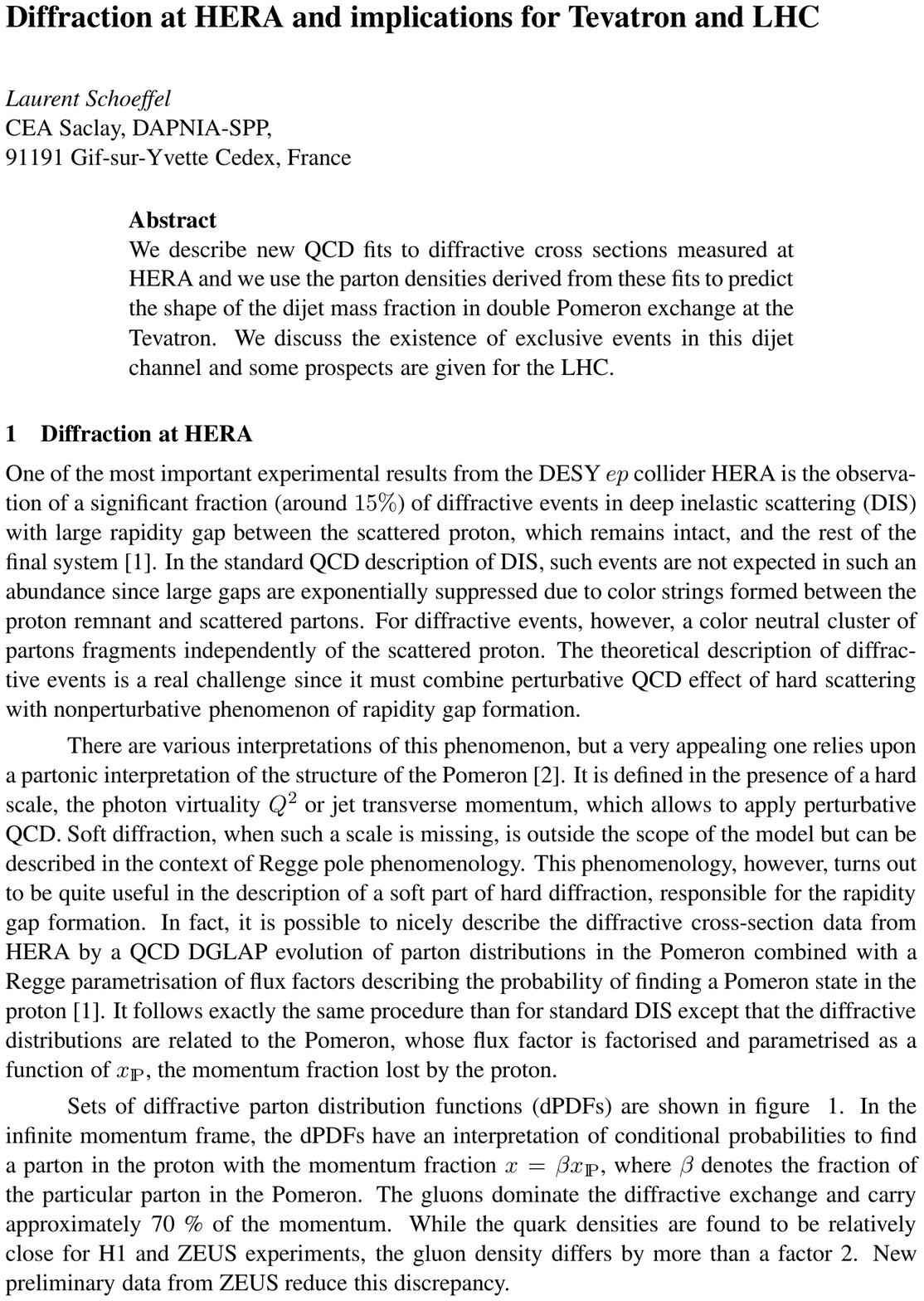} 

\coltoctitle{Search for exclusive events using the dijet mass fraction} 
\coltocauthor{O Kepka, C Royon}
\Includeart{\CAUT}{\CTIT}{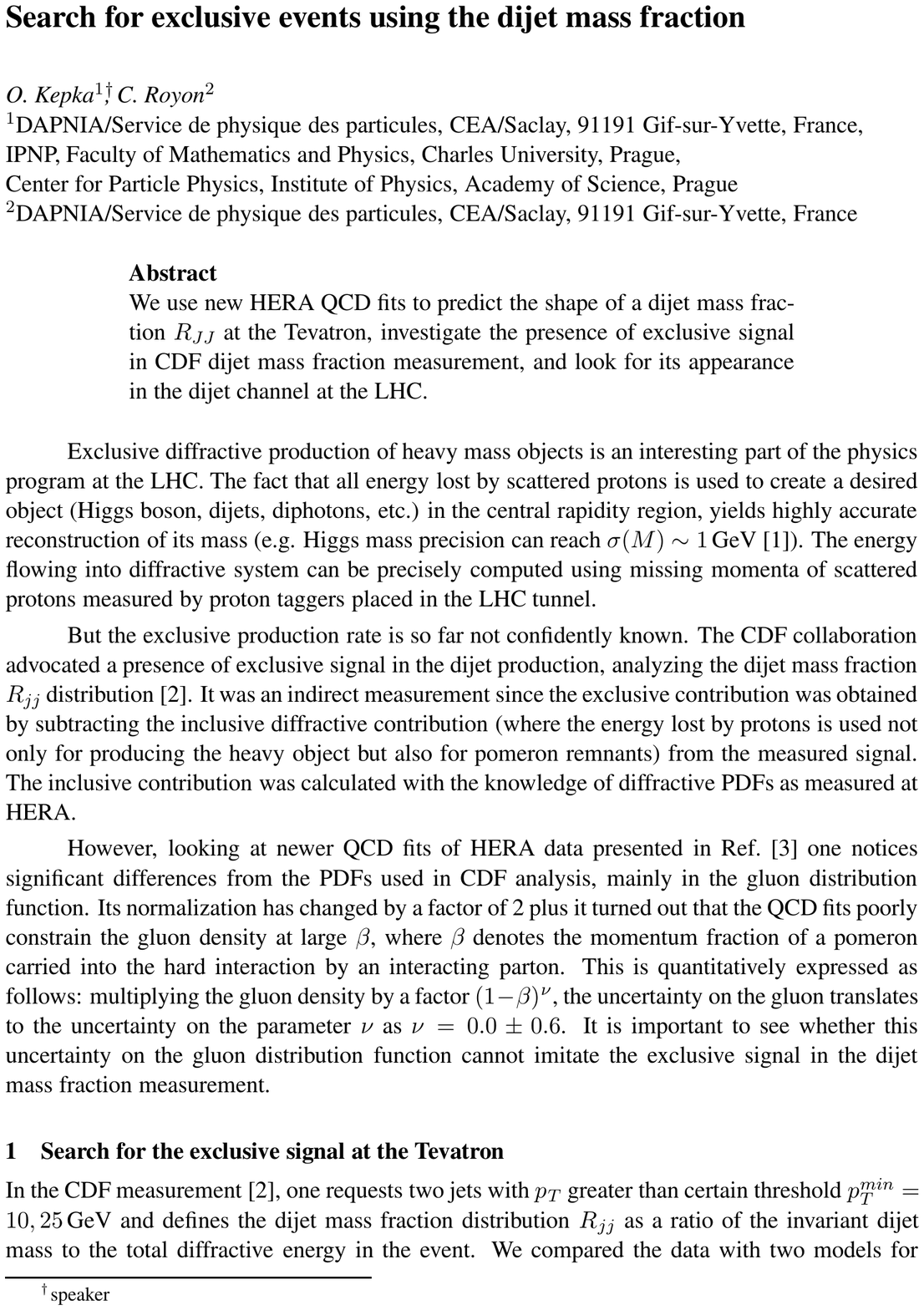} 

\coltoctitle{Diffractive production of quarkonia} 
\coltocauthor{A Szczurek}
\Includeart{\CAUT}{\CTIT}{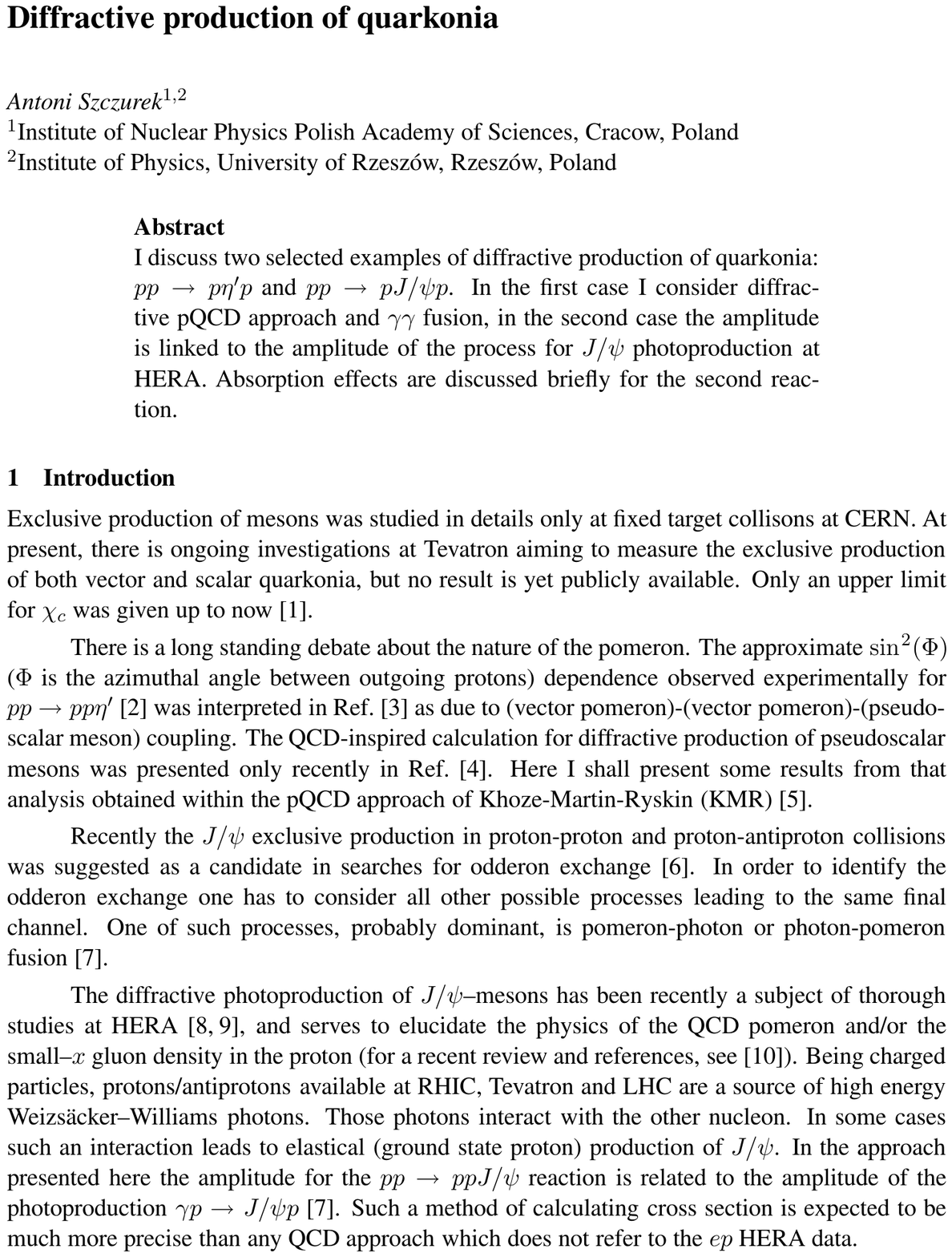} 

\coltoctitle{Rapidity gap survival in the black-disk regime} 
\coltocauthor{L Frankfurt, C.E Hyde, 
M. Strikman, C. Weiss} 
\Includeart{\CAUT}{\CTIT}{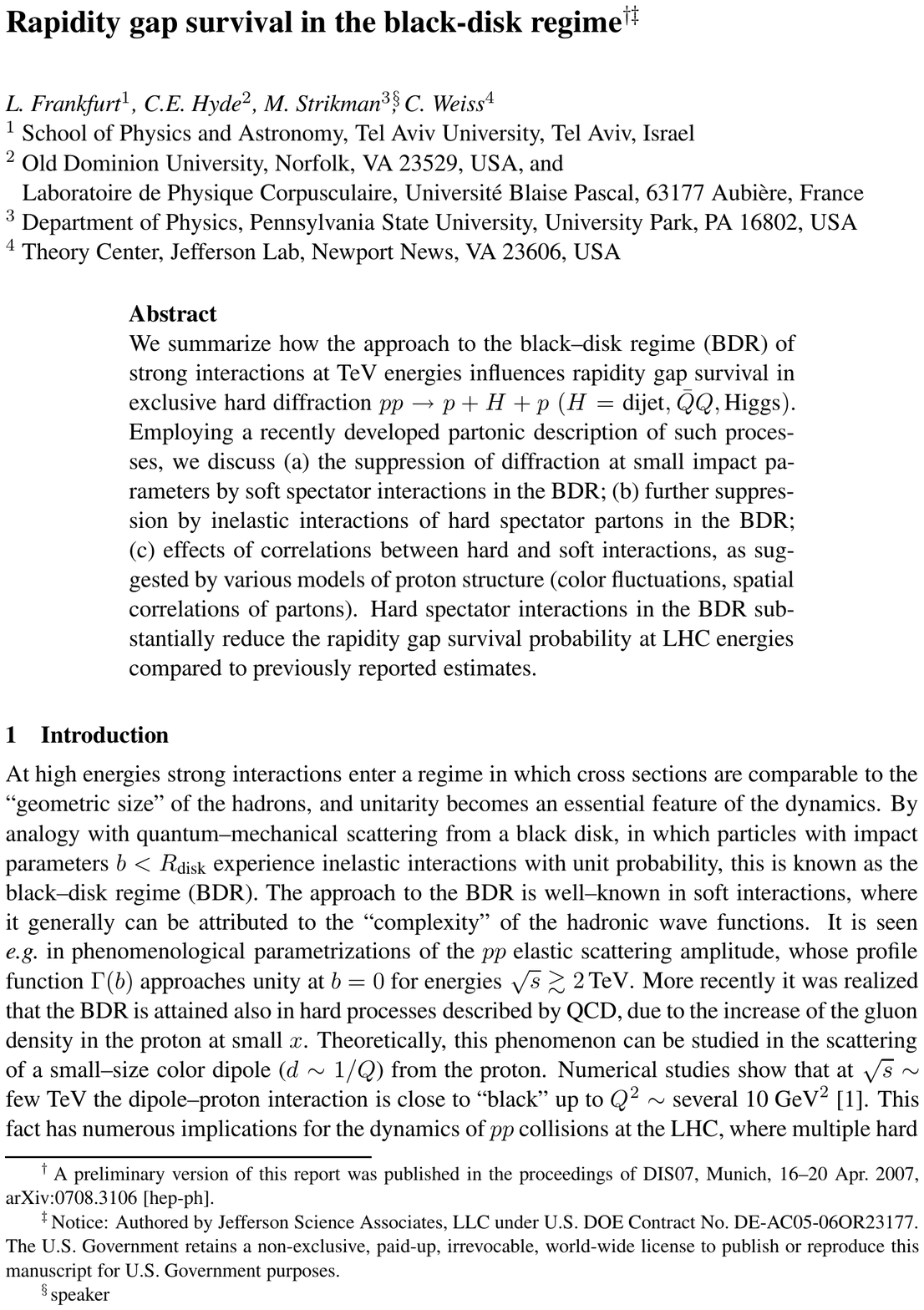} 

\coltoctitle{Pomeron intercept and slope: the QCD connection} 
\coltocauthor{K Goulianos }
\Includeart{\CAUT}{\CTIT}{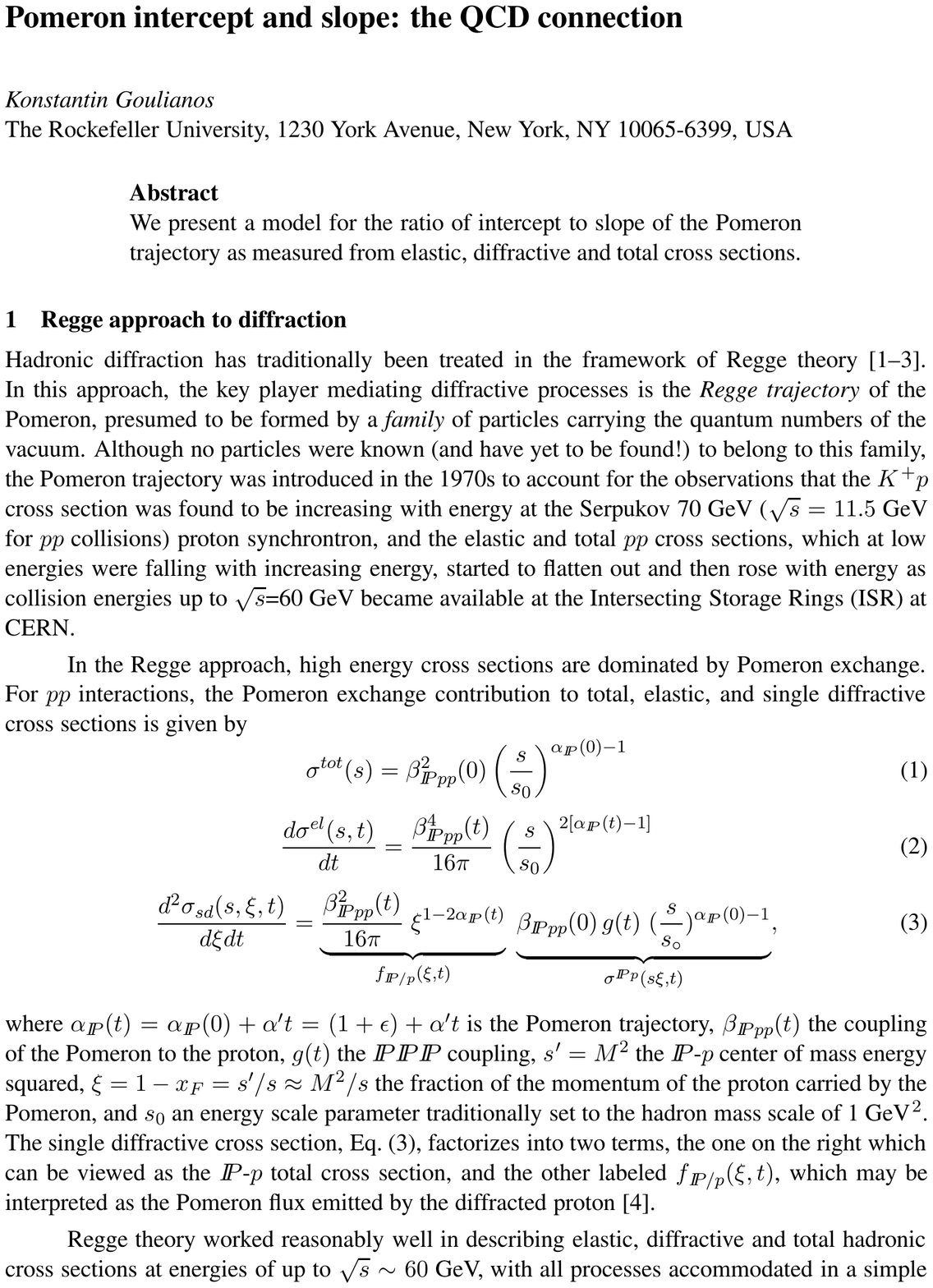} 

\coltoctitle{Exclusive $J/\psi$ and $\Upsilon$ hadroproduction  as a probe of the QCD Odderon} 
\coltocauthor{L Szymanowski}
\Includeart{\CAUT}{\CTIT}{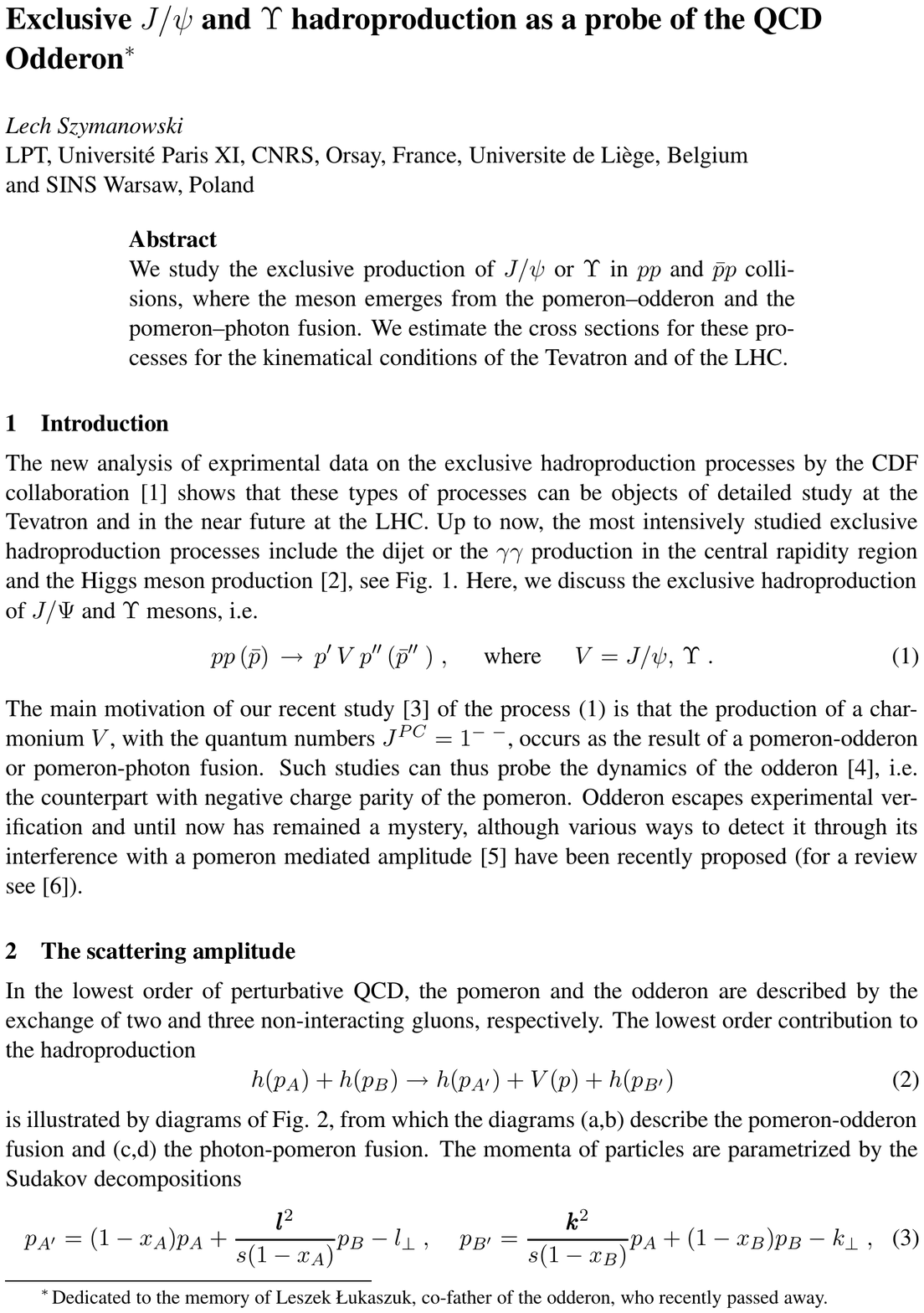} 

\coltoctitle{The soft and the hard pomerons: elastic scattering and unitarisation} 
\coltocauthor{J-R Cudell,
A Lengyel, E Martynov, O.V Selyugin}
\Includeart{\CAUT}{\CTIT}{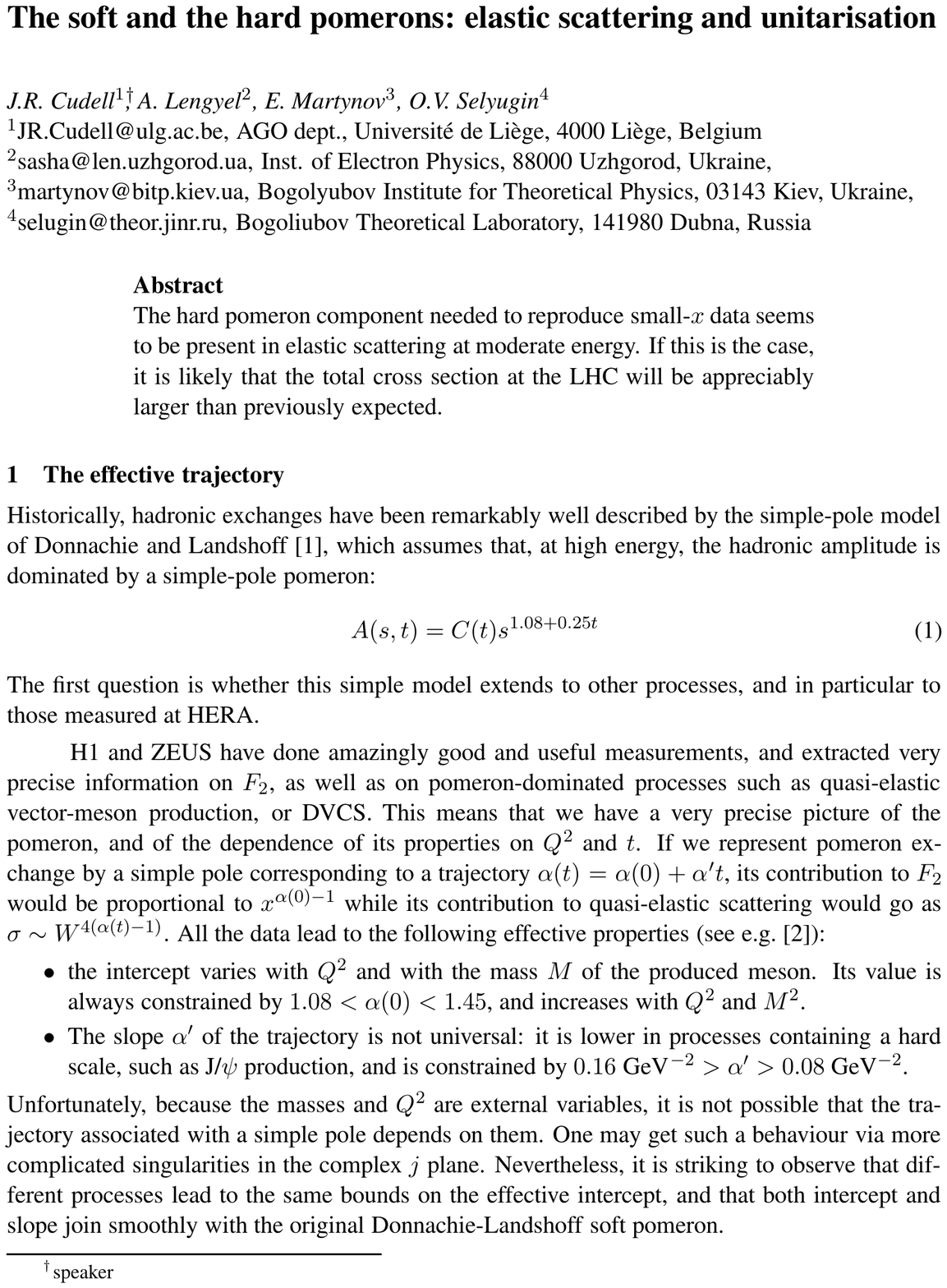} 

\coltoctitle{$pp$ elastic ecattering at LHC in a nucleon-structure model} 
\coltocauthor{M Islam,  J Ka$\check{s}$par R. J Luddy}
\Includeart{\CAUT}{\CTIT}{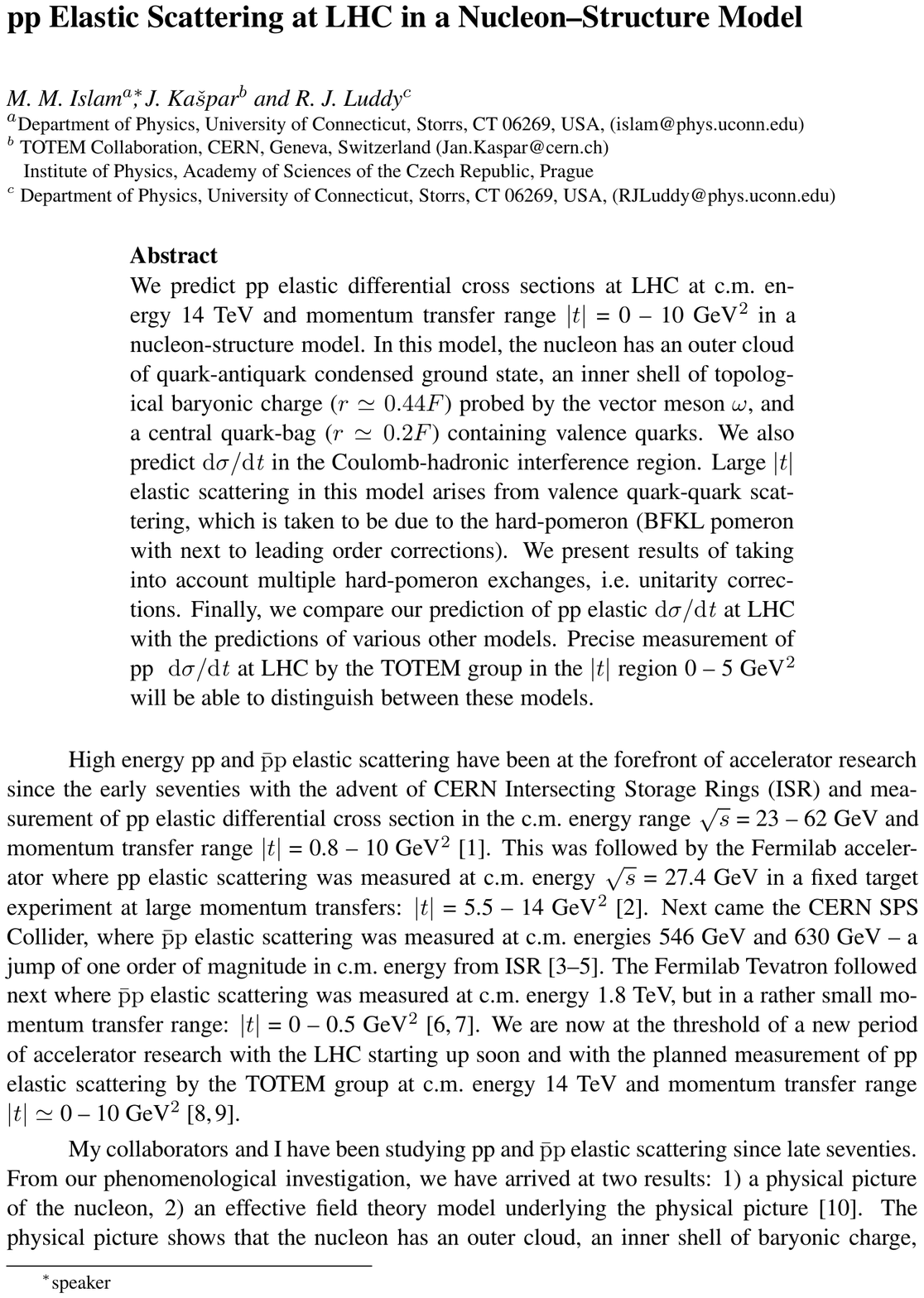} 

\coltoctitle{To the theory of high-energy nucleon collisions} 
\coltocauthor{V Kundrat, Jan Ka\v{s}par, 
Milo\v{s} Lokaj\'{i}\v{c}ek}
\Includeart{\CAUT}{\CTIT}{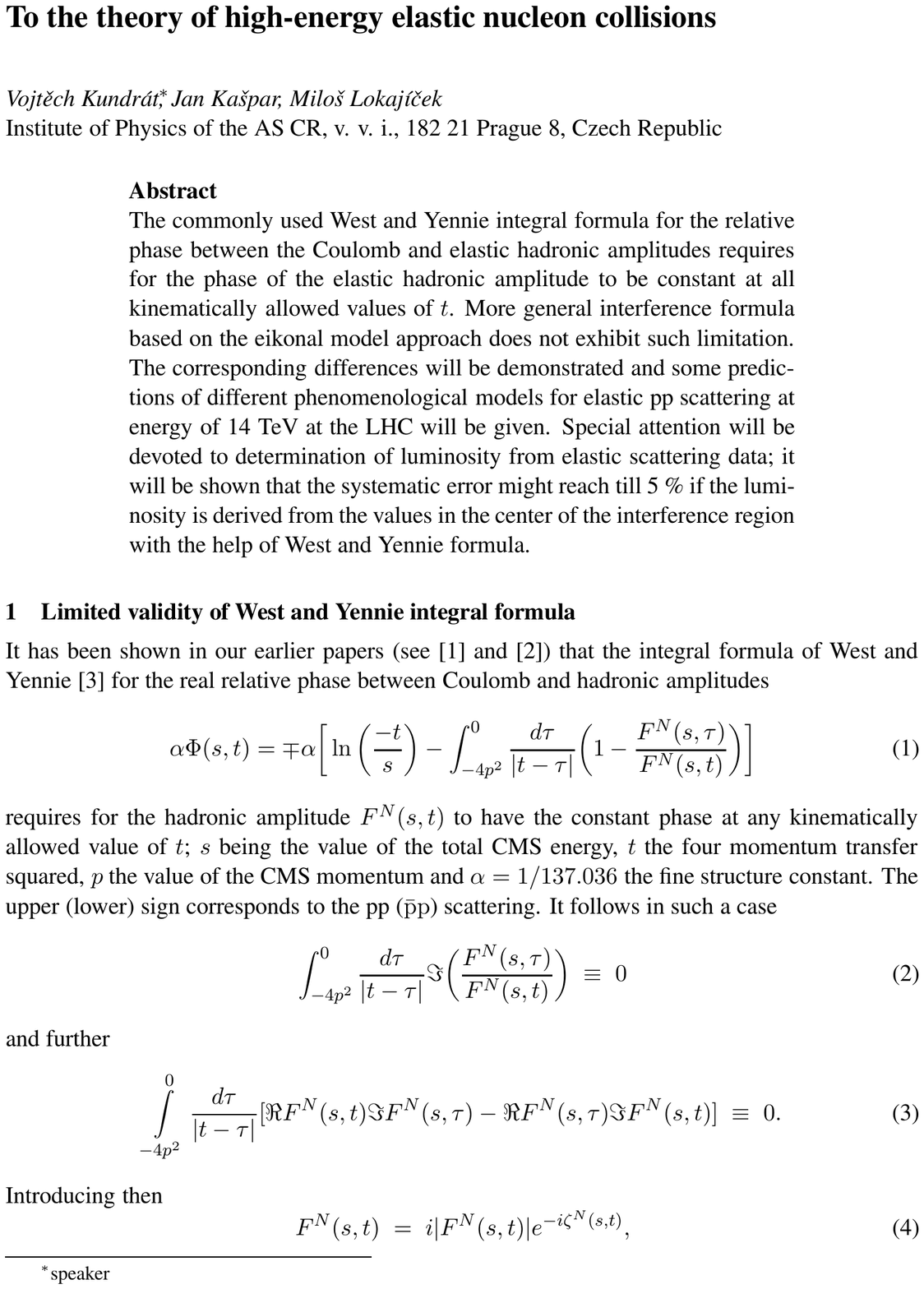} 

\coltoctitle{Saturation effects in elastic scattering at the LHC} 
\coltocauthor{O Selyugin, J.-R Cudell}
\Includeart{\CAUT}{\CTIT}{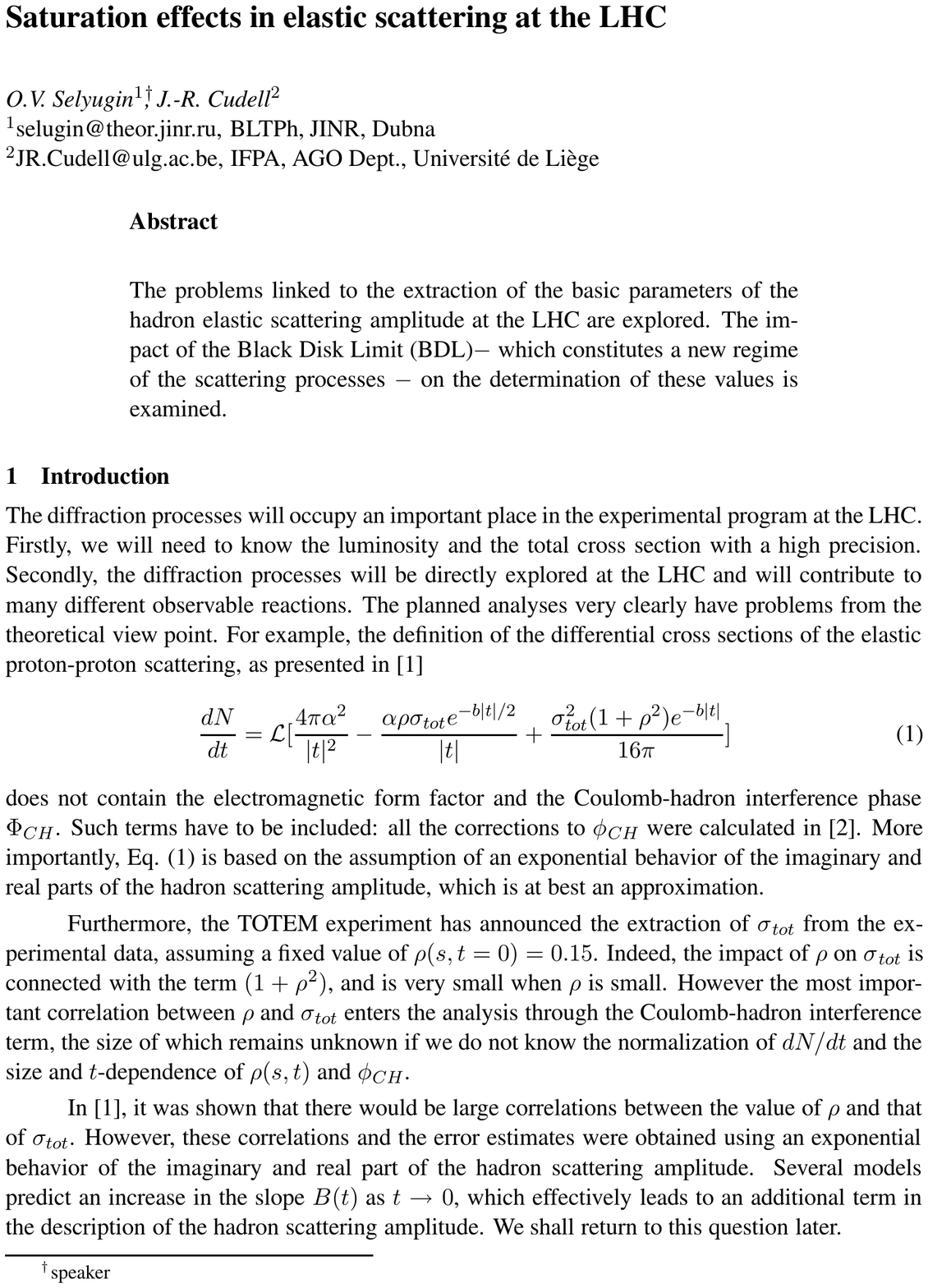} 

\coltoctitle{Elastic $pp$ and $\bar pp$ scattering in the models of unitarized pomeron} 
\coltocauthor{E Martynov}
\Includeart{\CAUT}{\CTIT}{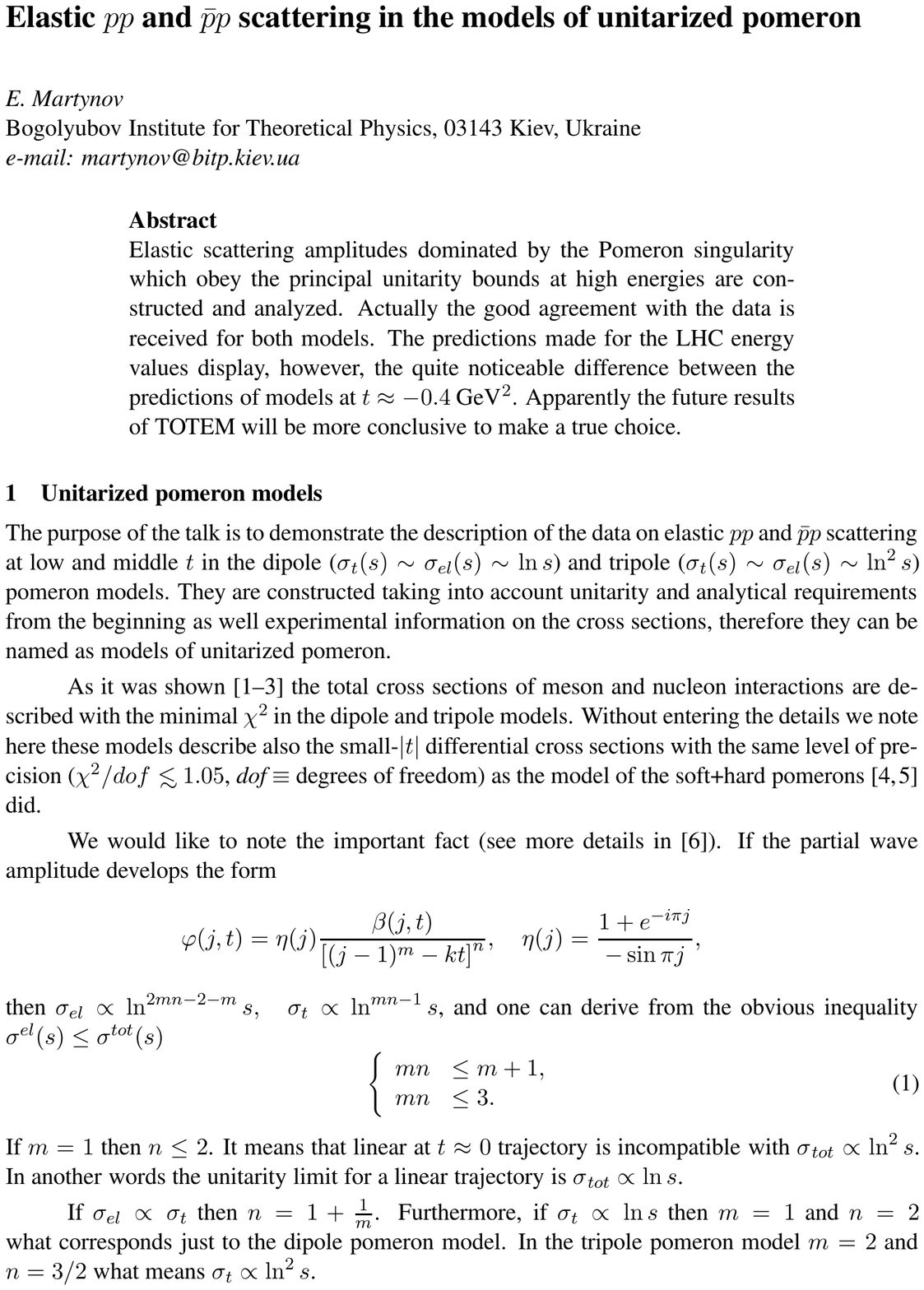}

\end{papers} 

\CLDP 

\part{Heavy-Ion Collisions} 
\label{hi} 
\newpage

\begin{papers} 

\coltoctitle{Forward Physics with BRAHMS in pp and dAu  collisions at RHIC} 
\coltocauthor{D R\"ohrich}
\Includeart{\CAUT}{\CTIT}{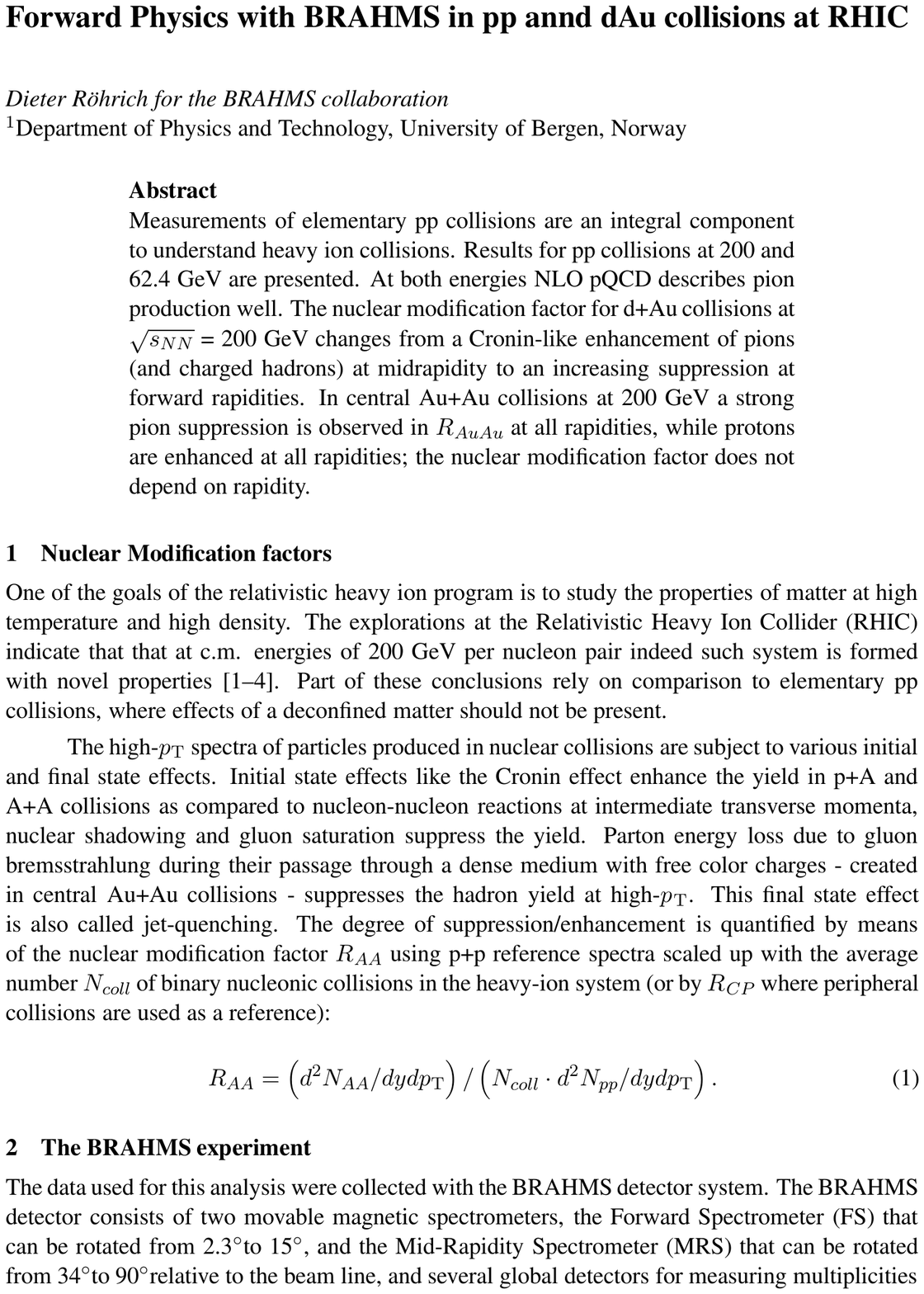}

\coltoctitle{Manifestations of gluon saturation at RHIC} 
\coltocauthor{J Albacete}
\Includeart{\CAUT}{\CTIT}{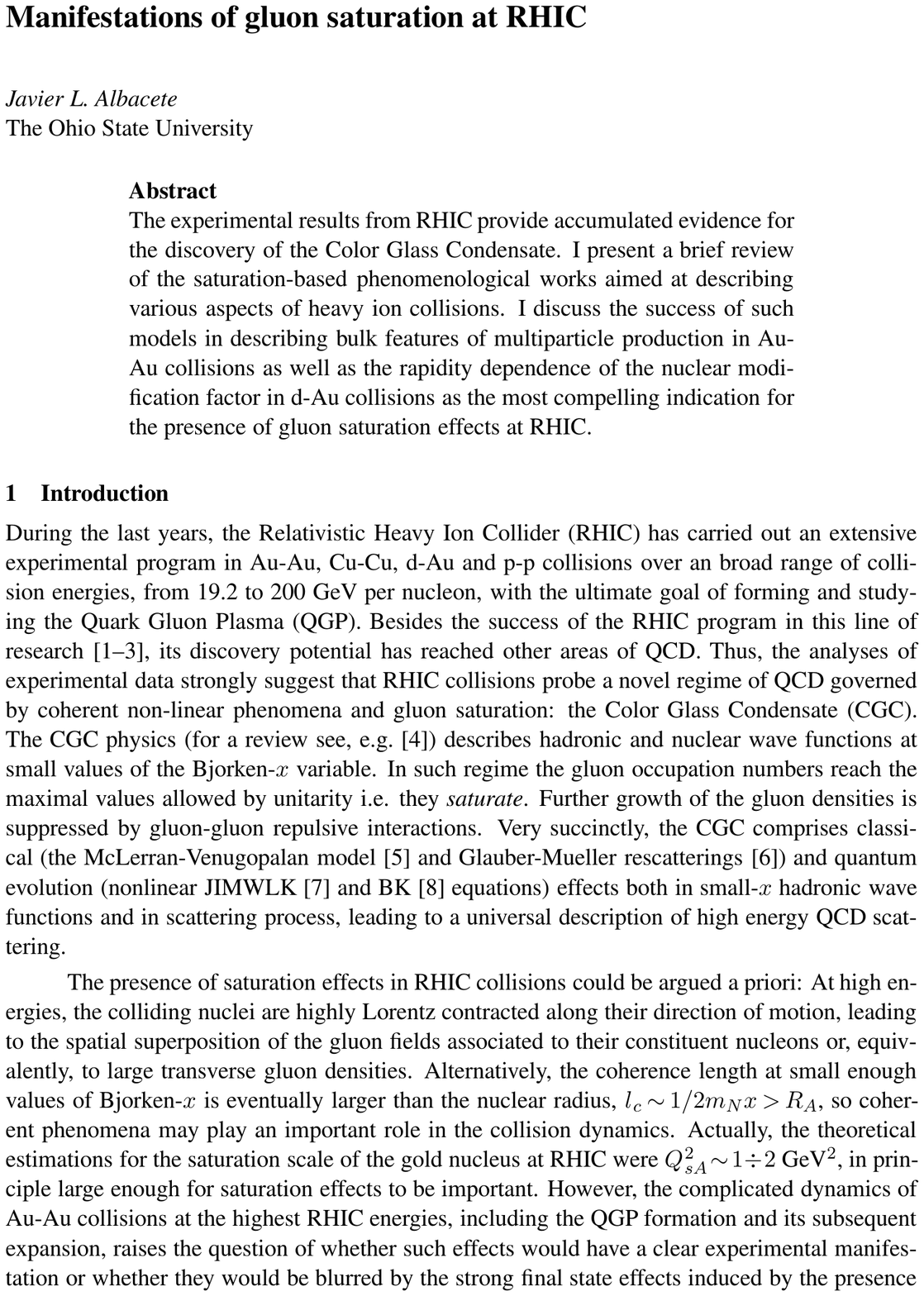}

\coltoctitle{Measurement of the cross section and the single transverse spin asymmetry
 in very forward neutron production from polarized $pp$ collisions at RHIC} 
\coltocauthor{M Togawa}
\Includeart{\CAUT}{\CTIT}{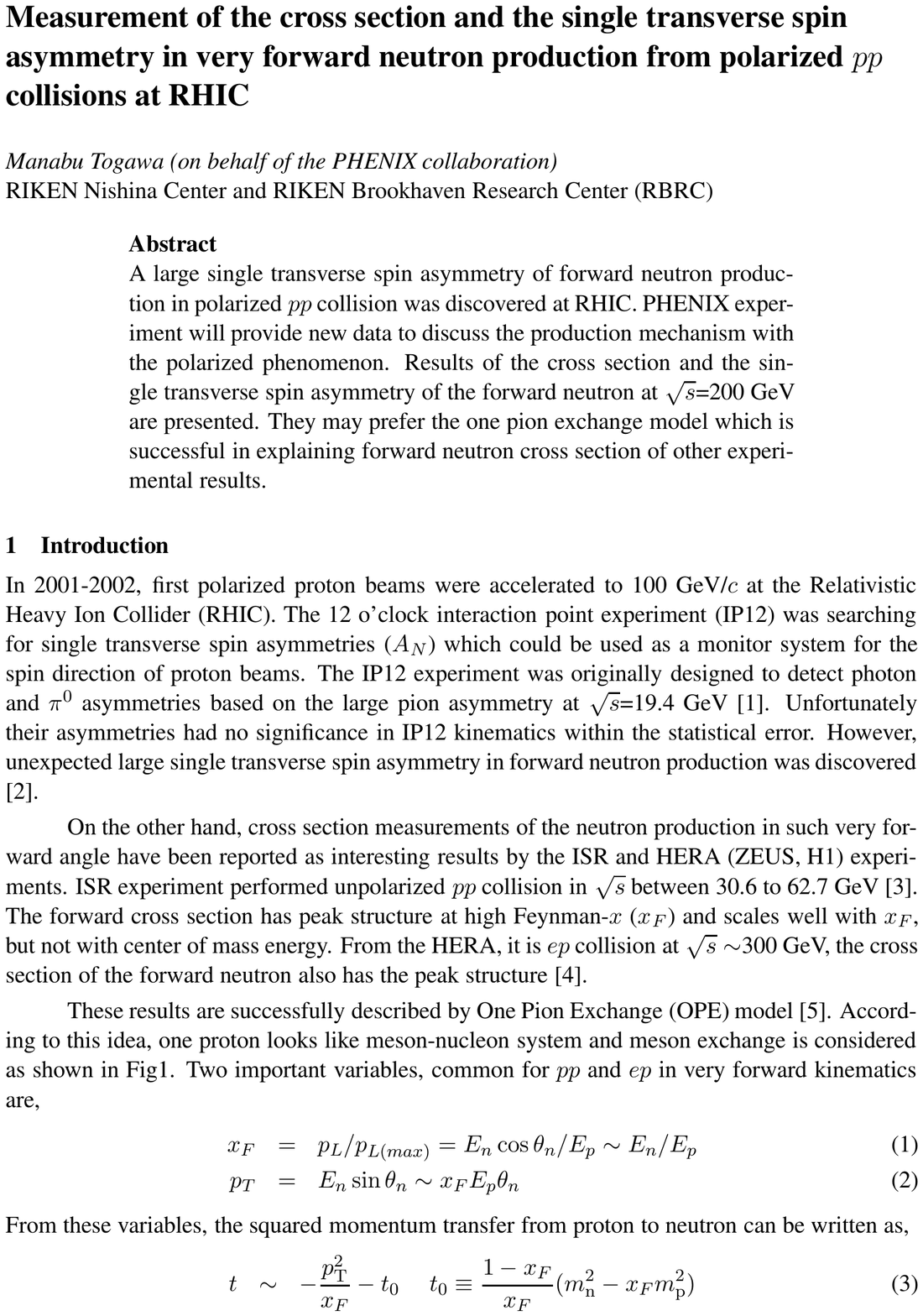}

\coltoctitle{Gluon saturation: from pp to AA } 
\coltocauthor{A Kovner}
\Includeart{\CAUT}{\CTIT}{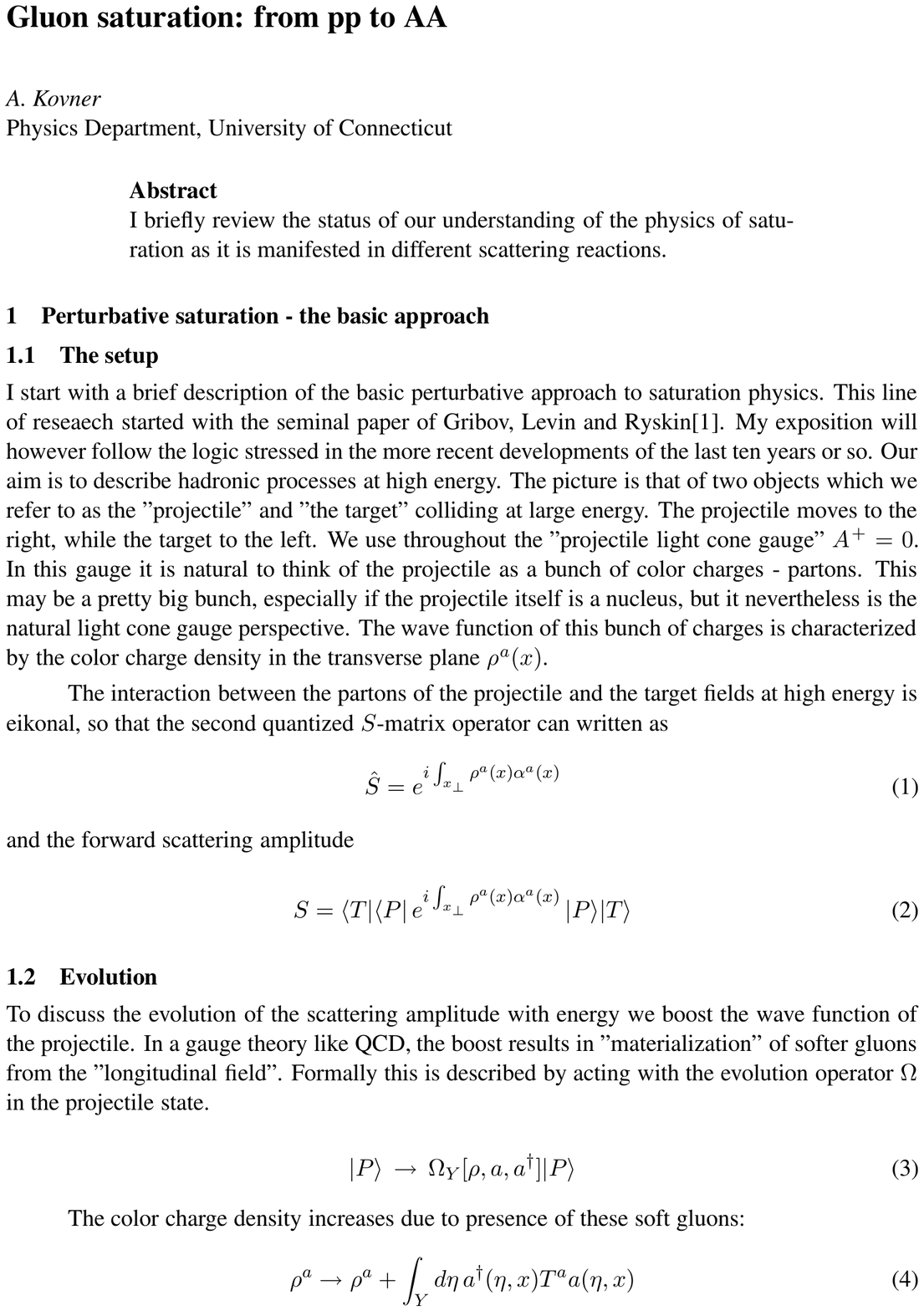}

\coltoctitle{Photoproduction in ultra-peripheral heavy-ion collisions} 
\coltocauthor{J Nystrand}
\Includeart{\CAUT}{\CTIT}{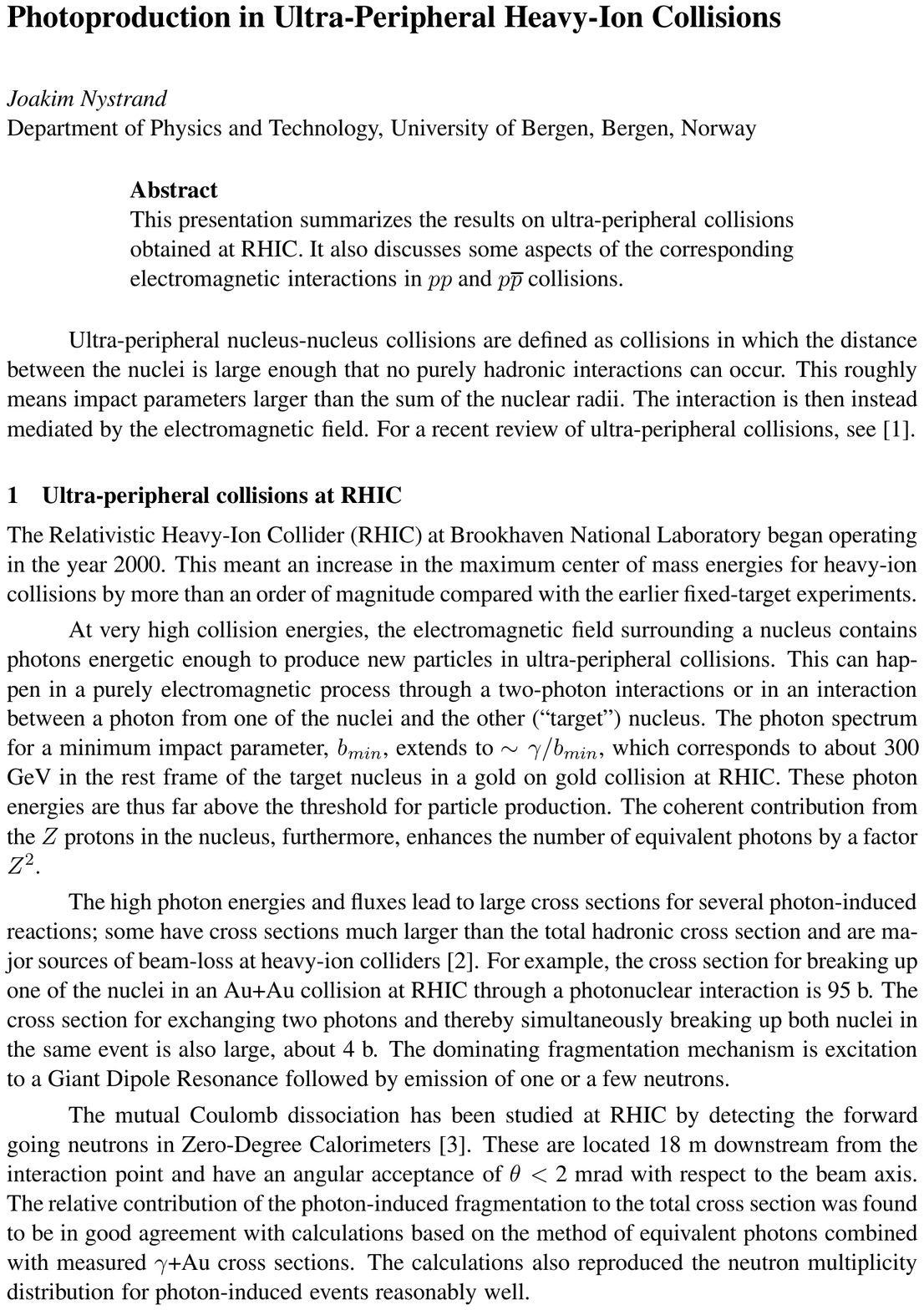}

\coltoctitle{Fractional energy losses in the black disk regime and BRAHMS effect} 
\coltocauthor{L Frankfurt, M Strikman}
\Includeart{\CAUT}{\CTIT}{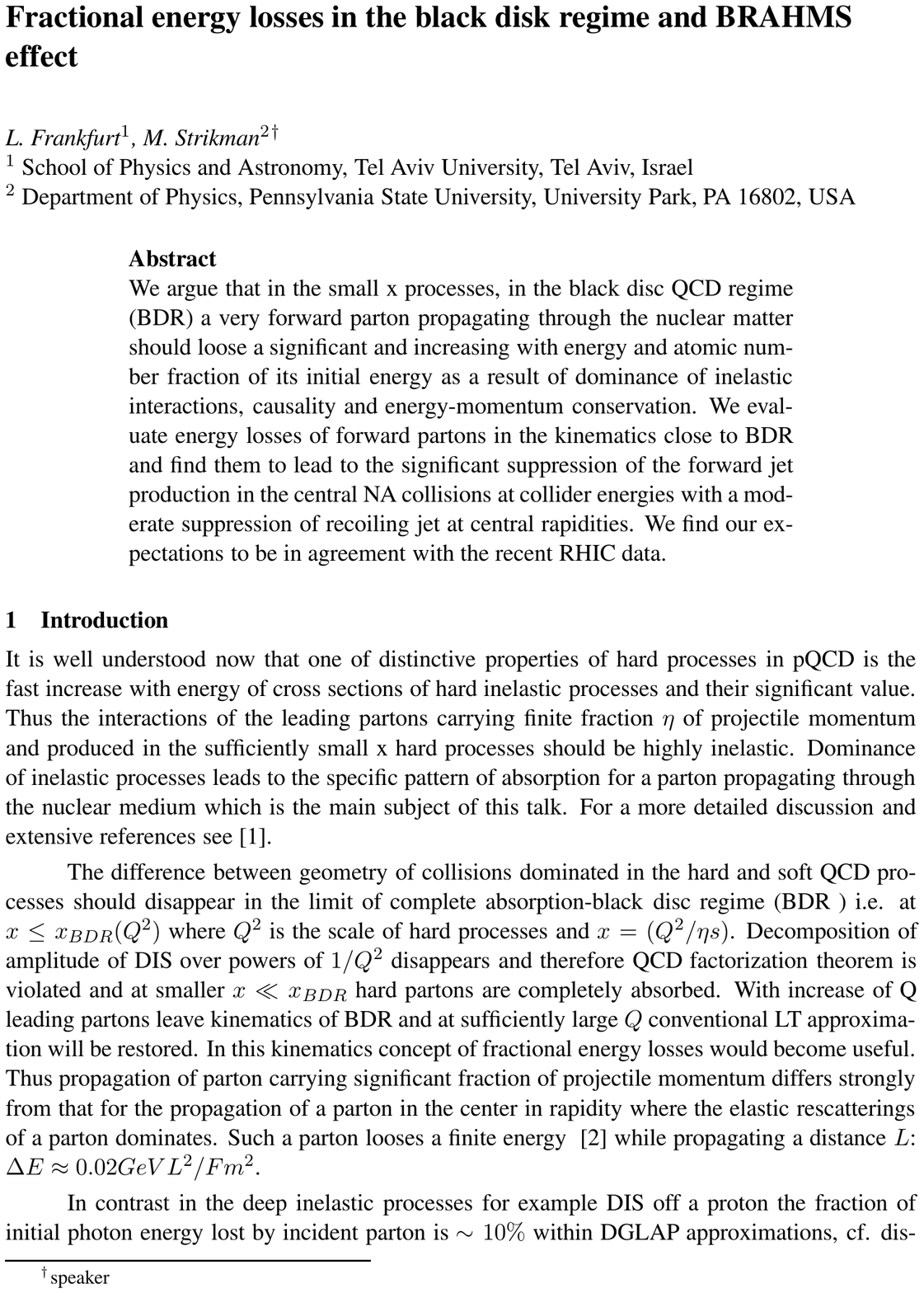}

\coltoctitle{The coherent inelastic processes on nuclei at ultrarelativistic energies} 
\coltocauthor{V.L Lyuboshitz , V V Lyuboshitz}
\Includeart{\CAUT}{\CTIT}{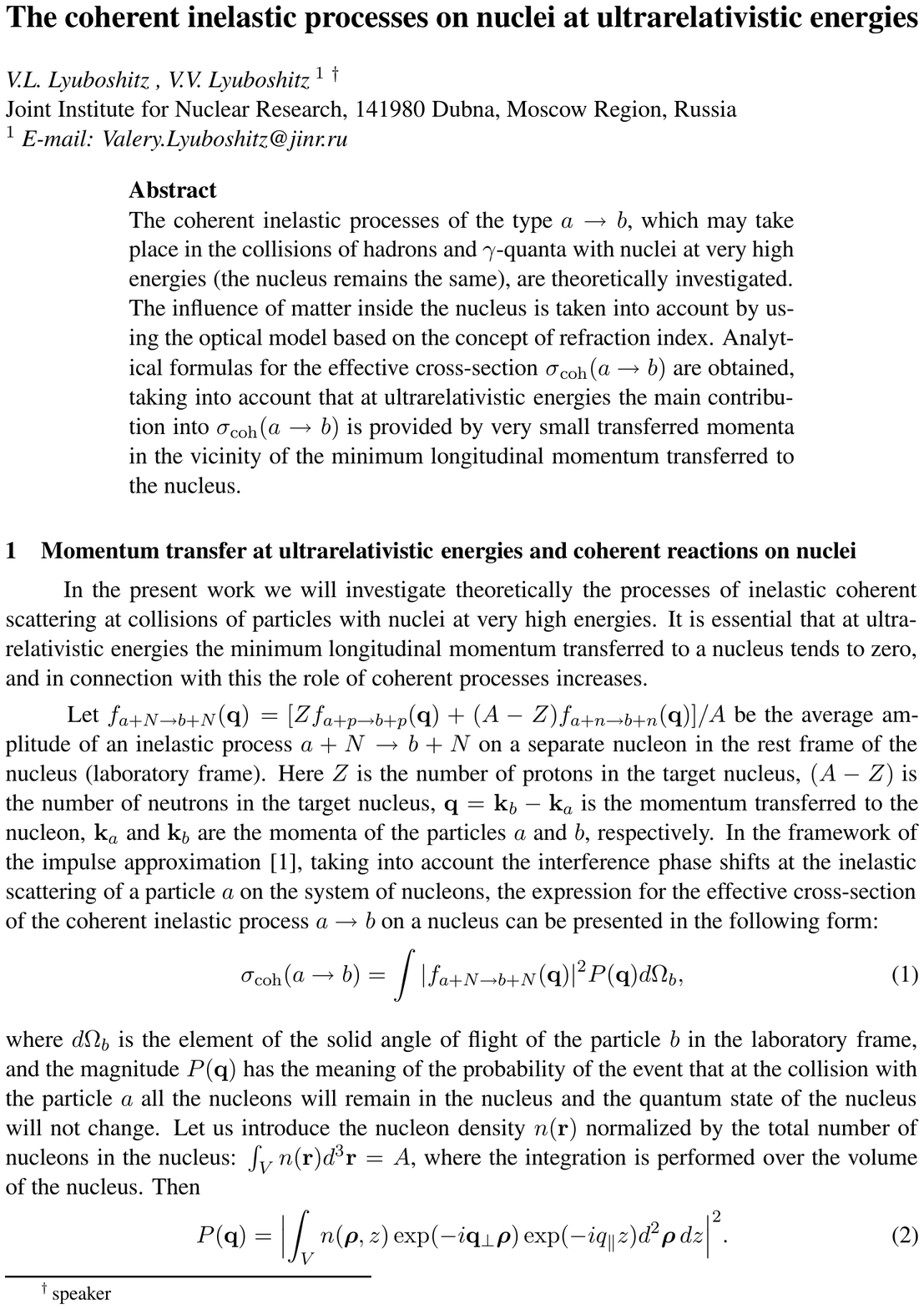}

\end{papers} 

\CLDP 

\part{Opportunities at future colliders} 
\label{future} 
\newpage

\begin{papers} 

\coltoctitle{Low $x$ and diffractive physics at future electron-proton/ion colliders} 
\coltocauthor{H Kowalski}
\Includeart{\CAUT}{\CTIT}{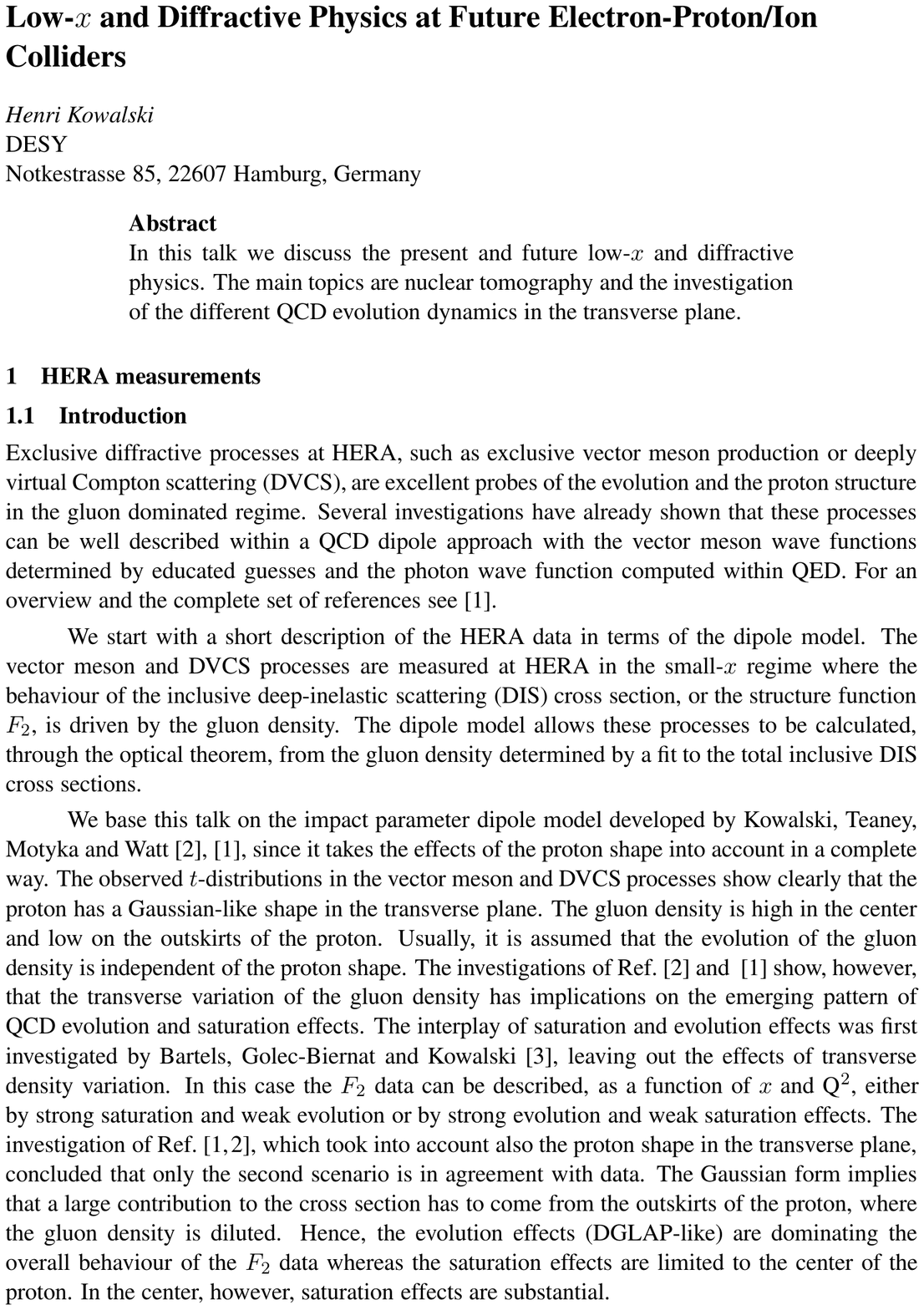} 

\coltoctitle{Low-$x$ physics at a future electron-proton/ion (EIC) collider facility} 
\coltocauthor{B Surrow}
\Includeart{\CAUT}{\CTIT}{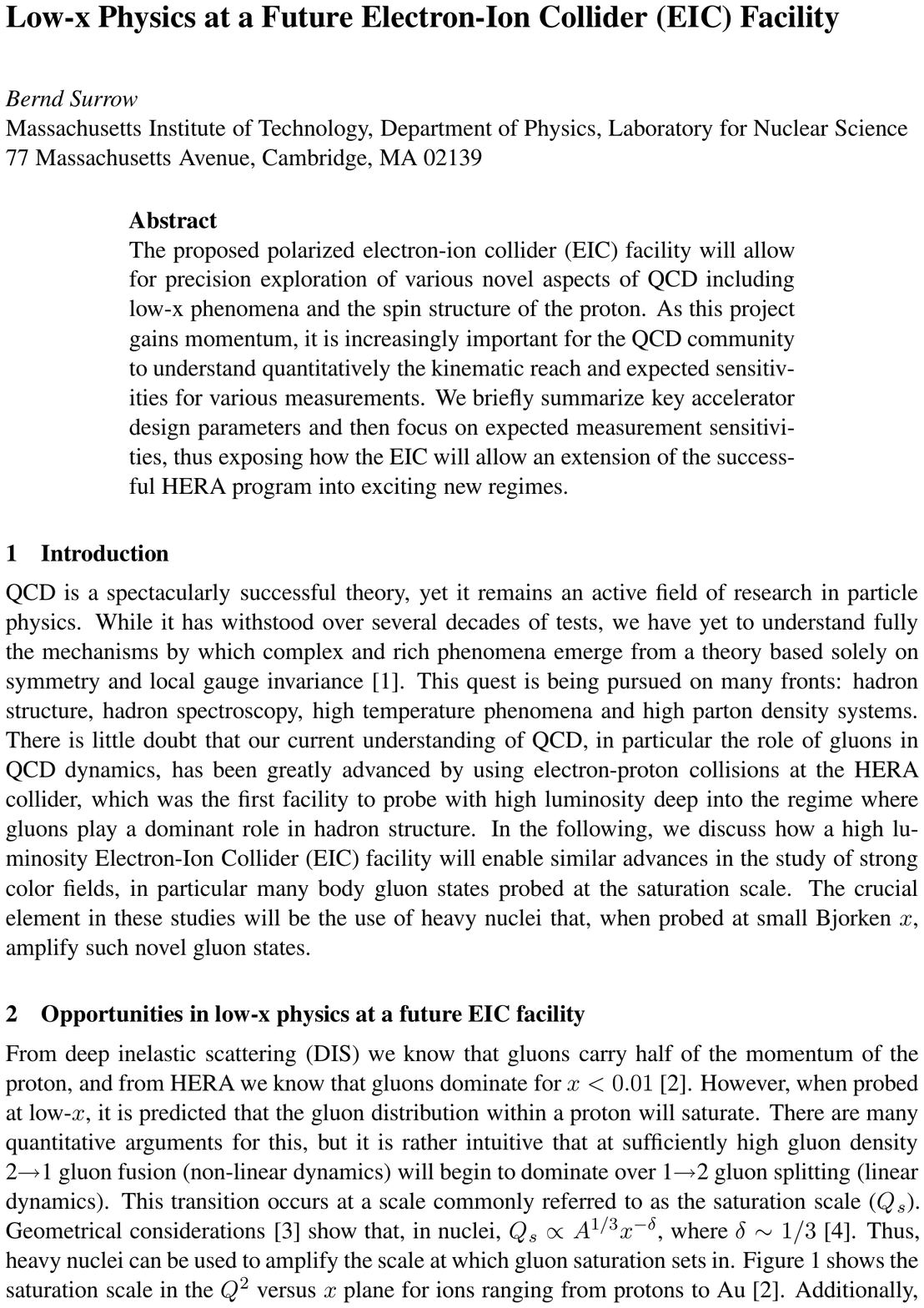}

\coltoctitle{The LHeC and its low $x$ Physics Potential } 
\coltocauthor{J Dainton, M Klein, P Newman}
\Includeart{\CAUT}{\CTIT}{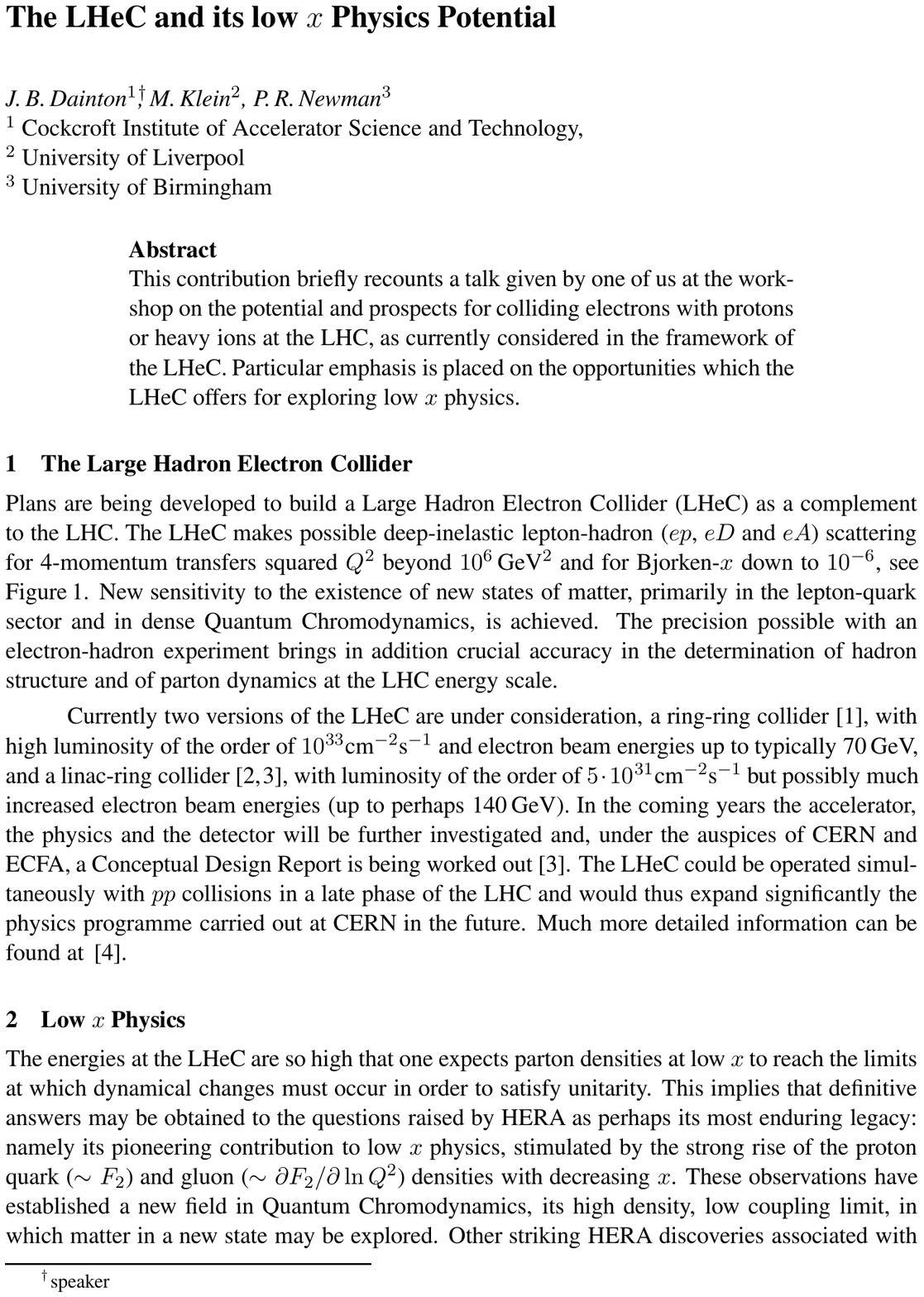} 

\coltoctitle{Perturbative QCD in the Regge limit: prospects at ILC} 
\coltocauthor{S Wallon}
\Includeart{\CAUT}{\CTIT}{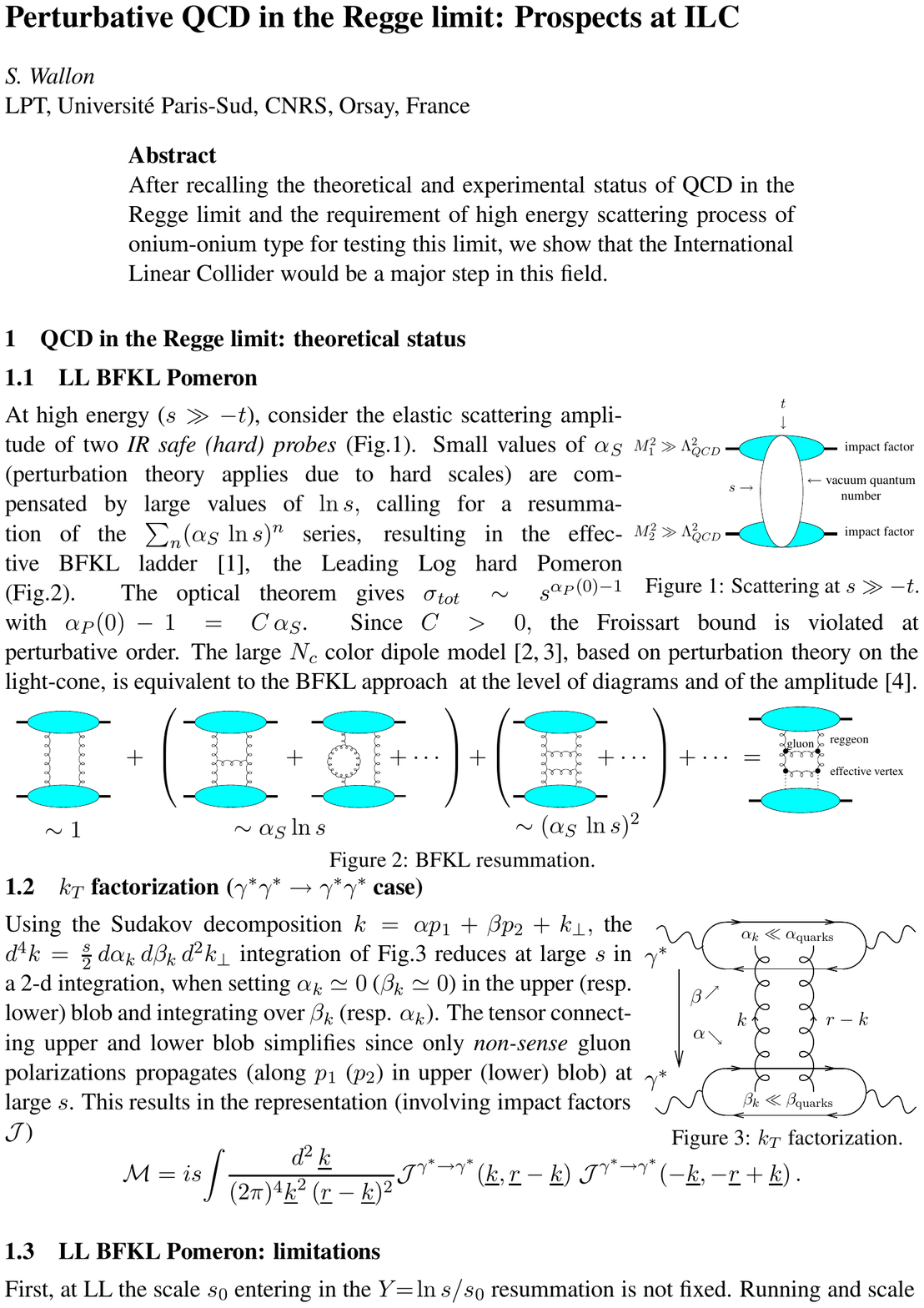}

\end{papers} 

\CLDP 

\part{Cosmic Rays and Astroparticle Physics} 
\label{cosmics} 
\newpage

\begin{papers} 

\coltoctitle{On the measurement of the proton-air cross section using cosmic ray
  data} 
\coltocauthor{R Ulrich,  J Bl{\"u}mer, R Engel, F Sch{\"u}ssler, M Unger}
\Includeart{\CAUT}{\CTIT}{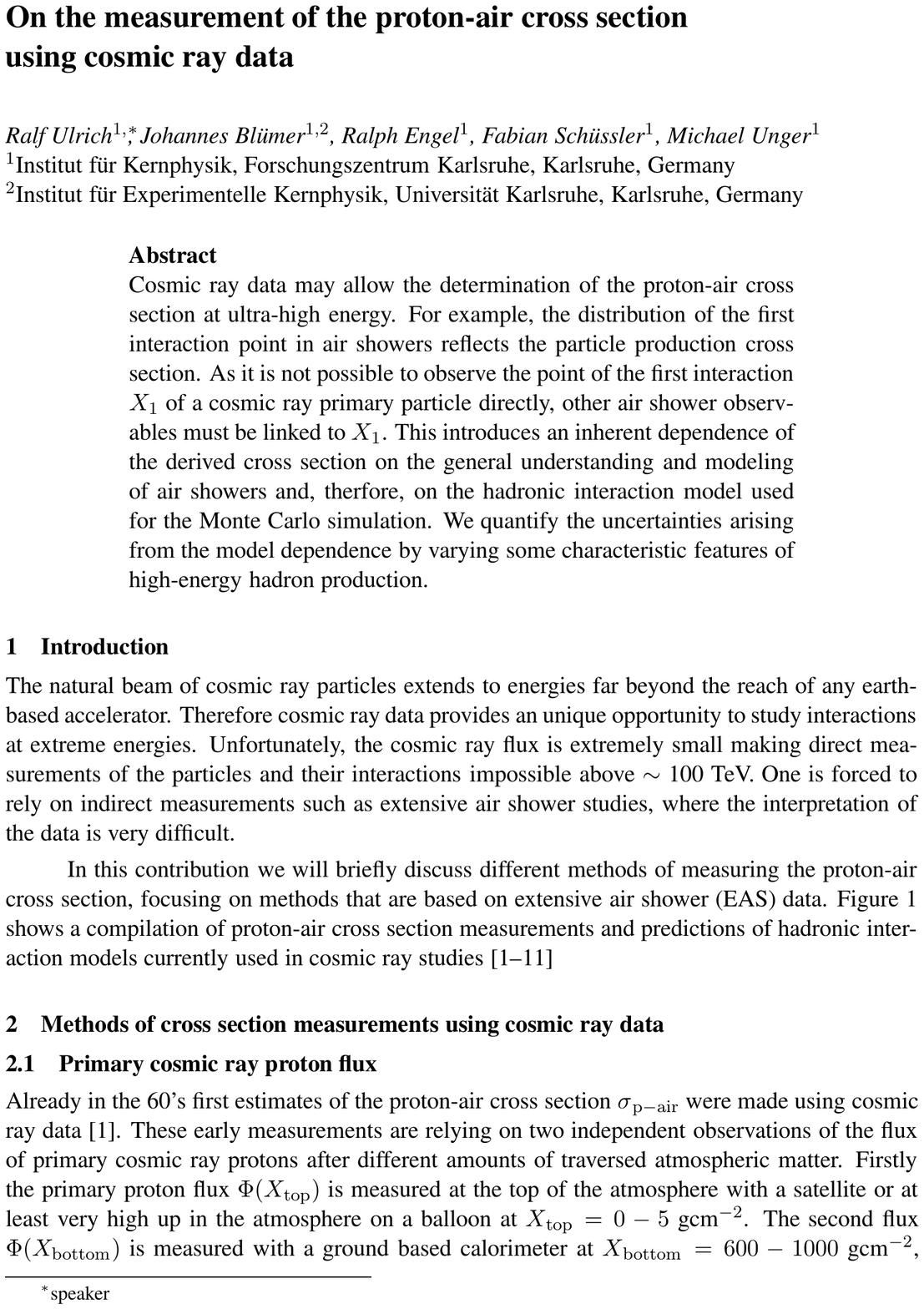}

\coltoctitle{Measuring proton-air Inelastic Cross-Section using Cosmic Ray Data
(not received)} 
\coltocauthor{K Belov}

\coltoctitle{EAS-TOP:The proton-air inelastic cross-section at $\sqrt{s} \approx $ 2 TeV} 
\coltocauthor{G. C Trinchero}
\Includeart{\CAUT}{\CTIT}{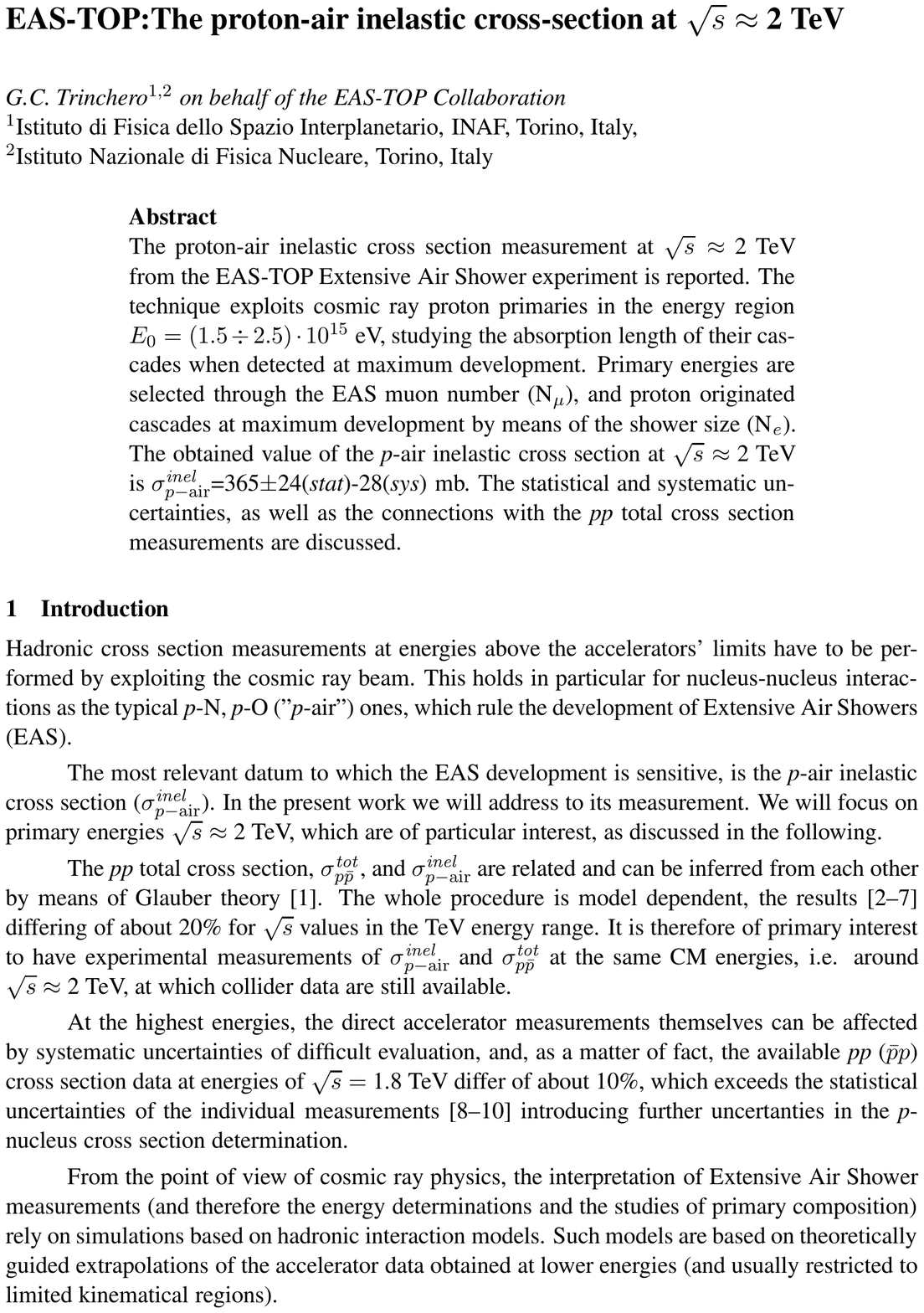} 

\coltoctitle{UHE Cosmic Rays above $10^{17}$ eV: the role of percolation} 
\coltocauthor{J Alvarez-Mu\~niz, 
P Brogueira, R Concei\c{c}\~ao, J Dias de Deus, 
M.C Esp\'\i rito Santo, M Pimenta}
\Includeart{\CAUT}{\CTIT}{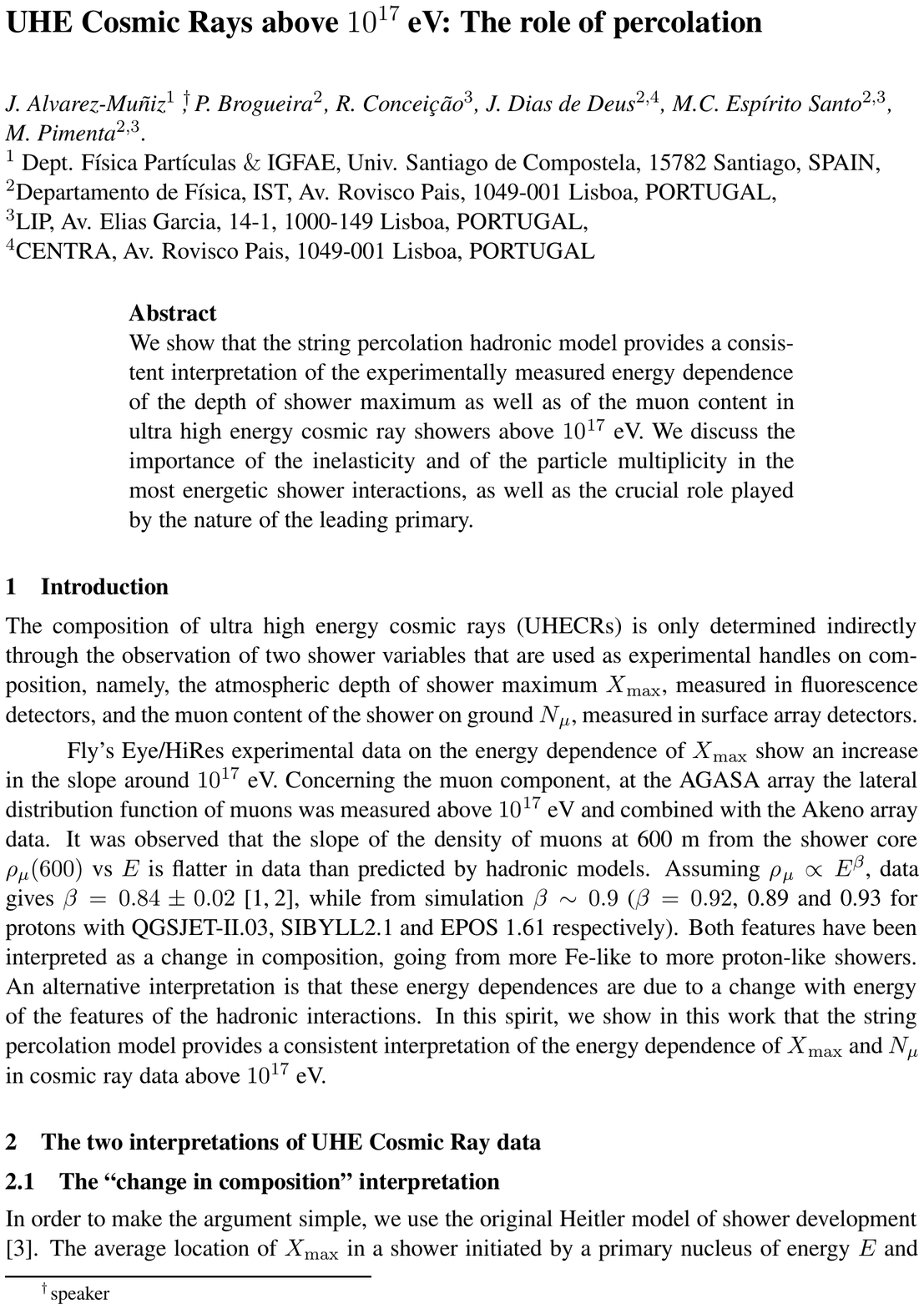} 

\coltoctitle{Extended air shower simulations based on EPOS} 
\coltocauthor{K Werner, T Pierog}
\Includeart{\CAUT}{\CTIT}{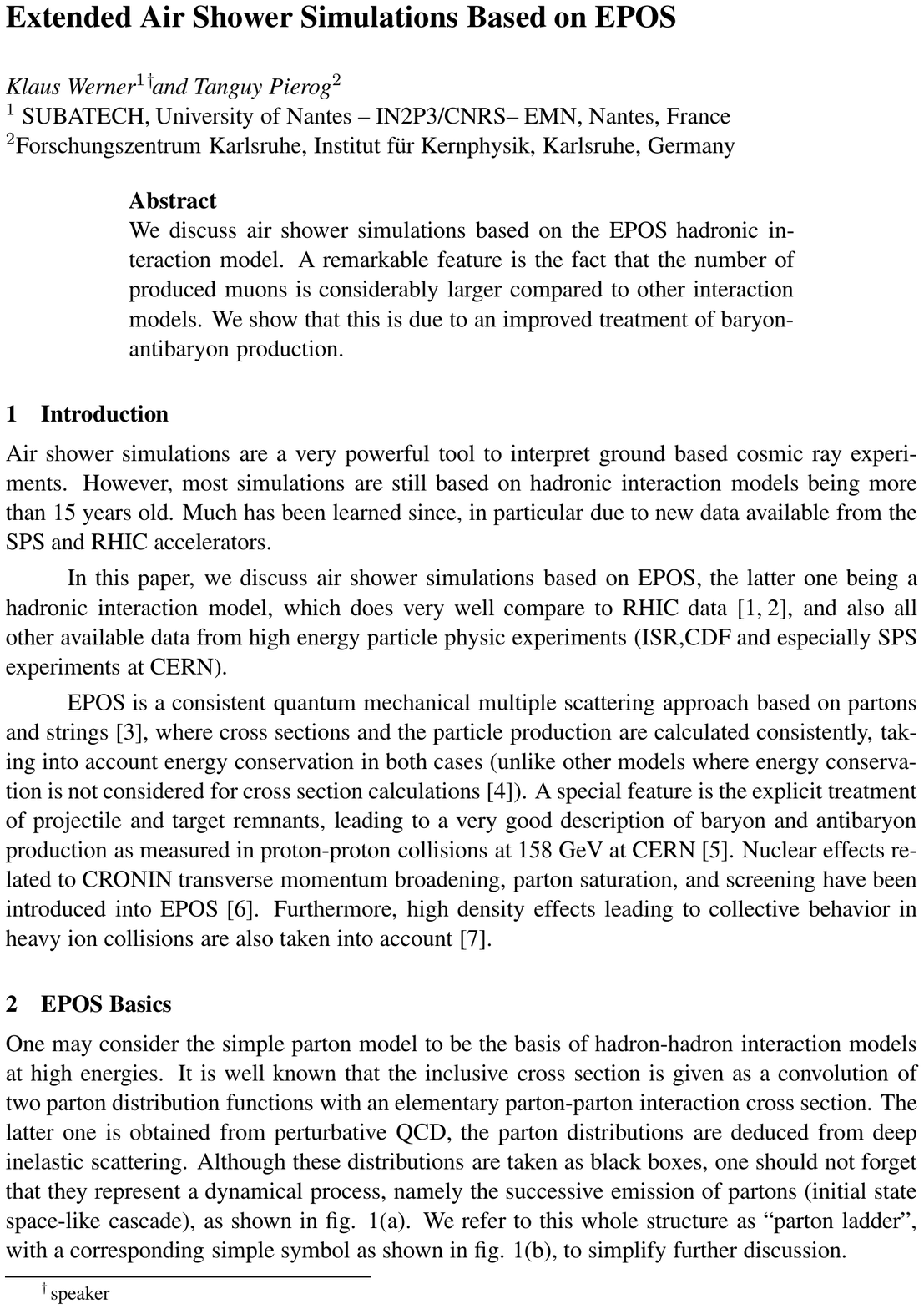} 

\coltoctitle{LHCf:  a LHC Detector for Cosmic Ray Physics} 
\coltocauthor{A Tricomi}
\Includeart{\CAUT}{\CTIT}{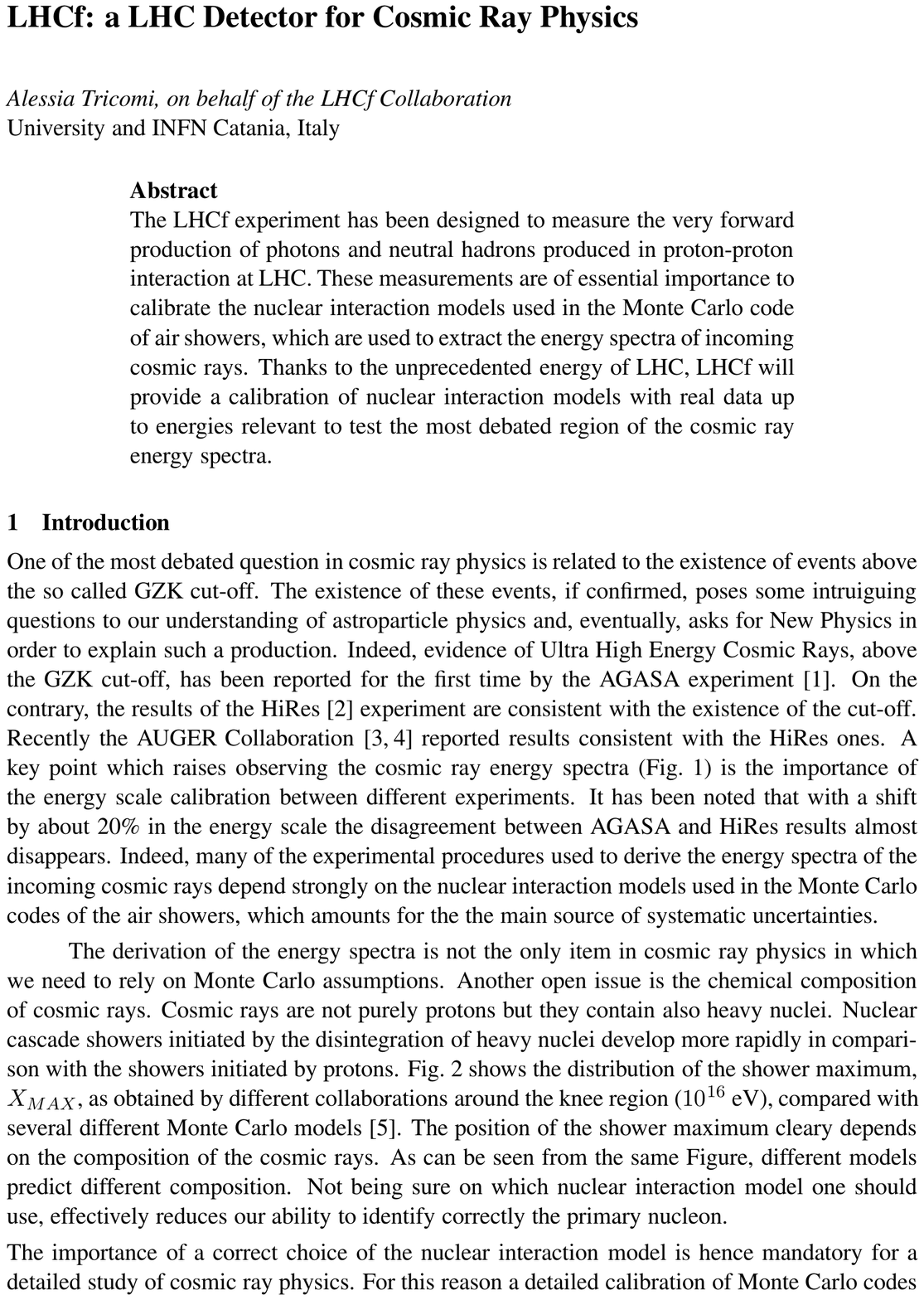} 

\end{papers} 

\CLDP 

\part{Theoretical Developments in High-Energy QCD} 
\label{theory} 
\newpage

\begin{papers} 

\coltoctitle{RHIC physics:  short overview} 
\coltocauthor{A Stasto}
\Includeart{\CAUT}{\CTIT}{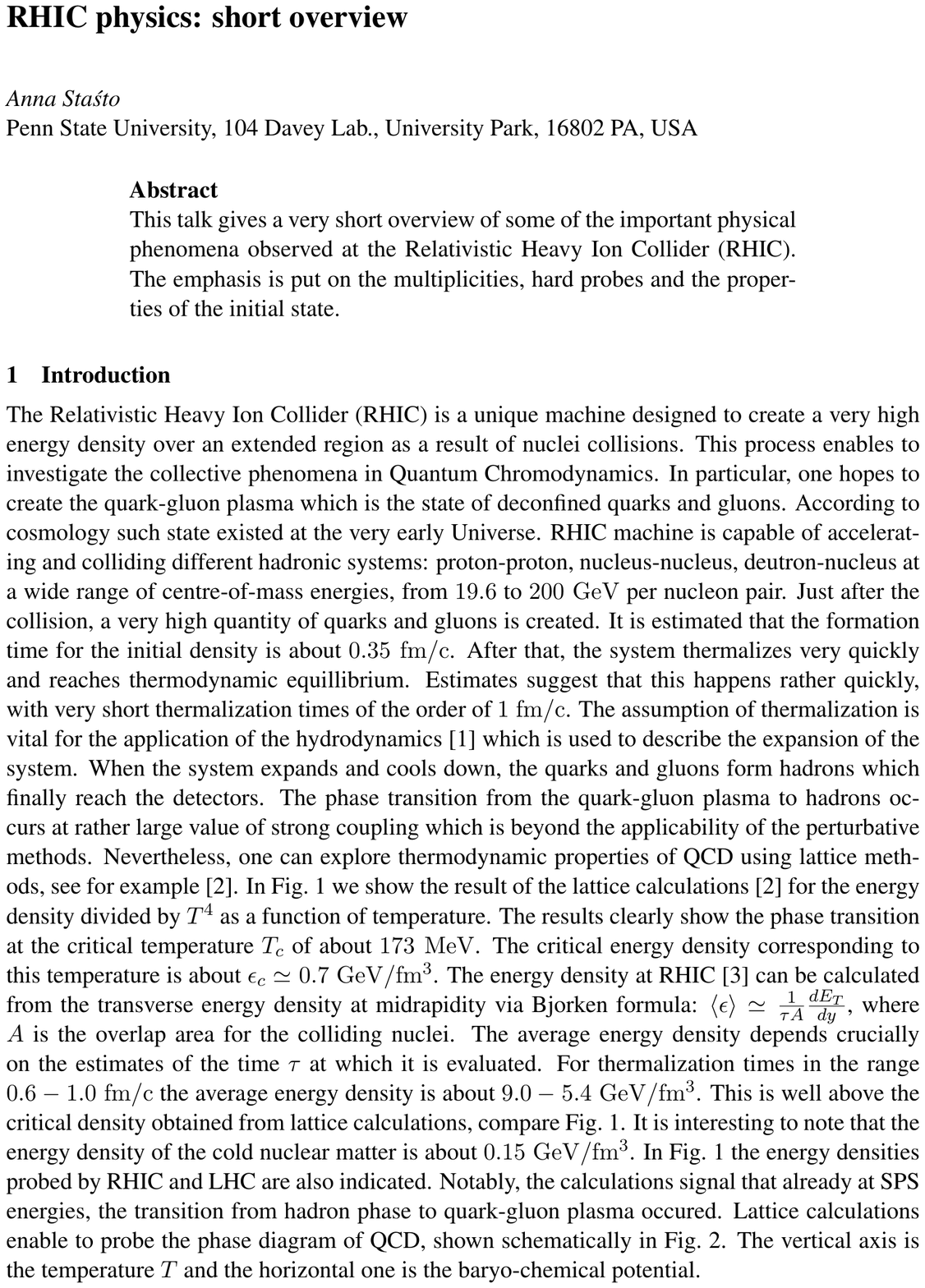} 

\coltoctitle{Nuclear shadowing and collisions of heavy ions} 
\coltocauthor{A Kaidalov}
\Includeart{\CAUT}{\CTIT}{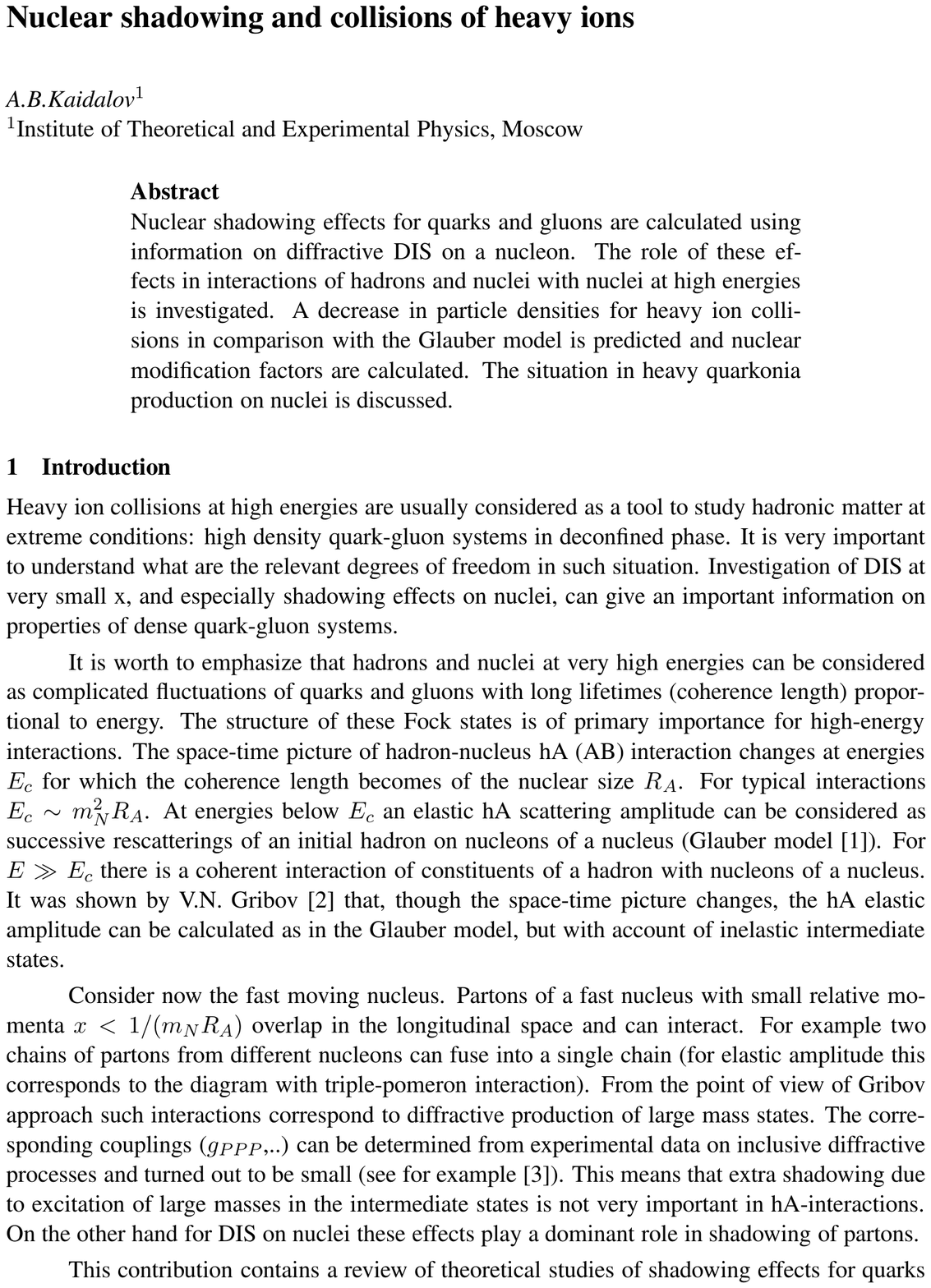} 

\coltoctitle{NLO jet production in $k_T$ factorization} 
\coltocauthor{J Bartels, A Sabio Vera, F Schwennsen}
\Includeart{\CAUT}{\CTIT}{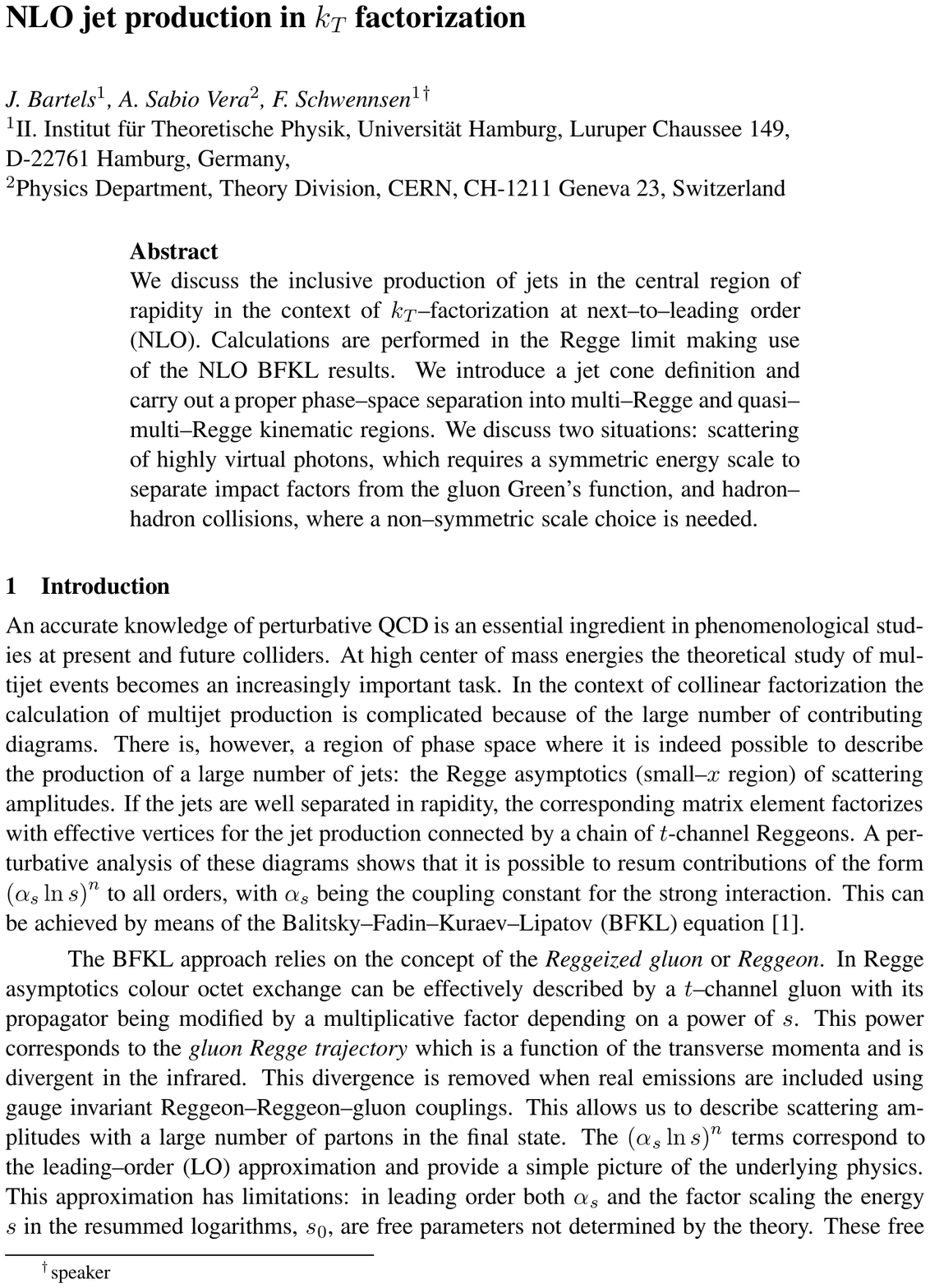} 

\coltoctitle{Connections between high-energy QCD and statistical physics} 
\coltocauthor{S Munier}
\Includeart{\CAUT}{\CTIT}{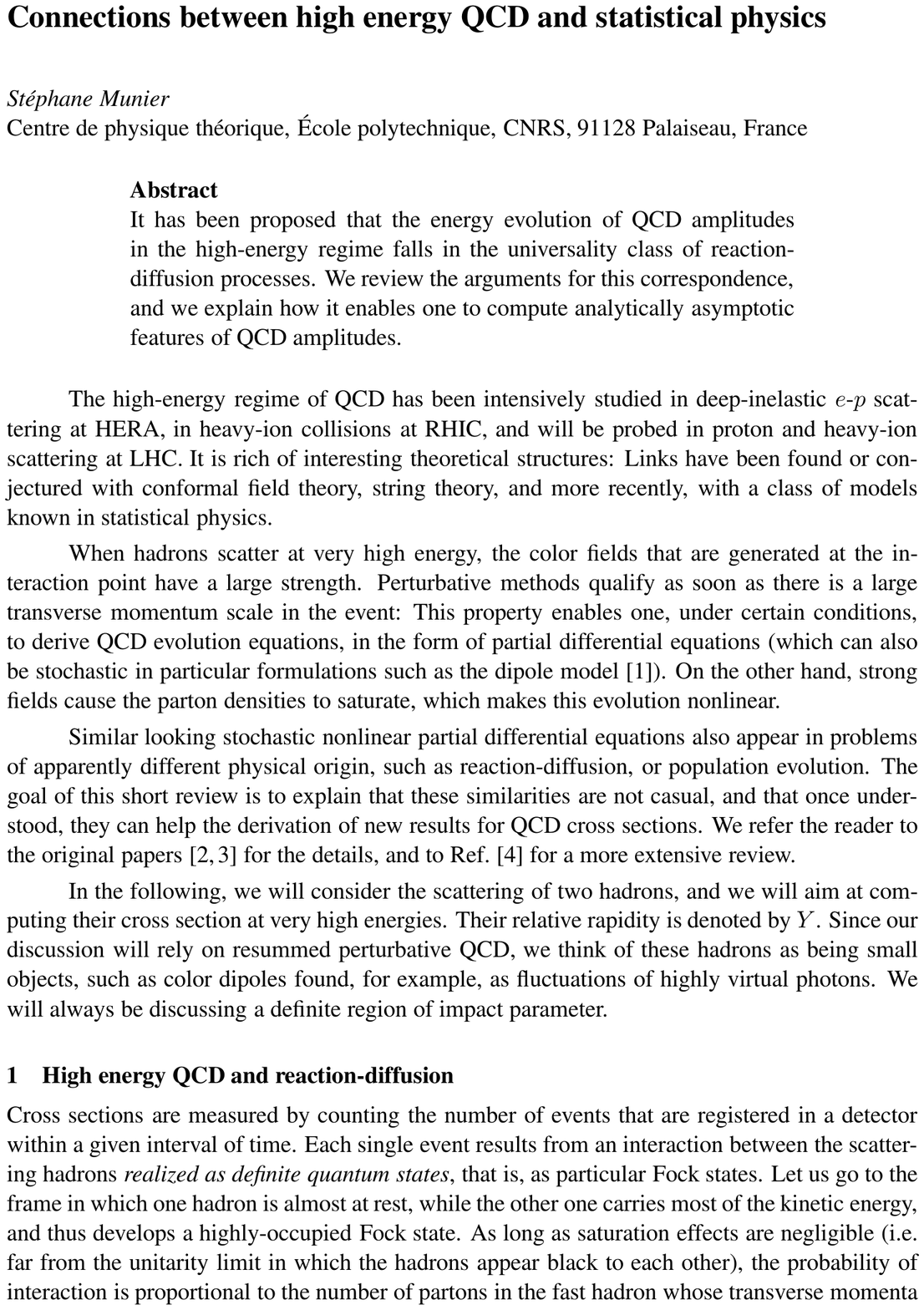} 

\coltoctitle{High energy QCD beyond the mean field approximation} 
\coltocauthor{A Shoshi}
\Includeart{\CAUT}{\CTIT}{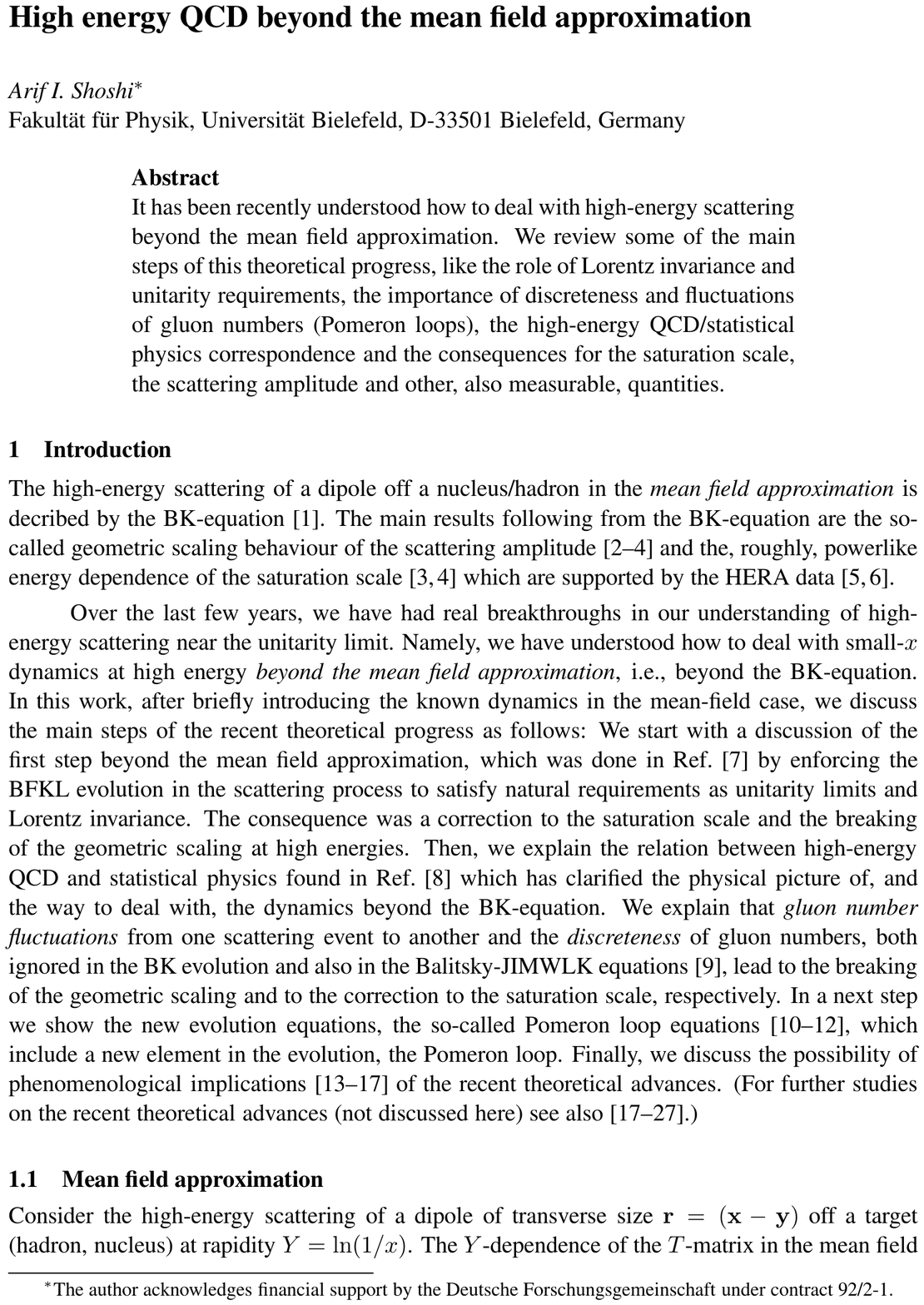} 

\coltoctitle{High-energy scattering and Euclidean-Minkowskian duality} 
\coltocauthor{E Meggiolaro}
\Includeart{\CAUT}{\CTIT}{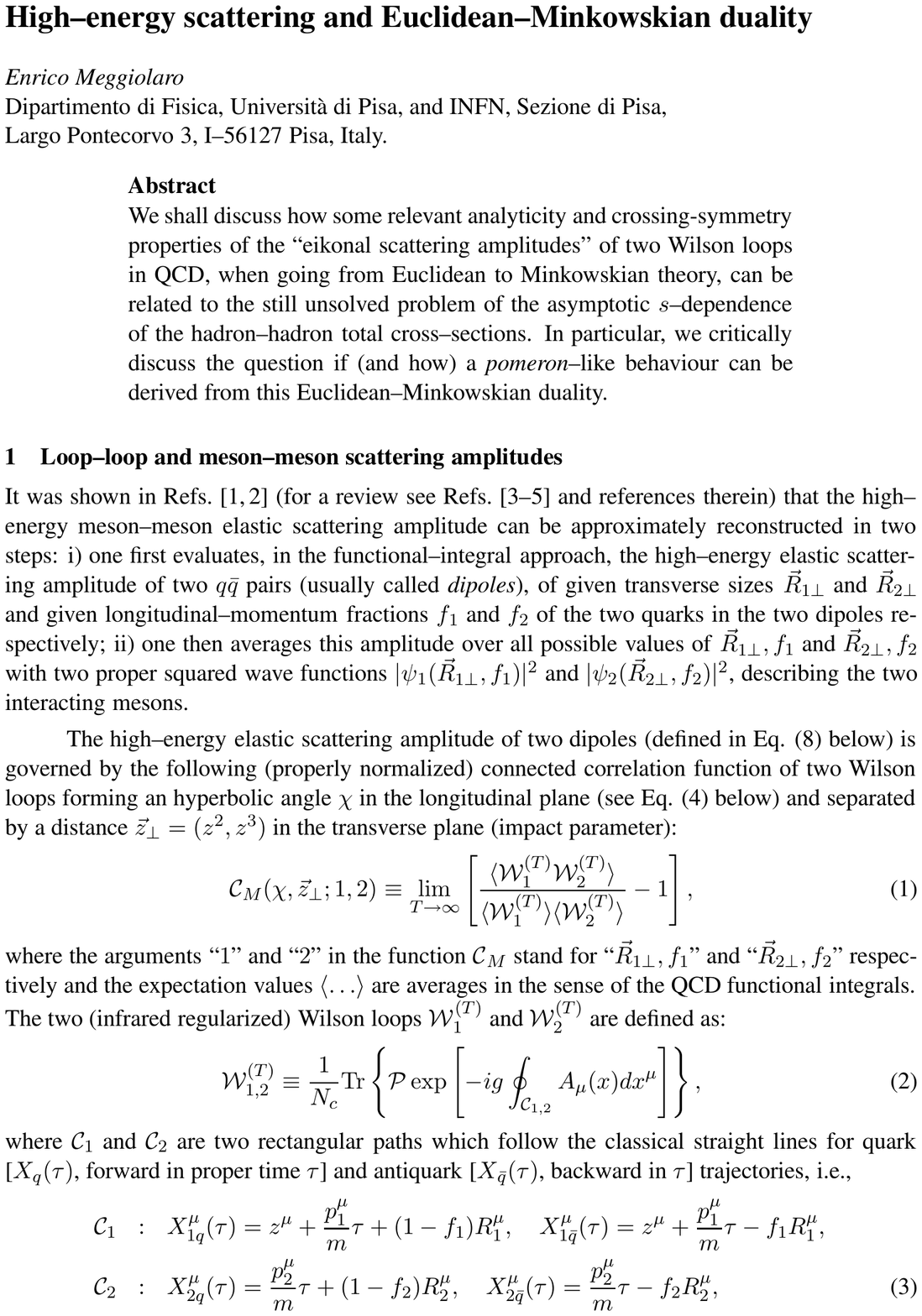} 

\coltoctitle{Insight into new physics with tagged forward protons at the LHC} 
\coltocauthor{V Khoze, A.D Martin, M.G Ryskin}
\Includeart{\CAUT}{\CTIT}{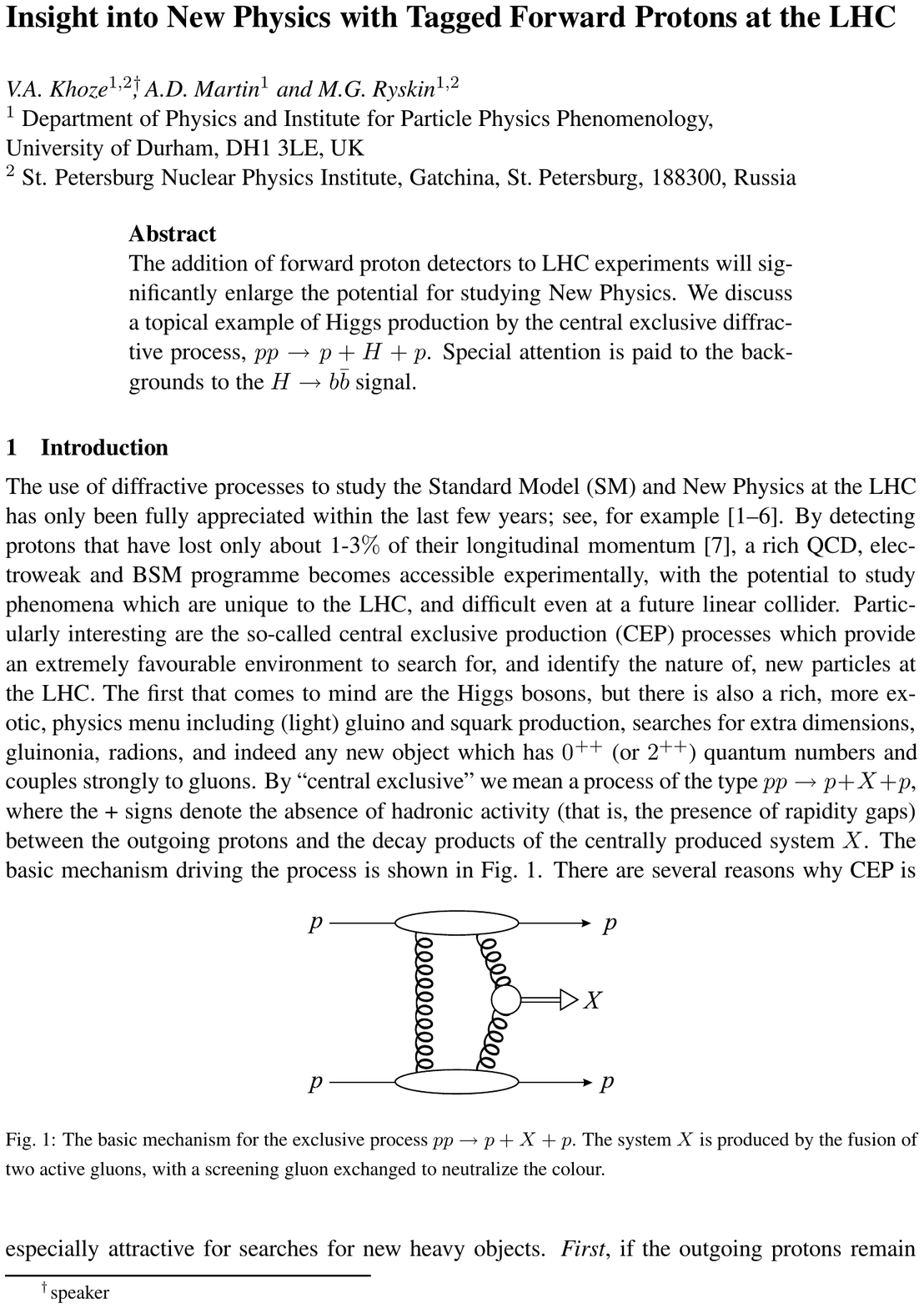} 

\coltoctitle{BFKL at NNLO} 
\coltocauthor{S Marzani, R.D Ball, P  Falgari, S Forte }
\Includeart{\CAUT}{\CTIT}{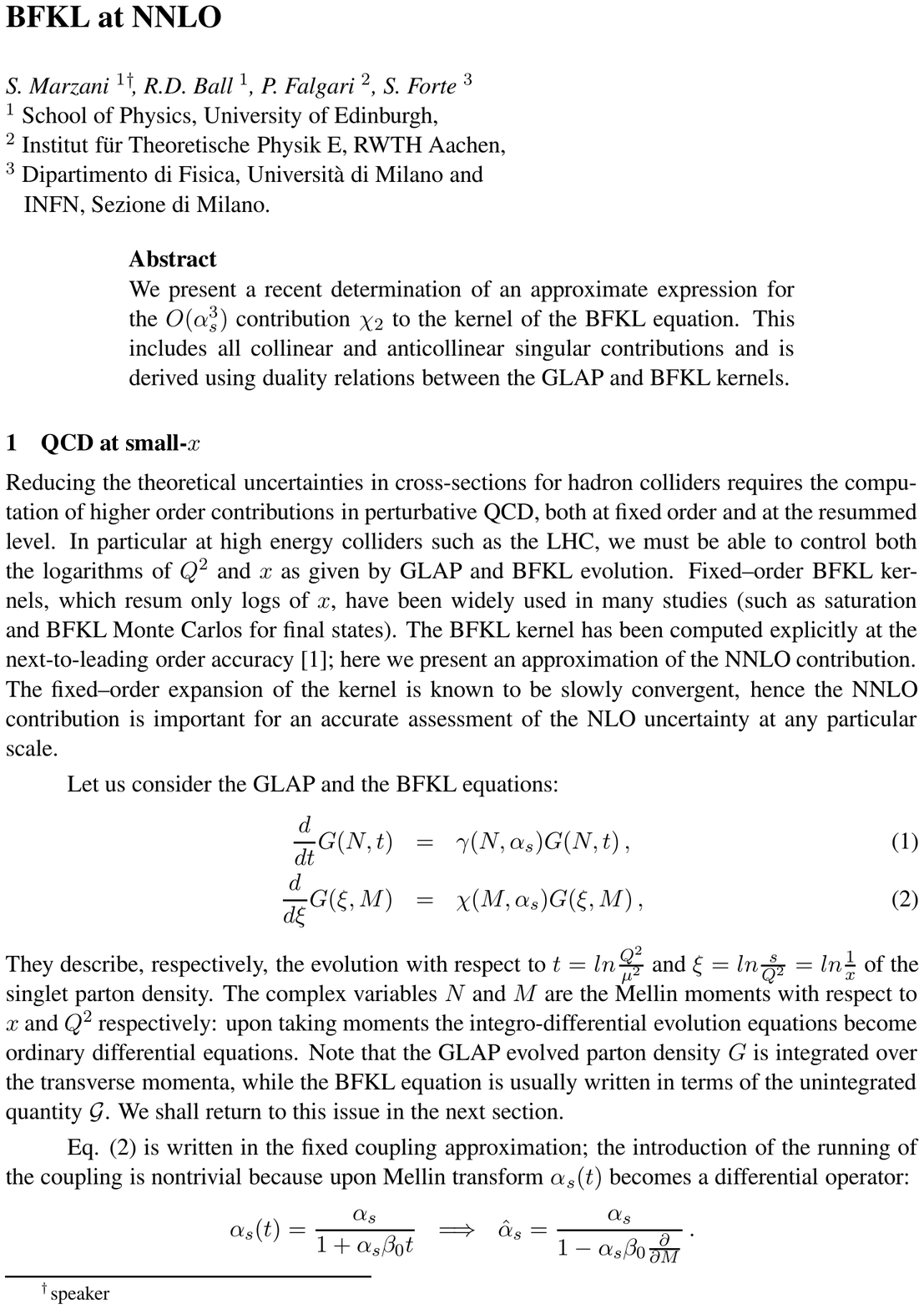} 

\coltoctitle{Unintegrated parton distributions and correlation functions} 
\coltocauthor{A Stasto}
\Includeart{\CAUT}{\CTIT}{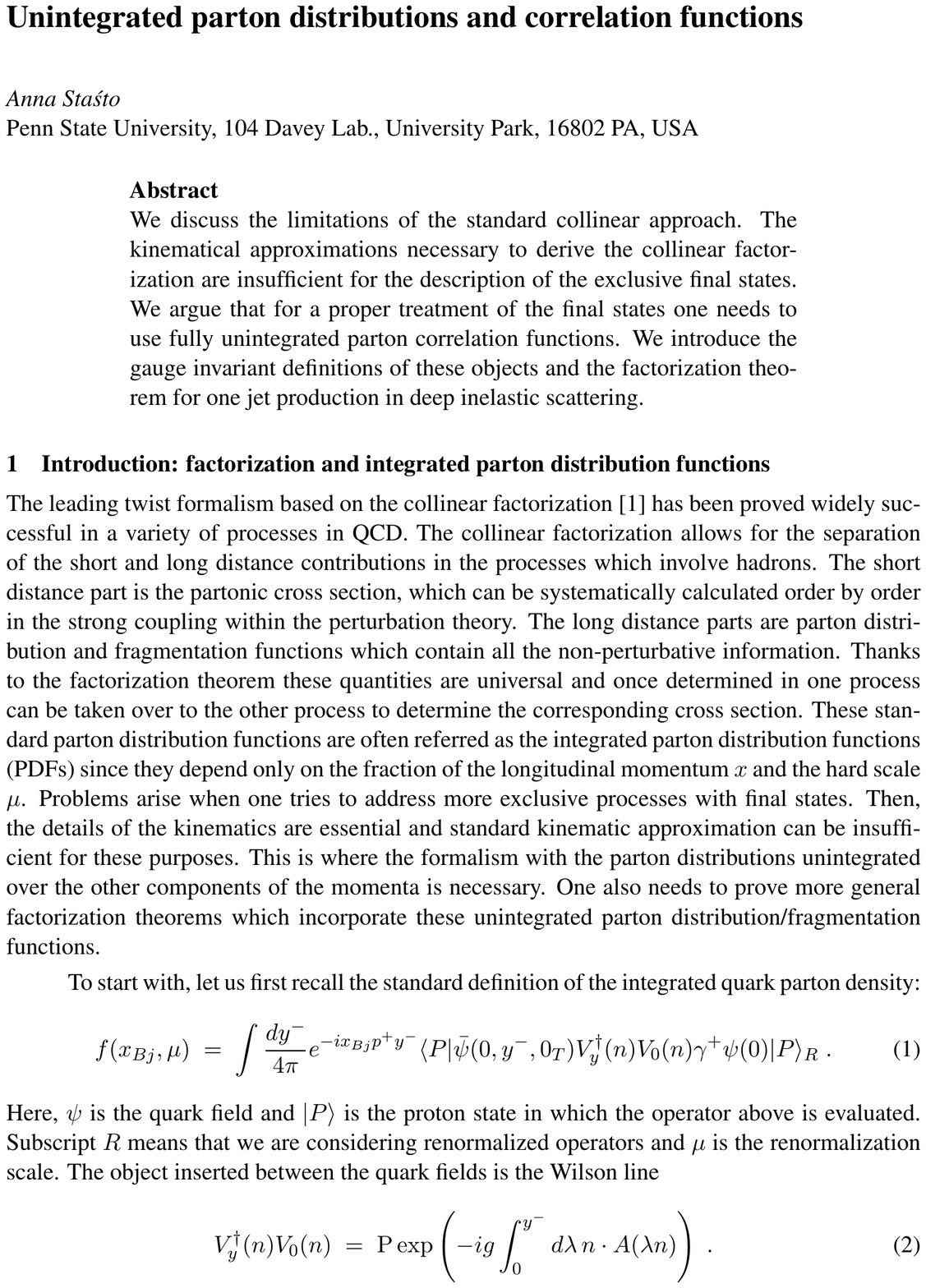} 

\coltoctitle{Nonlinear QCD at high energies } 
\coltocauthor{E Levin}
\Includeart{\CAUT}{\CTIT}{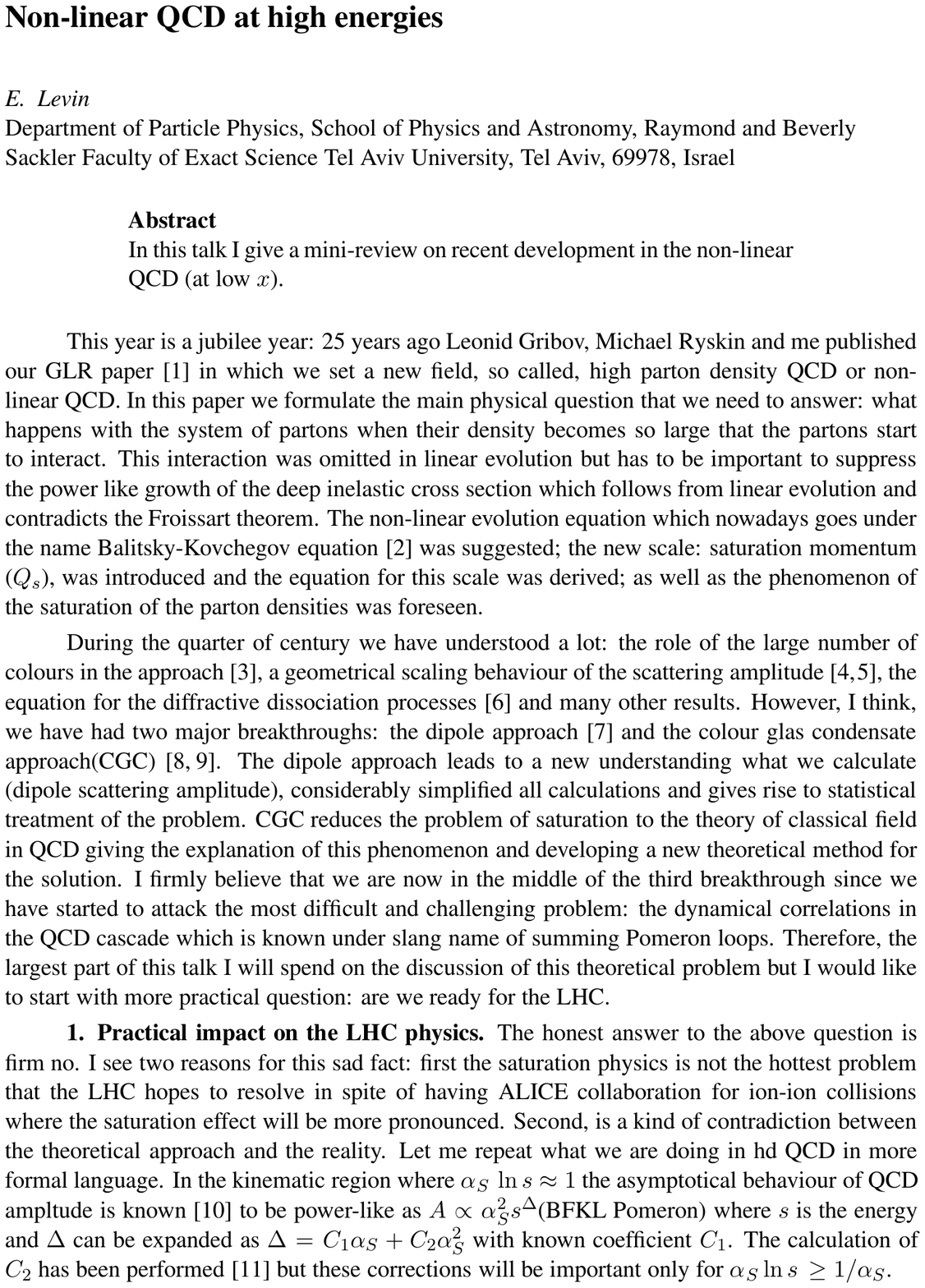} 

\coltoctitle{Small $x$ QCD and multigluon states: a color toy model} 
\coltocauthor{G P Vacca, P. L Iafelice}
\Includeart{\CAUT}{\CTIT}{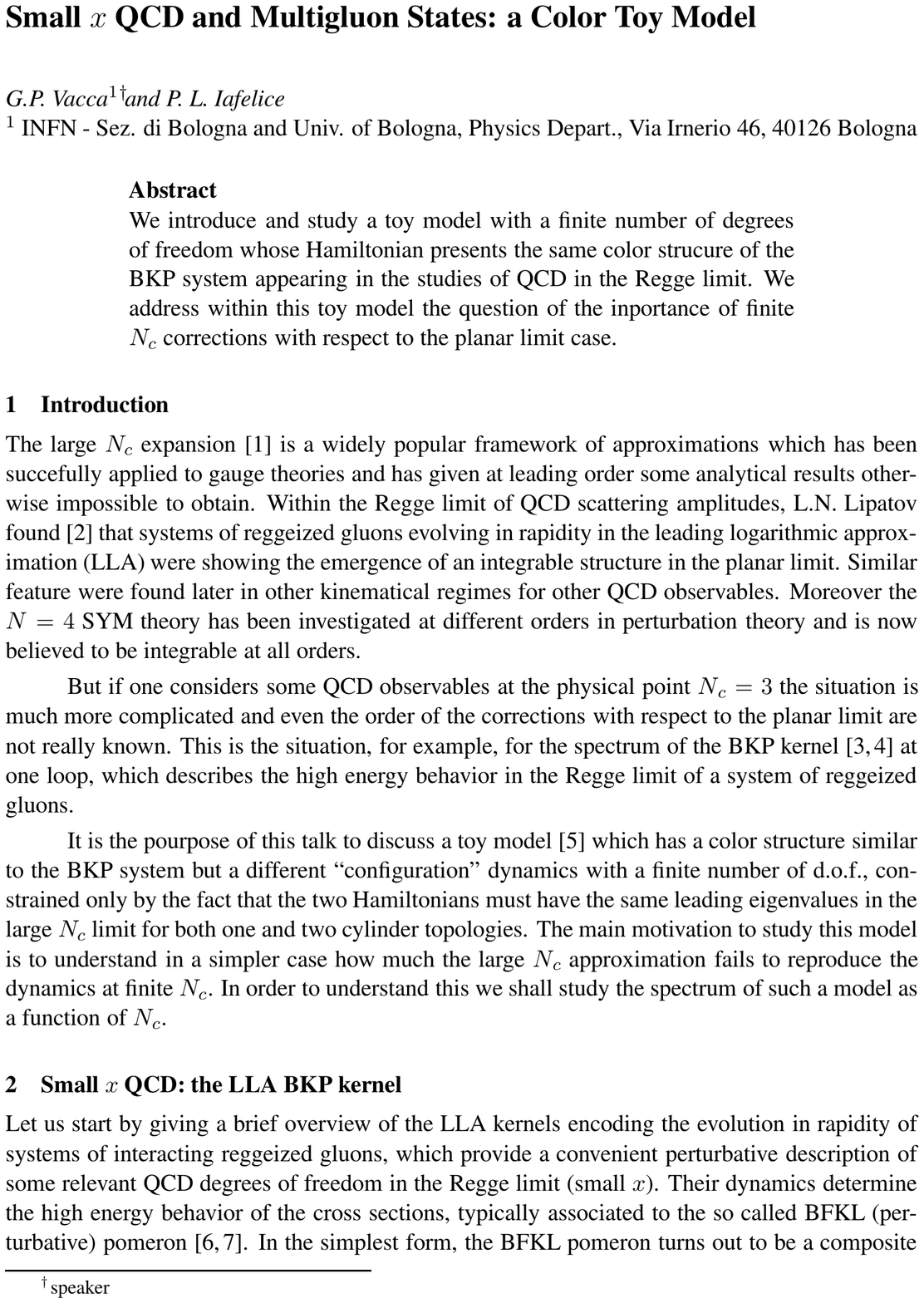} 

\coltoctitle{The Reggeon $\to$ 2 Reggeons $+$ particle vertex in the Lipatov effective action formalism} 
\coltocauthor{M Braun, M.I Vyazovsky}
\Includeart{\CAUT}{\CTIT}{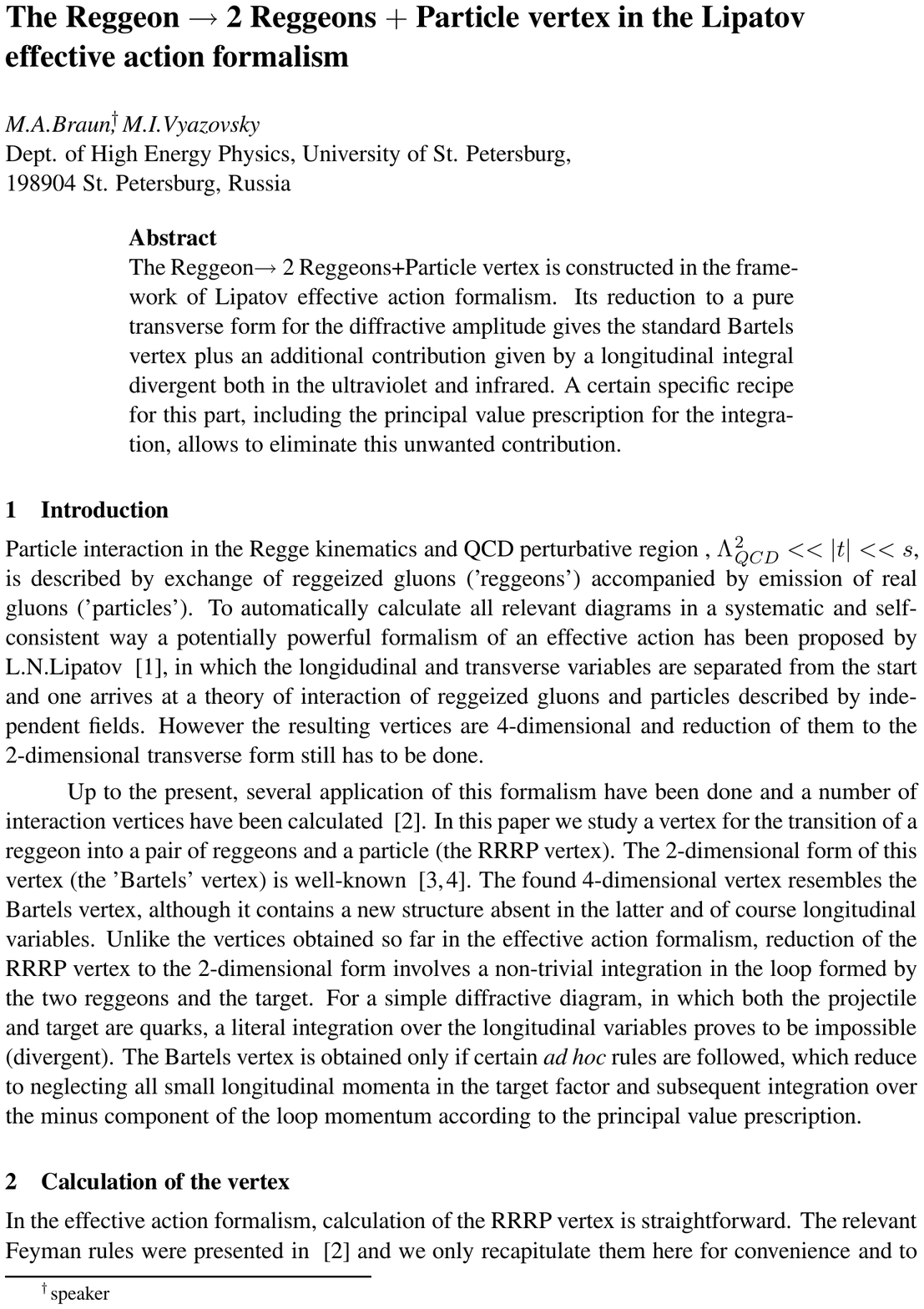} 

\coltoctitle{On the behaviour of $R_{pA}$ at high energy} 
\coltocauthor{M Kozlov, A Shoshi, B.W Xiao}
\Includeart{\CAUT}{\CTIT}{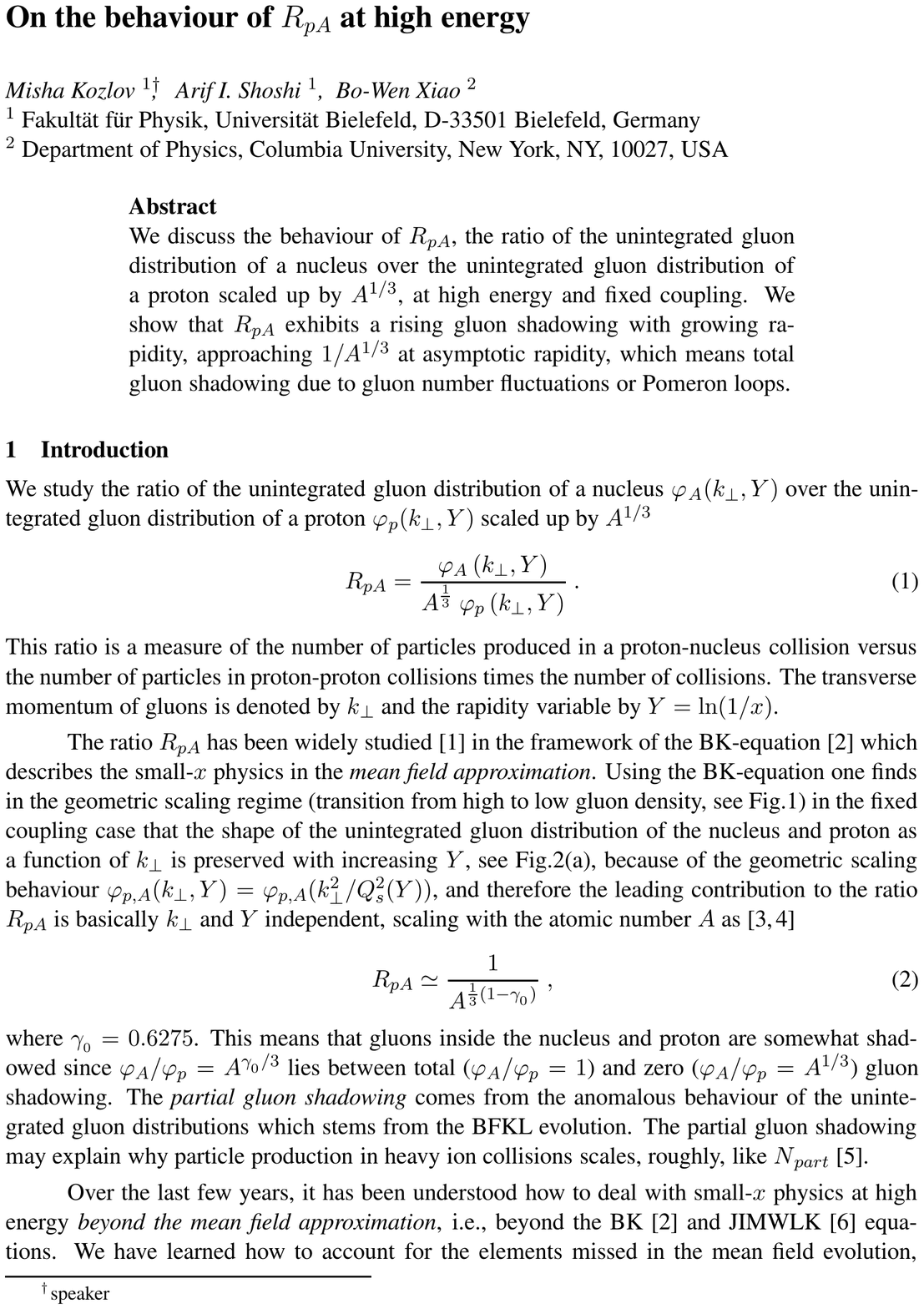} 

\coltoctitle{The coordinate representation of NLO BFKL and the dipole picture} 
\coltocauthor{V Fadin}
\Includeart{\CAUT}{\CTIT}{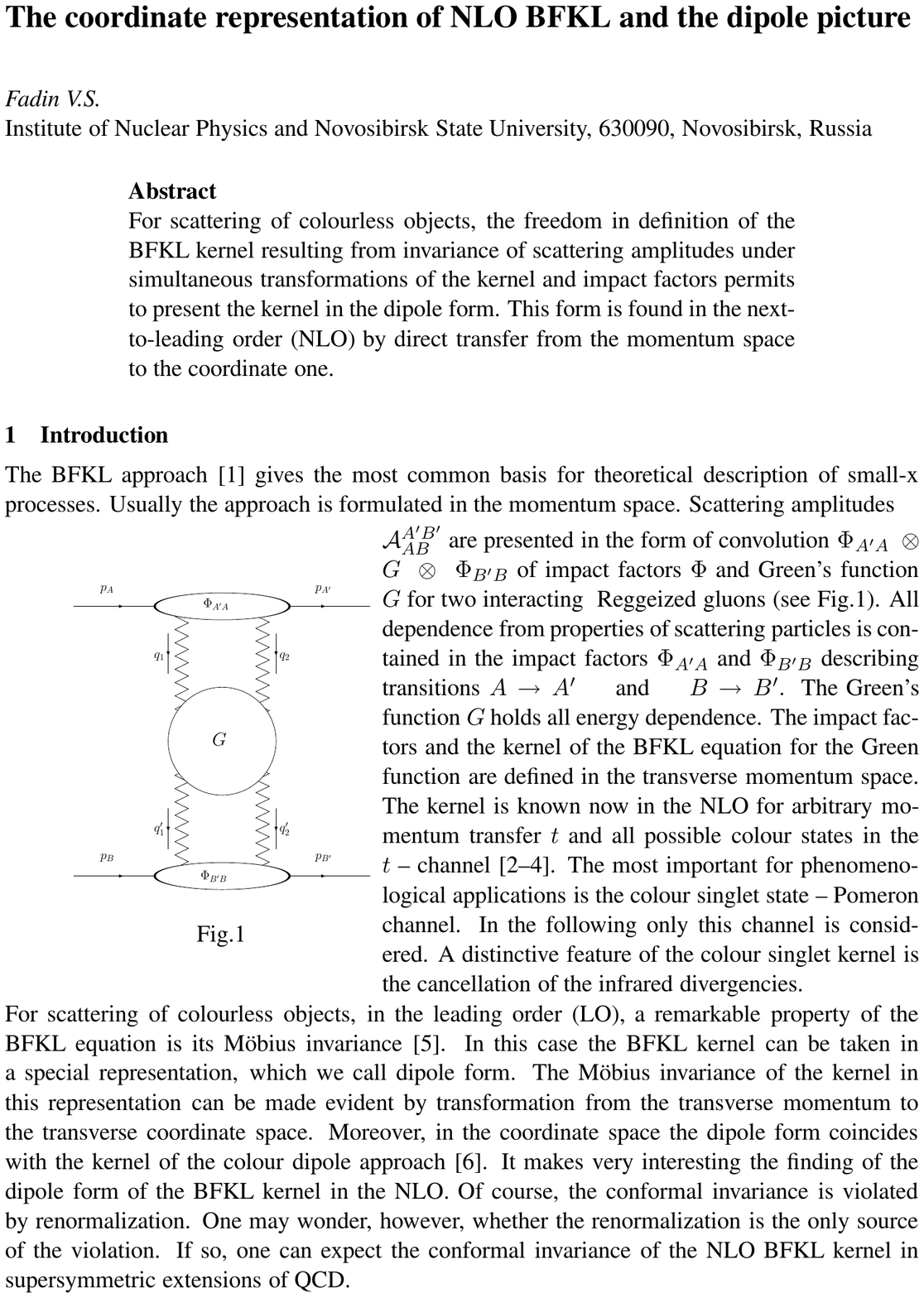} 

\coltoctitle{Angular decorrelations in Mueller-Navelet jets and DIS} 
\coltocauthor{A Sabio Vera, F Schwennsen}
\Includeart{\CAUT}{\CTIT}{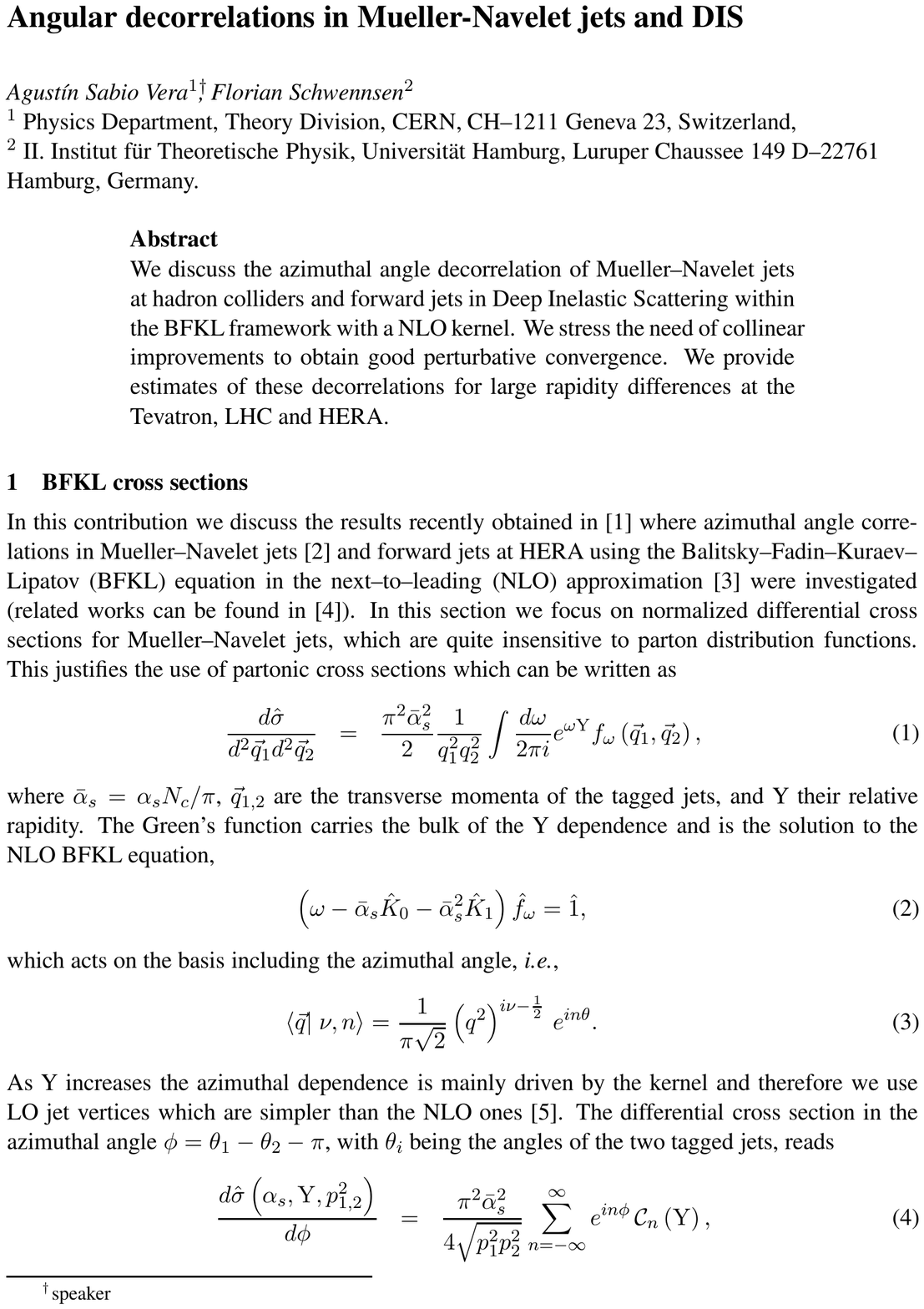} 

\coltoctitle{Breakdown of coherence?} 
\coltocauthor{M Seymour}
\Includeart{\CAUT}{\CTIT}{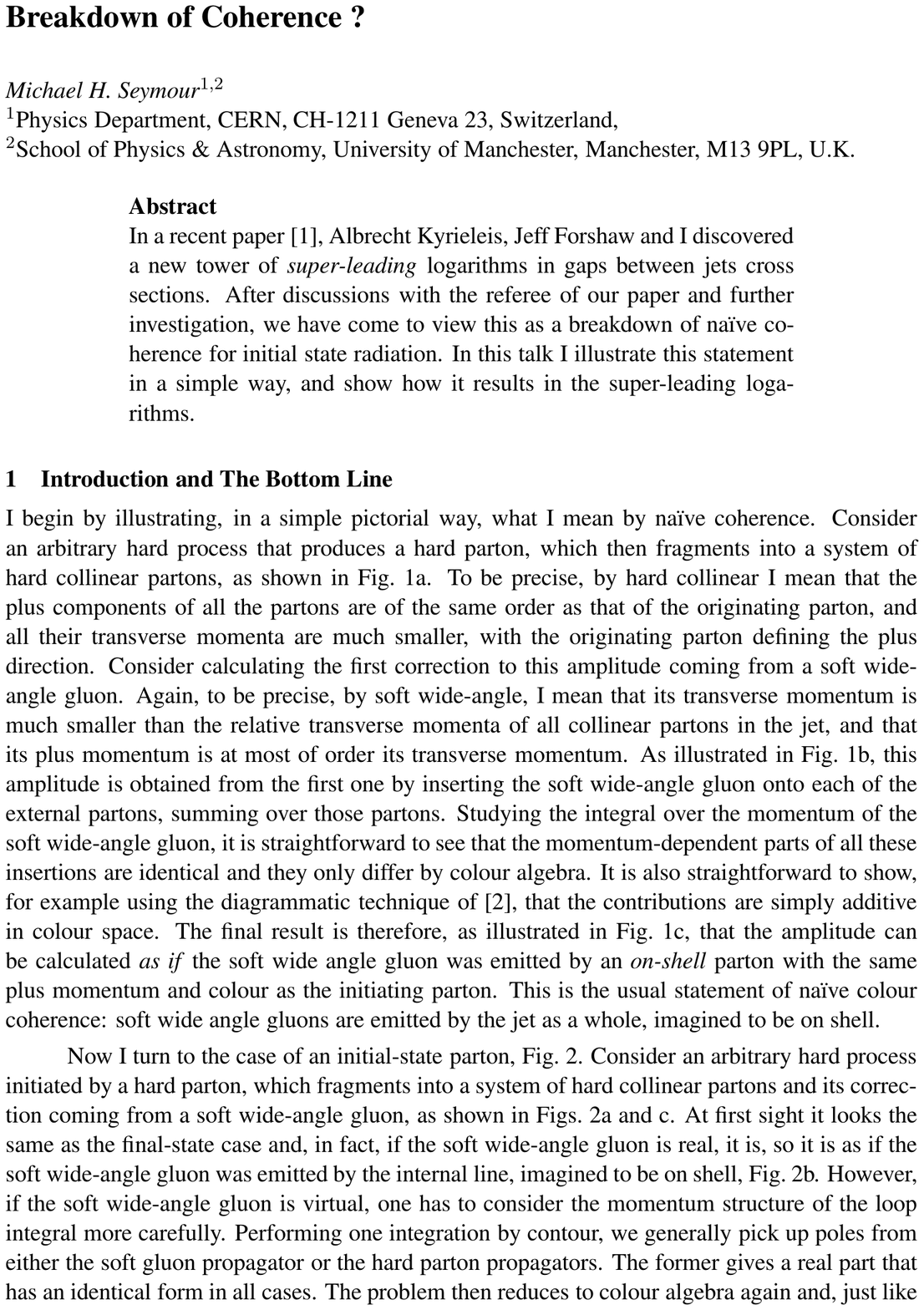} 

\coltoctitle{BFKL equation and anomalous dimensions in $N=4$ SUSY} 
\coltocauthor{L Lipatov}
\Includeart{\CAUT}{\CTIT}{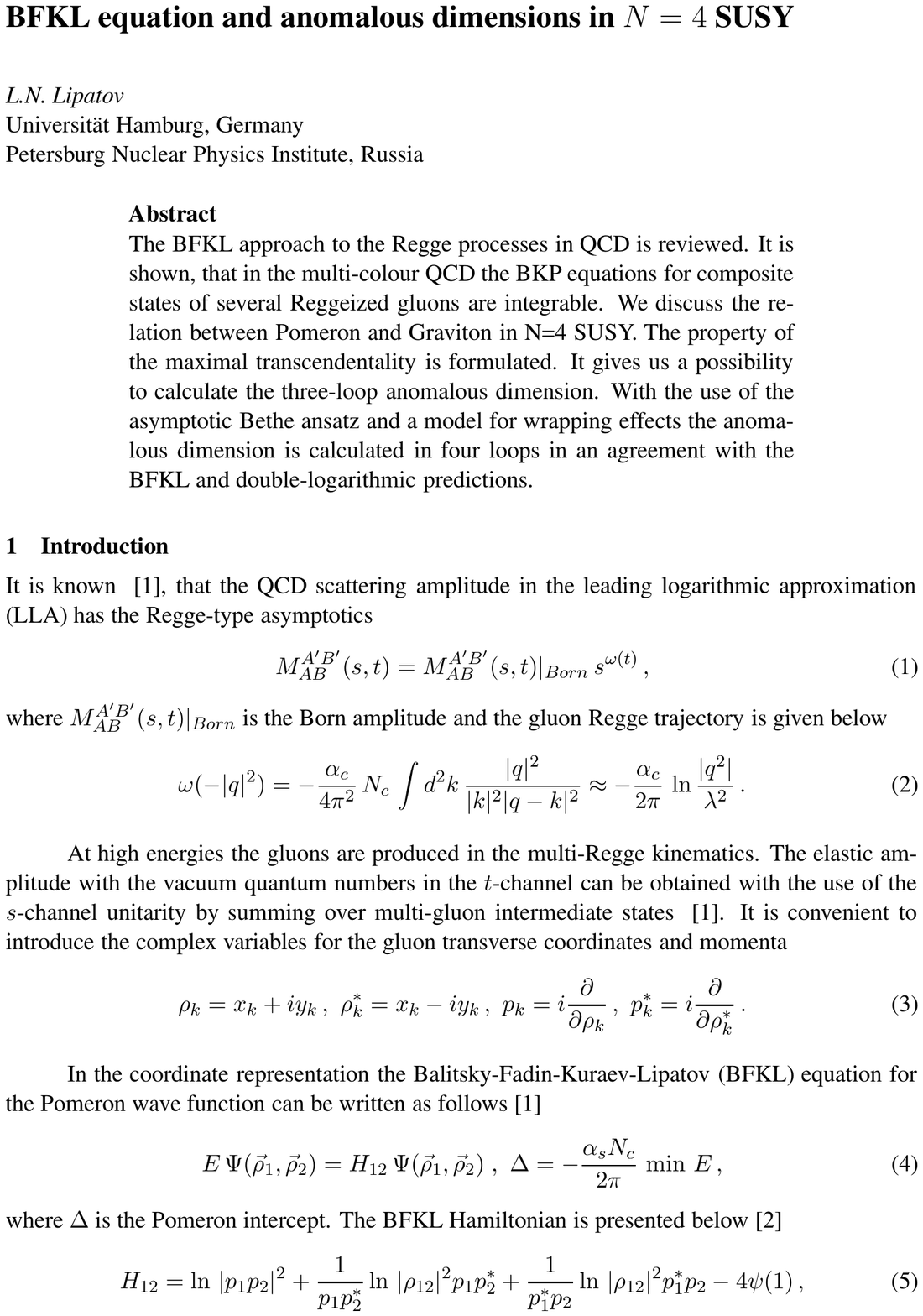} 

\coltoctitle{QCD anomalous dimensions made simple(r) (not received)} 
\coltocauthor{G Marchesini}

\coltoctitle{Gluon saturation and black hole criticality} 
\coltocauthor{L {\'A}lvarez-Gaum{\' e}, 
C G{\' o}mez,  
A Sabio~Vera, A Tavanfar, M.~A V{\'a}zquez-Mozo}
\Includeart{\CAUT}{\CTIT}{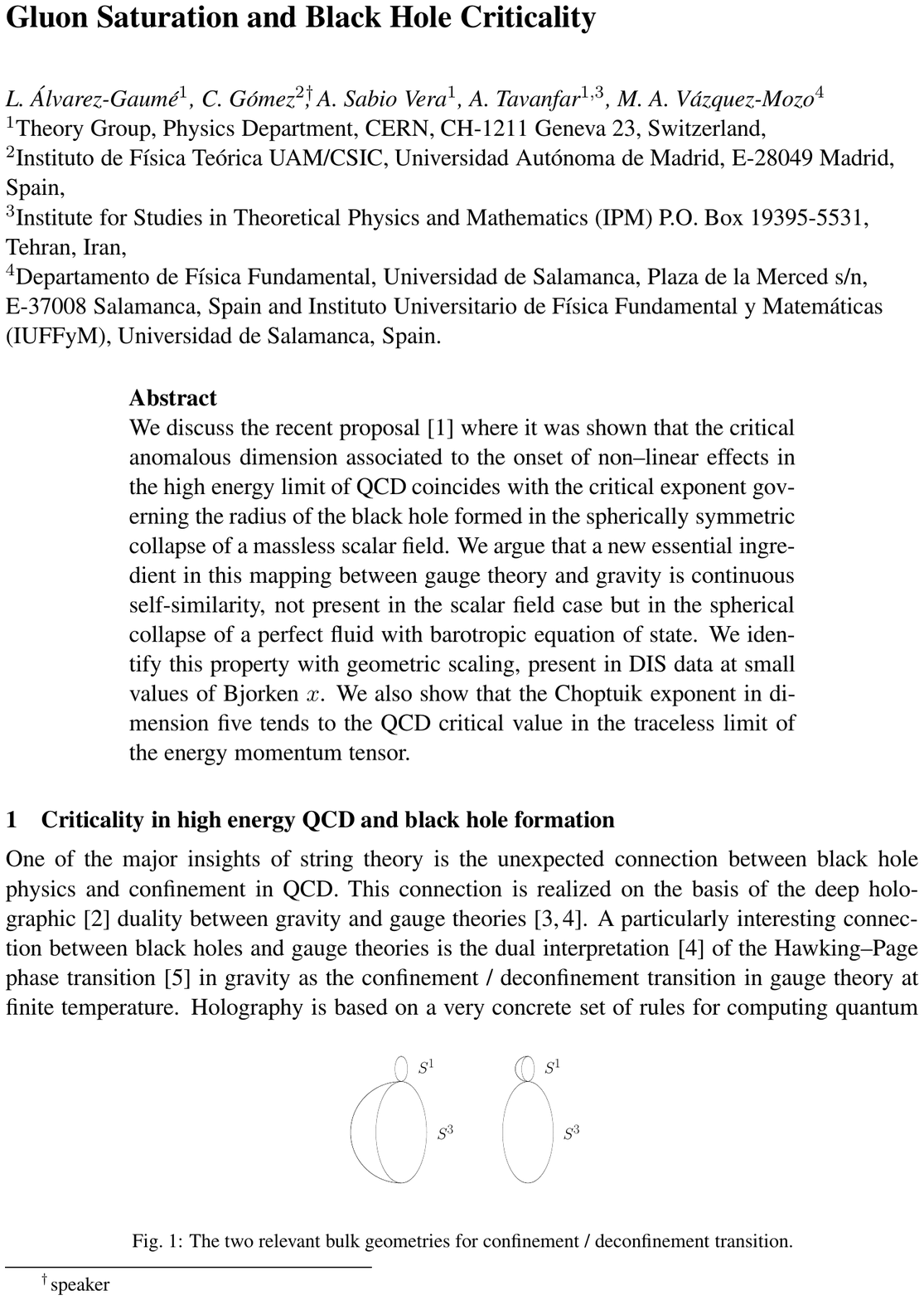}

\end{papers} 

\CLDP 
\part{Summary} 
\label{summaries} 
\newpage

\begin{papers} 

\coltoctitle{Experimental summary} 
\coltocauthor{H Abramowicz}
\Includeart{\CAUT}{\CTIT}{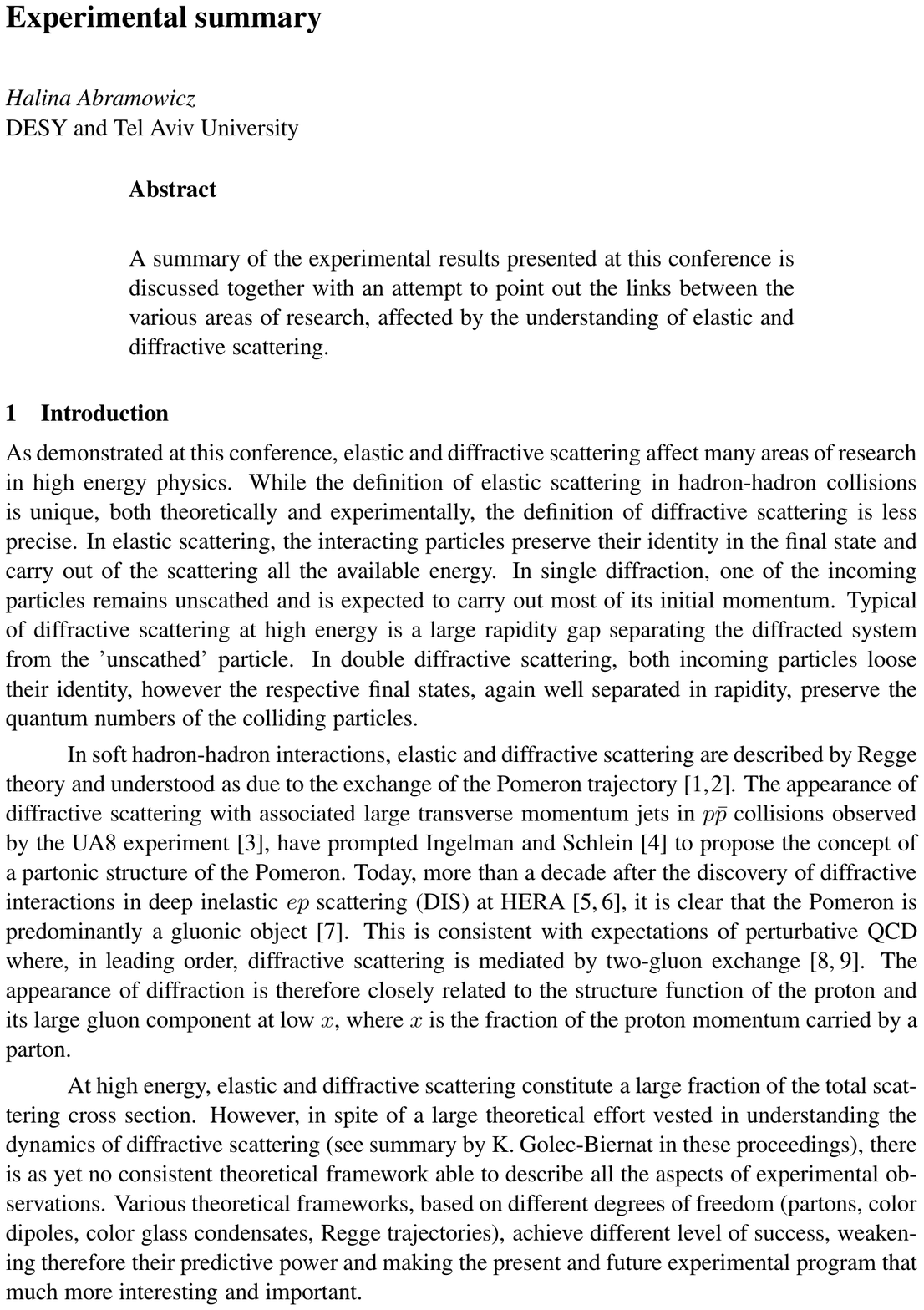} 

\coltoctitle{Theoretical summary} 
\coltocauthor{K Golec-Biernat}
\Includeart{\CAUT}{\CTIT}{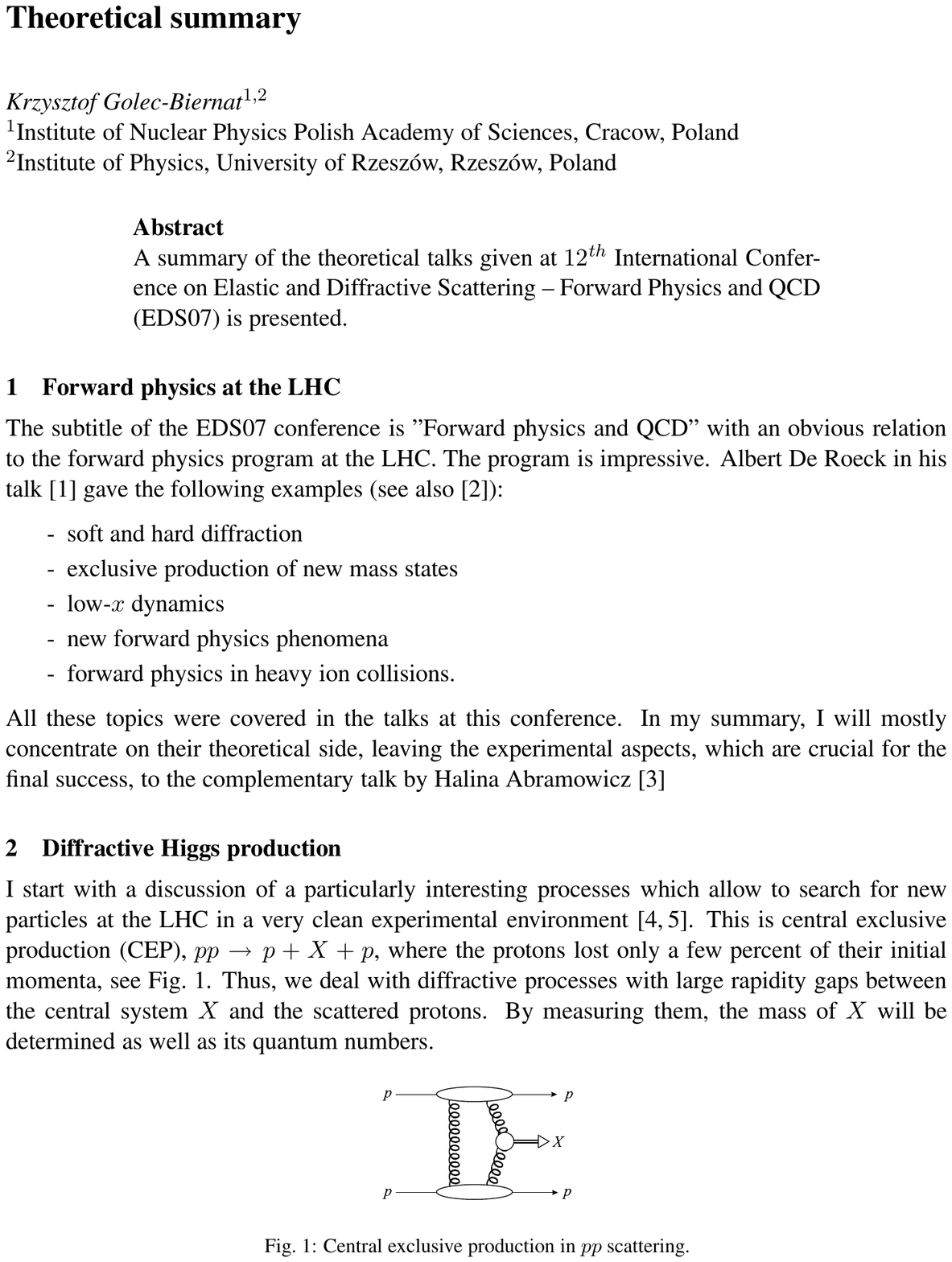} 

\end{papers} 

\end{document}